\documentclass{article}
\usepackage[utf8]{inputenc}
\usepackage{eufrak}
\usepackage[breaklinks=true, pdftex, 
pdfborder={0 0 0}, colorlinks=true, linkcolor=blue, citecolor=blue, urlcolor=blue, 
bookmarks=false, pdfpagelabels=false]{hyperref}
\usepackage{fullpage}
\usepackage{times}
\usepackage{cite,url,doi}
\hypersetup{colorlinks=true, linkcolor=blue!80!black, urlcolor=blue!80!black, citecolor=blue!80!black}
\usepackage{setspace}
\usepackage{amsmath,amssymb,amsfonts}
\usepackage{dsfont}
\usepackage{color}
\usepackage{amsthm}
\usepackage{algorithmic}
\usepackage[noresetcount,vlined,boxed]{algorithm2e}
\usepackage{graphicx,graphics}
\usepackage{textcomp}
\usepackage{xcolor}
\usepackage{bm}
\usepackage{upgreek}
\usepackage{comment}
\usepackage{mathrsfs} % for \mathscr
\usepackage[it,small,belowskip=-15pt]{caption}
\usepackage[top=1in,bottom=1in,left=0.95in,right=0.95in]{geometry} % Set margins
\def\BibTeX{{\rm B\kern-.05em{\sc i\kern-.025em b}\kern-.08em
		T\kern-.1667em\lower.7ex\hbox{E}\kern-.125emX}}
\usepackage[T1]{fontenc}
\usepackage[numbers,compress,sort]{natbib} %,sort

\newcommand{\abs}[1]{\left\lvert#1\right\rvert}
\newcommand{\ket}[1]{\left\lvert\left.#1\right\rangle\right.}
\newcommand{\bra}[1]{\left.\left\langle#1\right.\right\rvert}
\newcommand{\intbracket}[1]{\left[\!\left[#1\right]\!\right]}
\newcommand{\accolade}[1]{\left\lbrace#1\right\rbrace}
\newcommand{\dmax}{\delta_{\max}}
\newcommand{\s}{\bm{s}}
\newcommand{\x}{\bm{x}}

\newcommand{\rd}{\mathbb{R}^d}
\newcommand{\Rq}{\mathbb{R}^q}

\newcommand{\rdq}{\mathbb{R}^{d\times q}}
\newcommand{\rqd}{\mathbb{R}^{q\times d}}
\newcommand{\rqq}{\mathbb{R}^{q\times q}}

\newcommand{\R}{\mathbb{R}}
\usepackage{stackengine}
\newcommand{\ubar}[1]{\stackunder[1.0pt]{$#1$}{\rule{.8ex}{.075ex}}}

\newcommand{\E}[2][]{\operatorname{\mathbb{E}}_{#1}\left[ #2\right]}
 
\newcommand{\mhatk}{\hat{m}_k} 
\newcommand{\mhatkOne}{\hat{m}_{k+1}}

\newcommand{\xk}{\bm{x}_k}
\newcommand{\sk}{\bm{s}_k}
\newcommand{\xkun}{\bm{x}_{k+1}}

\newcommand{\vi}{\bm{v}}
\newcommand{\Qk}{\bm{Q}_k}
\newcommand{\QkOne}{\bm{Q}_{k+1}} 
\newcommand{\QkRandom}{\bm{{\stackunder[0.6pt]{$Q$}{\rule{1.3ex}{.075ex}}}}_k}

\newcommand{\QkTranspose}{\bm{Q}_k^{\top}}
 
\newcommand{\QkRandomTranspose}{\bm{{\stackunder[0.6pt]{$Q$}{\rule{1.3ex}{.075ex}}}}_k^{\top}}
\newcommand{\Uk}{\bm{U}_k} 
\newcommand{\ftheta}{f_{\ubar{\theta}}}

\newcommand{\gkhat}{\bm{\hat{g}_k}}
\newcommand{\dk}{\delta_k}
\newcommand{\Hkhat}{\bm{\hat{H}}_k}

\newcommand{\normii}[1]{{\left\lVert#1\right\rVert}_{2}}

\newcommand{\fok}{f_k^{{\bm{0}}}}
\newcommand{\epsok}{\tilde{\varepsilon}_k}
\newcommand{\Fok}{{\stackunder[0.6pt]{$f$}{\rule{.8ex}{.075ex}}}_k^{{\bm{0}}}}

\newcommand{\fsk}{f_k^{\s}}

\newcommand{\Fsk}{{\stackunder[0.6pt]{$f$}{\rule{.8ex}{.075ex}}}_k^{\s}}

\newcommand{\dkun}{{\delta_{k+1}}}

\newcommand{\Q}{\bm{Q}}

\newcommand{\aok}{a^0_k}
\newcommand{\aokOne}{a^0_{k+1}}
\newcommand{\akOne}{\bm{a}_{k+1}}
\newcommand{\ak}{\bm{a}_k}
\newcommand{\akTransp}{\bm{a}_k^\top}
\newcommand{\akOneTransp}{\bm{a}_{k+1}^\top}
\newcommand{\siik}{\bm{s}^i_k}
\newcommand{\siikOne}{\bm{s}^i_{k+1}}
\newcommand{\szerok}{\bm{s}^0_k} 
\newcommand{\sonek}{\bm{s}^1_k} 
\newcommand{\stwok}{\bm{s}^2_k} 
 
\newcommand{\squk}{\bm{s}^q_k}

\newcommand{\SSk}{\mathbb{S}_k} 
\newcommand{\SSkOne}{\mathbb{S}_{k+1}}
\newcommand{\SSkOneqplus}{\mathbb{S}_{k+1}^{q+1}}

\newcommand{\Lk}{\bm{L}_k}

\newcommand{\ef}{\varepsilon_f}
\newcommand{\accoladekinN}[1]{{\left\lbrace#1\right\rbrace}_{k\in\mathbb{N}}}

\newcommand{\bl}{\color{black}}

\newcommand{\prob}[1]{\mathbb{P}\left(#1\right)}
\newcommand{\Qrandom}{\bm{{\stackunder[0.6pt]{$Q$}{\rule{1.3ex}{.075ex}}}}}
\newcommand{\QrandomTranspose}{\bm{{\stackunder[0.6pt]{$Q$}{\rule{1.3ex}{.075ex}}}}^{\top}}

\newcommand{\Urandom}{\bm{{\stackunder[0.7pt]{$U$}{\rule{1.1ex}{.075ex}}}}}

\newcommand{\qtd}{q \times d}

\DeclareMathOperator*{\argmin}{arg\,min}

\newtheoremstyle{maxcutstyle}  % Custom style
{}                           % Space above
{}                           % Space below
{\itshape}                   % Body font
{}                           % Indent amount
{\bfseries}                  % Theorem head font
{.}                          % Punctuation after theorem head
{ }                          % Space after theorem head
{}                           % Theorem head spec (can be left empty for default)

\theoremstyle{maxcutstyle}
\newtheorem*{maxcut}{MaxCut}
\newtheorem{definition}{Definition}[section]
\newtheorem{theorem}{Theorem}[section]
\newtheorem{lemma}{Lemma}[section]

\usepackage{orcidlink}
\title{A Noise-Aware Scalable Subspace Classical Optimizer for the Quantum Approximate Optimization Algorithm\thanks{This material was based upon work supported by the U.S.\ Department of Energy, Office of Science, Office of Advanced Scientific Computing Research Applied Mathematics and ARQC programs under Contract Nos.\ DE-AC02-06CH11357 and DE-AC02-05CH11231.}}
\author{
	\href{mailto:kdzahini@anl.gov}{K. J. Dzahini \orcidlink{0000-0002-1515-4251}}\thanks{Argonne National Laboratory, 9700 S. Cass Avenue, Lemont, IL 60439, USA. %(\href{https://www.anl.gov/profile/kwassi-joseph-dzahini}{www.anl.gov/profile/kwassi-joseph-dzahini}).
	}
	\and
	\href{mailto:jmlarson@anl.gov}{J. M. Larson \orcidlink{0000-0001-9924-2082}}\footnotemark[2]
	\and 
	\href{mailto:mmenickelly@anl.gov}{M. Menickelly \orcidlink{0000-0002-2023-0837}}\footnotemark[2]
	\and
	\href{mailto:wild@lbl.gov}{S. M. Wild \orcidlink{0000-0002-6099-2772}}\thanks{Lawrence Berkeley National Laboratory, 1 Cyclotron Road, Berkeley, CA 94720, USA.% ( \href{https://wildsm.github.io/}{wildsm.github.io/}).
	}
}
\date{\today}
\begin{document}

%\linenumbers % To make it easier for reviewers

\maketitle

%\vspace*{-0.5cm}

\noindent
{\bf Abstract:} We introduce ANASTAARS, a noise-aware scalable classical optimizer for variational quantum algorithms such as the quantum approximate optimization algorithm (QAOA). ANASTAARS leverages adaptive random subspace strategies to efficiently optimize the ansatz parameters of a QAOA circuit, in an effort to address challenges posed by a potentially large number of QAOA layers. ANASTAARS iteratively constructs random interpolation models within low-dimensional affine subspaces defined via Johnson--Lindenstrauss transforms. This adaptive strategy allows the selective reuse of previously acquired measurements, significantly reducing computational costs associated with shot acquisition. Furthermore, to robustly handle noisy measurements, ANASTAARS incorporates noise-aware optimization techniques by estimating noise magnitude and adjusts trust-region steps accordingly. Numerical experiments demonstrate the practical scalability of the proposed method for near-term quantum computing applications.

$ $

%$ $\\
%{\bf Mathematics Subject Classification} 90C15 $\cdot$ 90C30 $\cdot$ 90C56 %$\cdot$ MSC4

%\tableofcontents
\clearpage

\clearpage
\section{Introduction}\label{secI}
Remarkable progress has been made in the design of noisy intermediate-scale quantum (NISQ) devices while large-scale, fault-tolerant quantum computers are still far from being available.
%\mmnote{That could be a controversial statement among believers in the early-fault-tolerant era.} \swnote{Matt - - which part "remarkable" or "far"??} \mmnote{Far. I've talked to people in MACH-Q who think we're in the early stages. I don't have an opinion, just suggesting caution.} 
%SW: I'm adding an additional qualifier: "large-scale"; no one will object.
%MM: Perfect. 
Interest in developing practical algorithms designed for NISQ devices is on the rise.
%~\cite{BleBraQuantReview2024}. 
Variational quantum algorithms (VQAs)~\cite{BhaCerVQA2022,BraLaRVQA2023,CerArrBabVQA2021,McCRomBabVQA2016} were introduced to exploit the capabilities of current quantum systems using a hybrid quantum-classical optimization process. 
In a VQA, the hybrid cycle involves executing a parameterized circuit on a quantum computer and using an optimizer on a classical machine to update the circuit parameters by minimizing a cost function based on the quantum circuit outputs. 
This approach allows VQAs to benefit from the use of shallow quantum circuits, making them less vulnerable to noise in NISQ devices~\cite{BleBraQuantReview2024}. 
VQAs exhibit promise for quantum simulations aimed at finding the ground state energies of complex molecules, as well as for solving difficult binary optimization problems such as 
%low autocorrelated binary sequences~[..] problems and 
MaxCut variants~\cite{ShaLotLarMaxCut2023,WanHadMaxCut2018}.
% VQAs have so far been applied in areas such as quantum chemistry simulations, machine learning, and optimization.

As one of the most promising VQAs, the quantum approximate optimization algorithm (QAOA)~\cite{QAOAFarGol2014} has attracted significant interest and has numerous and far-reaching applications. 
As emphasized in~\cite{BleBraQuantReview2024}, in addition to MaxCut problems, QAOA is well suited for finding good approximate solutions to various optimization problems, %including Maximum Independent Set (MIS), Binary Linear Least Squares (BLLS), Multi-Knapsack, Max E3LIN2, Binary Paint Shop Problem (BPSP), and more broadly, 
most broadly encapsulated by quadratic unconstrained binary optimization  problems (see~\cite{BleBraQuantReview2024} and references therein). 
QAOA is intended to find approximate solutions to difficult combinatorial optimization problems on quantum computers by encoding the problem's Hamiltonian into a quantum circuit and utilizing adiabatic time evolution to optimize the circuit's variational parameters. 
An approximate solution is then obtained by measuring the QAOA circuit using the parameters identified by a classical numerical optimization algorithm, which aims to optimize a problem-specific and well-designed cost function. 
However, the optimization methods employed must be both robust and efficient in navigating a cost function landscape that is typically periodic and nonconvex, since these are typical characteristics of the loss landscape~\cite{LarMenShiQAOA2024,ShaSafLarMultistart2019}. 
Moreover, quantum measurements are inherently stochastic, meaning the cost function values used by QAOA are typically statistical estimates derived from repeated measurements (shots) of the quantum circuit's output state, which inevitably introduce stochastic noise into the optimization process.

In this manuscript our focus is on the classical numerical optimization method, which plays a crucial role in the QAOA framework. Choosing the right optimization method is critical for improving QAOA performance. Current strategies for enhancing QAOA parameters can be broadly classified into three categories~\cite{BleBraQuantReview2024}:
%\mmnote{I just noticed how many times this particular citation gets used, consider finding others or just cite it less! }: 
machine learning (ML), gradient-based, and derivative-free approaches, with the last referring to optimization methods that do not require any derivative information from the user or the cost function oracle. Although ML methods can speed up QAOA optimization by leveraging correlations and patterns among parameters, they may encounter scalability challenges, particularly for more complex problems, and often require numerous training instances to achieve good performance. 
On the other hand, commonly used gradient-based methods such as stochastic gradient descent or {\sf BFGS} can be more robust to noise and problem variations~\cite{PellowJarman2021, MMSWMX23}. 
%\mmnote{BFGS robust to noise?! No way}. 
%SW: The literature says both this and the converse. For example, the cited Ref (and Ref 1, which cites it) found this... now cited.
% MM: I appreciate the "can". :-) 
However, gradient-based methods can be computationally expensive and sensitive to noise in NISQ devices and often require a large number of measurements to compute via the parameter shift rule~\cite{Wierichs2022}.
%\swnote{Can someone please add a canonical reference for this?}. 
Some researchers have attempted to adaptively select shot counts in the computation of such noisy gradients \cite{kubler2020adaptive, arrasmith2020operator}. 
As an alternative, given the difficulty in obtaining accurate gradients, practitioners often rely on derivative-free optimization (DFO) methods~\cite{AuHa2017,CoScVibook,LaMeWi2019} as the classical optimizer in VQAs~\cite{LarMenShiQAOA2024}. Examples of such methods include the popular model-based trust-region (MBTR) algorithm {\sf BOBYQA}~\cite{bobyqa}, as well as {\sf COBYLA}~\cite{cobyla94,Powe98a,nloptWebsite}, {\sf NEWUOA}~\cite{Po2006}, {\sf SBPLX}~\cite{Rowa90a}, {\sf PRAXIS}~\cite{Bren73b}, and {\sf Nelder--Mead}~\cite{NeMe65a},  which were recently benchmarked as QAOA optimizers by Shaydulin et al.~\cite{ShaSafLarMultistart2019}. 
Their study 
%in~\cite{ShaSafLarMultistart2019} 
revealed that QAOA's performance deteriorates significantly as the number of layers
%,~$p$, 
increases, indicating the difficulty of parameter optimization even at relatively shallow circuit depths. 
The software package {\sf scikit-quant}~\cite{ScikitQuant}, on the other hand, integrates various DFO optimizers, which the authors have found to be effective for optimizing variational parameters within the VQA framework. 
We note that although none of the aforementioned DFO methods are specifically designed for stochastic optimization, they are often found to be far more efficient than their stochastic counterparts. 
However, 
%by design, 
any method designed for a deterministic problem will not resample a cost function at the same parameter setting more than once. Consequently, deterministic methods are inevitably prone to becoming ``stuck'' in the presence of stochastic noise. 
These observations motivated the development of {\sf ANATRA}~\cite{LarMenShiQAOA2024}, an algorithm for noisy DFO problems that encompass those presented by VQAs. {\sf ANATRA} was shown to outperform many state-of-the-art deterministic and stochastic optimizers on QAOA problems, particularly in high-noise environments.

The performance of QAOA is typically assessed through the ratio of the expectation value of the globally optimal QAOA solution to the globally optimal value of the original optimization problem.
This ratio asymptotically improves as the number of layers, $p$, increases, as QAOA recovers the aforementioned adiabatic evolution when~$p$ approaches infinity~\cite{BleBraQuantReview2024}. 
The latter observation suggests that creating greater depth circuits by increasing $p$, which entails solving correspondingly higher-dimensional optimization problems,  would guarantee better ratios.
However, performance guarantees of DFO methods, including all of those mentioned above, scale polynomially in dimension $p$.
The situation is exacerbated by stochastic noise in a QAOA framework, since performance guarantees for stochastic DFO are considerably weakened.
These observations shape our primary goal in this work, which is to develop a stochastic DFO algorithm capable of solving \textit{high-dimensional} stochastic problems. 

In this work we introduce {\sf ANASTAARS}, a Noise-Aware Stochastic Trust-region Algorithm using Adaptive Random Subspaces.
{\sf ANASTAARS} is based on the existing {\sf STARS}~\cite{DzaWildSub2022} algorithmic framework, achieving scalability by optimizing random models that approximate the cost function within low-dimensional affine subspaces.
Limiting model construction to subspaces significantly reduces per-iteration costs in terms of function evaluations. 
In contrast to the {\sf STARS} algorithmic framework and previous work involving random subspaces with fixed dimensions  (see~\cite{carfowsha2022Rand,carfowsha2022aRandomised,RobRoy2022RedSpace,shao2022Thesis} and the RSDFO(-GN) framework \cite{CRsubspace2021}),   {\sf ANASTAARS} employs an \emph{adaptive} subspace dimension selection strategy similar to that used in DF-BGN~\cite{CRsubspace2021}, as well as subspace variants of POUNDerS \cite{menickelly2023avoiding, menickelly2024augmenting}. 
Instead of generating a completely new poised set of interpolation points at each iteration, the proposed method updates the model by generating only a few or even a single new interpolation point, reusing past points (and their corresponding function values) from strictly lower-dimensional subspaces in such a way that the resulting set remains poised.
This approach not only introduces a novel way to reduce per-iteration costs in terms of function evaluations but also avoids constructing random models in fixed-dimension subspaces, resulting in an efficient optimization process through the use of adaptive subspace models. 
Furthermore, to address the observation that model-based methods perform well when the signal-to-noise ratio is high, as emphasized in~\cite{LarMenShiQAOA2024}, {\sf ANASTAARS} incorporates a strategy from the noise-aware numerical optimization literature by utilizing an estimate of the noise level in function evaluations.

%\textcolor{blue}{To be completed.}
%SW" ???

\section{Background}\label{sec2Backgr}
\subsection{The Quantum Approximate Optimization Algorithm}
Introduced by Farhi et al.~\cite{QAOAFarGol2014}, QAOA is a VQA aimed at solving combinatorial optimization problems by preparing a parameterized quantum circuit, designed so that high-quality solutions to the optimization problem correspond to modes of measurements. 
Alongside an initial state $\ket{\bm{\psi}_0}$ and a hyperparameter~$p$ representing the number of layers, the circuit is determined by operators~$\bm{H}_P$ and~$\bm{H}_M$, respectively the problem Hamiltonian and the mixer Hamiltonian. The prepared quantum state is
\begin{align}\label{circuitEq}
\ket{\bm{\psi}(\bm{\gamma},\bm{\beta})}=e^{-i\beta_p\bm{H}_M}e^{-i\gamma_p\bm{H}_P}\cdots e^{-i\beta_1\bm{H}_M}e^{-i\gamma_1\bm{H}_P}\ket{\bm{\psi}_0},
\end{align}
with $\bm{\gamma}=(\gamma_1,\dots,\gamma_p)^\top$ and $\bm{\beta}=(\beta_1,\dots,\beta_p)^\top$ denoting free parameters. In the QAOA circuit with properly tuned parameters, 
$\ket{\bm{\psi}(\bm{\gamma},\bm{\beta})}$ converges to the ground state of the problem Hamiltonian~$\bm{H}_P$ as the number of layers~$p$ increases toward infinity.
By design of the cost function, the energy of the problem Hamiltonian approaches the optimal value of the optimization problem's objective function. 
A classical optimization algorithm is used to tune the variational parameters $\bm{\beta}$ and $\bm{\gamma}$ by solving the problem
\begin{align}\label{ExpecEnergy}
\underset{\bm{\gamma},\bm{\beta}}{\min}\accolade{ \bra{\bm{\psi}(\bm{\gamma},\bm{\beta})}\bm{H}_P\ket{\bm{\psi}(\bm{\gamma},\bm{\beta})}},
\end{align}
where the cost function in~\eqref{ExpecEnergy} represents the expectation of the energy. 
The cost function 
%\eqref{ExpecEnergy} 
can only be sampled in practice.
During each QAOA iteration, the expected value 
\begin{align*}
f(\gamma_1,\dots,\gamma_p,\beta_1,\dots,\beta_p)=\bra{\bm{\psi}(\bm{\gamma},\bm{\beta})}\bm{H}_P\ket{\bm{\psi}(\bm{\gamma},\bm{\beta})}
\end{align*}
is estimated from multiple measurements of $\ket{\bm{\psi}(\bm{\gamma},\bm{\beta})}$ and provided to the classical optimizer as the cost function. 
In other words, if $\x=(\gamma_1,\dots,\gamma_p,\beta_1,\dots,\beta_p)^{\top}\in\rd$ with $d=2p$, the classical optimizer aims to solve the stochastic optimization problem
\begin{align}\label{stochProb}
\underset{\x\in\rd}{\min} f(\x)\quad\text{with}\quad f(\x)=\E[\ubar{\theta}]{\ftheta(\x)},
\end{align}
while having access only to the noisy function values $\ftheta(\x)$, where $\ubar{\theta}$ is a random variable modeling the stochasticity from the aforementioned measurements. 
Note that each individual measurement corresponds to a complete execution of the circuit (a shot). 

%The most intriguing and promising applications focus on scenarios with a limited number of shots for the stochastic objective.
A problem frequently considered in the context of QAOA is the weighted MaxCut. We define an instance of MaxCut as follows.
\begin{maxcut}
Consider an undirected graph $G=(V,E)$ where $V$ is the set of vertices identified with $V=\left[n\right]$ and $E$ is the set of edges. 
Denote by $w_{uv}$ the weight associated with each edge $(u,v)\in E$, which connects the vertices~$u$ and~$v$. Find $\s:=(s_1,\dots,s_{n})\in\accolade{-1,1}^{n}$  that maximizes
\begin{align*}
h(\s)=\sum_{(u,v)\in E}
\frac{w_{uv}}{2}(1-s_u s_v).
\end{align*}
\end{maxcut}
The Hamiltonian $\bm{H}_P$ encoding the MaxCut problem on qubits is obtained by mapping spin variables $s_i$ onto the spectrum of Pauli-$\bm{Z}$ matrices; that is, 
\begin{align*}%\label{HpDef}
\bm{H}_P=\sum_{(u,v)\in E}
\frac{w_{uv}}{2}\left(\bm{I}-\bm{Z}_u\bm{Z}_v\right),
\end{align*}
where $\bm{Z}_u$ is the Pauli-$\bm{Z}$ operator acting on the $u$th qubit. The
mixer Hamiltonian $\bm{H}_M$ is defined as
\begin{align}\label{HmDef}
\bm{H}_M=\sum_{i\in V}\bm{X}_i,
\end{align}
where $\bm{X}_i$ is the Pauli-$\bm{X}$ operator acting on the $i$th qubit. 
The initial state $\ket{\bm{\psi}_0}$ is chosen as the ground state of the mixer Hamiltonian, which for $\bm{H}_M$ in Equation~\eqref{HmDef} is
\begin{align*}
\ket{\bm{\psi}_0}=\ket{+}^{\otimes n}:=\frac{1}{\sqrt{2^n}}\sum_{\mathbf{x}\in \accolade{0,1}^n}\ket{\mathbf{x}}.%, \quad\text{where}\ n=\abs{V}.
\end{align*}

%\mmnote{I'm going back and forth mentally on whether or not this section (II.A) is too many details for an audience that probably already knows QAOA. }
%SW: It is totally OK and important for deining the notation understanding precisely what the dimension p stands for and setting the stage for the asymptotics.
% MM: Opinion appreciated. 

\subsection{Subspace Model-Based Stochastic Trust-Region Methods}\label{sec2SecB}
As discussed in Section~\ref{secI}, DFO methods for deterministic cost functions can be more shot-efficient than those tailored for stochastic optimization. %\mmnote{At this point, it only just dawned on me that we are definitely missing \cite{kubler2020adaptive}, an adaptive sampling algorithm specifically for VQAs from Lukasz' group. }
% MM: Resolved in previous section. 
Some of the most successful approaches are adaptations of the model-based trust-region (MBTR) framework~\cite{CoScVibook}, with notable implementations pioneered by Michael Powell. For noise-free settings such methods typically require a number of evaluations that is polynomial in the dimension $d$; methods for noisy problems often require significantly more samples. 
Our primary concern in designing a novel algorithm is to be deliberately noise-aware, 
since MBTR methods, traditionally developed for deterministic problems, eventually get ``stuck" in the presence of noise, as discussed in Section~\ref{secI}.
Dzahini and Wild~\cite{DzaWildSub2022} took a step in this direction with the design of {\sf STARS}, where scalability is achieved by iteratively constructing and minimizing stochastic models that approximate the cost function within low-dimensional random subspaces, in contrast to earlier methods~\cite{BaScVi2014,chen2018stochastic,shashaani2018astro} that utilize ``full-space'' models. 

%\mmnote{Proposal: Defer the following discussion of the additional complications posed by stochasticity to AFTER you've already explained the at-least-linear sampling cost of constructing models in the deterministic case. }
% SW: This is difficult because the "Noisy" measurements here are stochastic; partial addressed in the previous paragraph, continue to iterate if this is still a major challenge.
Indeed, it has been established that in a simple stochastic noise framework \cite{menickelly2023latency, ha2025two}, 
%where noisy gradient values are unavailable, 
obtaining good full-space models at each iteration~$k$ of 
%{\sf STORM} 
a standard model-based method
(applied to~\eqref{stochProb}) requires at least $\Omega((d+1)\max\accolade{\dk^{-2},\dk^{-4}}(1-\alpha^{1/(d+1)}))$ function evaluations, for fixed $\alpha\in (0,1)$ and $\dk>0$ a measure of locality, with~$d$ replaced by~$q\ll d$ in the above quantity for {\sf STARS} subspace models. This shows in particular the low per-iteration cost of building the subspace models compared with full-space ones, especially since~$q$ can be chosen independently of~$d$. 
%\mmnote{This paragraph is too heavy, and you haven't described what $\dk$ is in this point of the text. A non-expert will be very lost. Also, to be more quantum specific, you could cite my past work with Otten about shot complexity of STORM-like methods in VQAs \cite{menickelly2023latency}}.

To solve Problem~\eqref{stochProb}, the {\sf STARS} framework, which inspires the algorithm proposed in the present work, operates as follows. 
At each iteration~$k$, given an incumbent point~$\xk\in\rd$ and a realization $\Qk\in\rdq$ of~a specific random matrix $\QkRandom$ (where the general random variable is denoted with an underbar), 
%\mmnote{$\Qk$ and $\QkRandom$ should differ somehow, the macros are currently identical. Oh no, now I see one is underlined. Maybe you need something more drastic.} 
an interpolation model $\mhatk$ is built on the affine subspace $\mathcal{Y}_k:=\accolade{\xk+\Qk\s:\s\in\Rq}$ using realizations of the noisy function~$\ftheta$. This model is typically quadratic and is given by
\begin{align}\label{quadModEq}
\mhatk(\s):=\mathfrak{f}_k+\gkhat^\top\s+\frac{1}{2}\s^\top\Hkhat\s \; \approx \; f(\xk+\Qk\s),
\end{align}
with $\gkhat\approx\QkTranspose\nabla f(\xk) \in\Rq$ and $\Hkhat\in\rqq$~respectively denoting the low-dimensional model gradient and Hessian. These two defining quantities and the scalar $\mathfrak{f}_k+$ consist of $\frac{1}{2}(q+1)(q+2)$ parameters, which are obtained by solving linear interpolation systems of this size; this size also corresponds to the number of estimates of $f$ that are employed and thus directly informs the number of shots employed in model construction. As we emphasize below, when $q\ll d$ there are significantly fewer model parameters (and hence fewer shots employed by any one model) than there are for full-space models (i.e., when $q=d$).

The model $\mhatk$ is used to approximate the cost function $f$ in the \emph{trust region} $\mathcal{B}(\xk,\dk;\Qk):=\left\lbrace\xk+\Qk\s\in \mathcal{Y}_k:\right.$ $\left.\normii{\s}\leq\dk\right\rbrace$, where $\dk>0$ denotes the trust-region radius.
%\mmnote{Someone who doesn't know trust-region methods, let alone DFO, should be made aware of the computational effort involved in computing $\gkhat, \Hkhat$. In particular, consider letting them know that in the deterministic case, this entails an at-least-linear, at-most-quadratic (in $q$) number of function evaluations. Follow-up: This is where you really need to hit home why subspaces could help in DFO. }
%SW: Addressed above.
% MM: Great! 
A trial step is obtained from the (approximate) solution of $\sk\approx\arg\min\accolade{\mhatk(\s):\s\in\Rq,\normii{\s}\leq \dk}$.
Estimates $\fok\approx f(\xk)$ and $\fsk\approx f(\xk+\Qk\sk)$ are computed using evaluations of the noisy function~$\ftheta$. The potential change in~$f$ due to the trial step~$\sk$ is assessed by comparing a preset value $\eta_1\in (0,1)$ with the ratio $\rho_k:=\frac{\fok-\fsk}{\mhatk(\bm{0})-\mhatk(\sk)}$ of the reduction in the estimates with the reduction in the model. 
If the ratio is greater than $\eta_1$, then the iteration is declared successful, in which case the incumbent is replaced by the trial step, and the trust-region radius is not decreased. 
Otherwise, the trust-region radius is decreased, and the incumbent remains unchanged. 
An instantiation of {\sf STARS} is presented in Algorithm~\ref{algoStoScalTR}.

As emphasized in~\cite[Section~3.1]{DzaWildSub2022}, the success of an algorithm in the {\sf STARS} framework requires that the subspace~$\mathcal{Y}_k$ defined by the span of the columns of~$\Qk$ must be aligned with the gradient $\nabla f(\xk)$ in order to ensure sufficient decrease in~$f$. 
By alignment, we mean that
%Indeed, for any $\nabla f(\xk)\neq 0$, it holds that 
\begin{equation}
\label{eq:bound}
\|\QkTranspose\nabla f(\xk)\| \ge \varepsilon \|\nabla f(\xk)\|,
\end{equation}
%\begin{align*}
%\nabla f(\xk)\ \text{is orthogonal to}\ \mathcal{Y}_k\ \Longrightarrow\ \QkTranspose\nabla f(\xk)=0.
%\end{align*}
%\mmnote{Around here is where we need to explain why well-alignment is still a reasonable concept in the presence of barren plateaus ... it seems to run counter to random subspaces, since the gradient is nearly 0 everywhere in parameter space. I'm not sure how to resolve this dissonance ... }
%SW: I have changed the description to the actual property being used, which is helpful to expose the gradient.
% MM: I have deeper philosophical issues, but to be clear, NO solver derived from a first-order method can overcome them here. In that sense, what we're claiming is fine, so long as the practitioner is broadly aware that something global (e.g., multistart QAOA like Jeff and Ruslan's past work) ought to be generating these local runs. 
%SW: Since the numerical results here are all from local runs...
%which means that the subspace model gradient $\gkhat$ approximating $\QkTranspose\nabla f(\xk)$ can be arbitrarily small even though the full-space gradient $\nabla f(\xk)$ itself is nonzero. 
which means that the subspace-reduced gradient norm is close to the full-space gradient norm in a relative sense.
If the converse of the event \eqref{eq:bound} were to occur sufficiently often, then an algorithm in the {\sf STARS} framework could not converge. 
Therefore, we assume that the random matrix $\QkRandom$ with realizations $\Qk\in\rdq$ determining the subspace satisfies the following so-called  well-alignment condition.
\begin{definition}(Simplified version of~\cite[Definition~3.2]{DzaWildSub2022})
For fixed $\upepsilon,\upbeta\in (0,1/2)$, a sequence $\accoladekinN{\QkRandom}$ of random matrices, with $\QkRandom\in\rdq$, is $(1-\upbeta)$-probabilistically $\upepsilon$-well aligned if, for any $\vi\in\rd$, 
\begin{align}\label{probWellAlign}
\prob{\normii{\QkRandomTranspose\vi}\geq \upepsilon\normii{\vi}}\geq 1-\upbeta.
\end{align}
Matrices satisfying~\eqref{probWellAlign} will be referred to as $WAM(\upepsilon,\upbeta)$.
\end{definition}
It follows from~\eqref{probWellAlign} that in the particular case $\vi=\nabla f(\xk)$ and $\upbeta$ small enough, with high probability, $\|\QkTranspose\nabla f(\xk)\|$ is bounded away from zero whenever $\nabla f(\xk)\neq 0$, as desired. 
Fortunately, the existence of random matrices satisfying~\eqref{probWellAlign} is guaranteed, for example, by the celebrated Johnson--Lindenstrauss lemma~\cite{JohnsonLind1984}, reported in~\cite{KaNel2014SparseLidenstrauss} and recalled next.
\begin{lemma}(Johnson--Lindenstrauss~\cite[Lemma~1]{KaNel2014SparseLidenstrauss})
For any integer $d>0$ and $0<\upepsilon,\upbeta<1/2$, there exists a probability distribution on $d\times q$ real matrices for $q=\Theta(\upepsilon^{-2}\log(1/\upbeta))$ such that for any $\vi\in\rd$ and any matrix $\Qrandom$ drawn from the aforementioned distribution
\begin{align}\label{JLTEq}
\prob{(1-\upepsilon)\normii{\vi}\leq \normii{\QrandomTranspose\vi}\leq (1+\upepsilon)\normii{\vi}}\geq 1-\upbeta.
\end{align}
\end{lemma}
\begin{algorithm}[htb]
\caption{Simplified {\sf STARS} algorithm~\cite{DzaWildSub2022}.}
\label{algoStoScalTR} 
{{
\textbf{[0] Initialization}\\
\hspace*{4mm}Choose algorithmic parameters and a starting point \scalebox{0.8}{$\bm{{\x}}_0\in\rd$}, fix a subspace dimension \scalebox{0.8}{$q\in\intbracket{1,d}$}, and set the\\ \hspace*{4mm}iteration counter \scalebox{0.8}{$k \gets 0$}.\\
%constants{\bl \footnotemark}  $\ \gamma>1$, $\eta_1\in(0,1)$, $\eta_2>0$, $j_{\max}\in\N$, $\varepsilon\in(0,1)$, $\beta\in(0,1)$, \\ 
%\hspace*{10mm}$c_1\geq 1$, initial trust-region radius $\delta_0>0$, and maximum trust-region radius\\
%\hspace*{10mm} $\dmaxx=\gamma^{j_{\max}}\delta_0$, starting point $\bm{{\x}}_0\in\rn$, and dimension $p\in\intbracket{1,n}$.\\ 
%\hspace*{5mm}Set the iteration counter $k \gets 0$.\\
%\textbf{[1] Construction of subspace model}\\
\textbf{[1] Subspace model, estimates, and updates}\\
%\hspace*{10mm}Generate $\Qk$: a realization of a
%%\footnotemark 
%{\em WAM}($1-\varepsilon,\beta$), using a distribution $\mathbb{Q}_{k}$.\\
\hspace*{4mm}Generate \scalebox{0.8}{$\Qk\in\rdq$} using Haar measure to build the model \scalebox{0.8}{$\mhatk:\Rq\to\R$}.\\
%\hspace*{5.5mm}$\mhatk:\Rq\to\R$ approximating $f(\x)$.\\
%in $\mathcal{B}(\xk,\dk;\Qk)$\\
%that is $(\kef,\keg;\Qk)$-fully linear 
%$\mhatk(\s)={\f}_k+\gkhat^{\top}\s+$\\
%\hspace*{10mm}$\frac{1}{2}\s^{\top}\Hkhat\s\approx f(\xk+\Qk \s)$, that locally approximates $f$ 
%in  $\mathcal{B}(\xk,c_1\dk;\Qk)$, using a .\\ 
%\hspace*{10mm}with the affine space defined by the columns of $\Qk$.\\
%\textbf{[2] Step calculation}\\
\hspace*{4mm}Minimize \scalebox{0.8}{$\mhatk$} in a trust region to obtain \scalebox{0.8}{$\sk$}.\\
%$\sk\approx {\argmin}\left\{ \mhatk(\s) : \; \s\in\rq:\normii{\s}\leq \dk\right \}$.\\
%\textbf{[3] Estimate computation}\\
\hspace*{4mm}Obtain accurate estimates \scalebox{0.8}{$\fok\approx f(\xk)$} and \scalebox{0.8}{$\fsk\approx f(\xk+\Qk\sk)$}.\\
%\textbf{[4] Updates}\\
\hspace*{4mm}Use \scalebox{0.77}{$\rho_k=\frac{\fok-\fsk}{\mhatk(\bm{0})-\mhatk(\sk)}$} and \scalebox{0.8}{$\normii{\gkhat}$} to check whether the iteration is successful or not.\\
\hspace*{4mm}Set \scalebox{0.8}{$\bm{\x}_{k+1}=\xk+\Qk\sk$} (\textbf{success}) and \scalebox{0.8}{$\bm{\x}_{k+1}=\xk$} (\textbf{failure}).\\
\hspace*{4mm}Update the trust-region radius. Set \scalebox{0.8}{$k \gets k+1$}, and go to~\textbf{[1]}.\\
%\hspace*{10mm}If $\rho_k\geq \eta_1$ and $\normii{\gkhat}\geq \eta_2\dk$ (\textbf{success}): \\
%\hspace*{16mm} set $\bm{\x}_{k+1}=\xk+\Qk\sk$ and $\dkun=\min\accolade{\gamma\dk,\dmaxx}$. \\
%\hspace*{10mm}Otherwise (\textbf{failure}): set $\bm{\x}_{k+1}=\xk$ and $\dkun=\gamma^{-1}\dk$.\\
%\hspace*{5.5mm}Update the counter $k \gets k+1$, and go to~\textbf{[1]}.\\
}}
\end{algorithm}

Various distributions, such as Gaussian matrices~\cite[Theorem~3.4]{DzaWildSub2022}, matrices generated by hashing and hashing-like approaches~\cite{DzaWildHashing2022,KaNel2014SparseLidenstrauss}, and Haar distribution orthogonal matrices~\cite{MeckesElizHaar2019} are commonly used for generating random matrices that satisfy~\eqref{probWellAlign}. 
The last strategy, used in the proposed {\sf ANASTAARS} algorithm, is presented in the result stated next. 
We first recall the Haar measure, which is the unique translation-invariant probability measure  on the compact group $\mathbb{O}(d)$ or orthogonal matrices.
\begin{theorem}\label{HaarTheor1}($\!\!$\cite[Theorem~2.2]{DzaWildSubDir24}, inspired by~\cite[Lemma~6.1]{MeckesElizHaar2019})
Let $\upepsilon,\upbeta\in (0,1/2)$ and $\Urandom:=(\ubar{U}_{ij})_{1\leq i,j\leq d}\in\mathbb{O}(d)$ be a Haar-distributed matrix. Let $\Urandom_{\qtd}\in\rqd$ be the matrix consisting of the first~$q$ rows of $\Urandom$, and define $\QrandomTranspose:=\sqrt{\frac{d}{q}}\Urandom_{\qtd}$. %\\
Then there exists an absolute constant $\eta>0$ such that if $q\geq \frac{4}{\eta}\upepsilon^{-2}\log(2/\upbeta)$, then for all $\vi\in\rd$,~\eqref{JLTEq} holds.
\end{theorem}

For theoretical purposes and practical efficiency, the estimates $\fok$ and $\fsk$ are required to be sufficiently close to their corresponding estimated function values by satisfying so-called $\ef$-accuracy conditions~\cite[Definition~3.12]{DzaWildSub2022},~\cite{audet2019stomads,chen2018stochastic,dzahini2020expected,dzahini2020constrained}. More precisely, given $\ef>0$ and~$\dk>0$,
%$\xk\in\rd$ and $\Qk\sk\in\rd$, 
the estimates $\fok\approx f(\xk)$ and $\fsk\approx f(\xk+\Qk\sk)$ must satisfy
%are called $\ef$-accurate, respectively for a given~$\dk$, if
\begin{align*}
	\abs{\fok-f(\xk)}\leq \ef\dk^2\quad\text{and}\quad \abs{\fsk-f(\xk+\Qk\sk)}\leq \ef\dk^2,
\end{align*}
in which case they are called $\ef$-accurate.
%\end{definition}
Given that $f(\xk)$ and $f(\xk+\Qk\sk)$ are not accessible in practice, the above conditions are satisfied only in probability. 
Such estimates are obtained by averaging samples or realizations of the noisy function.
For example, for an appropriate sample size~$\pi_k$, $\fok=\frac{1}{\pi_k}\sum_{\ell=1}^{\pi_k}f_{\theta_\ell^{\bm{0}}}(\xk)$, where the~$\theta_\ell^{\bm{0}}$ are realizations of independent random samples of~$\ubar{\theta}$. 
For details on how random estimates $\Fok$ and $\Fsk$ with respective realizations $\fok$ and $\fsk$ can be generated, see~\cite[Section~2.3]{audet2019stomads},~\cite[Section~5]{chen2018stochastic}, and~\cite[Section~5.1]{dzahini2020constrained}.
On the other hand, the model $\mhatk$ is required to be fully linear (see, e.g.,~\cite[Definition~3.10]{DzaWildSub2022}),
a condition that amounts to the model satisfying similar error bounds as a first-order Taylor model. 
Full linearity is achieved, for instance, and as employed in this paper (see Section III), by using a linear model interpolating function values on a set of points of appropriate geometry. 
%that is, $\abs{f(\xk+\Qk\s)-\mhatk(\s)}\leq \kappa_{ef}\dk^2$ and $\normii{\QkTranspose\nabla f(\xk+\Qk\s)-\nabla\mhatk(\s)}\kappa_{eg}\dk$, for all $\normii{\s}\leq\dk$ and some $\kappa_{ef},\kappa_{eg}>0$. \mmnote{Full linearity is an opaque concept. It should generally be motivated by Taylor expansions, but this is going to take up space ...}. Section... \mmnote{Should this point to something in the text?} offers further details on the practical construction of subspace models.

%\begin{align*}
%\abs{f(\xk+\Qk\s)-\mhatk(\s)}\leq \kappa_{ef}\dk^2
%\end{align*}
%with fully linearity referring to a measure of local accuracy.
\section{Subspace Model Construction Leading to Adaptive Subspace Selection}\label{Sec3ModelBuild}
%Previous model-based algorithms utilizing random subspaces maintain a fixed subspace dimension throughout the framework, generate new interpolation points and calculate the corresponding function values or estimates at each iteration. 
A key contribution of this paper is the development of strategies that enable the reuse of interpolation points (and their corresponding function values, thus reducing the number of shots required by the algorithm) from an unsuccessful iteration~$k$ to construct higher-dimensional subspace models in the subsequent iteration~$k+1$.
%\mmnote{The motivation currently doesn't stress enough why reuse could be beneficial. In general,the  per-iteration shot costs involved in constructing interpolation models should be more heavily spotlighted for someone who doesn't think about model-based DFO. Follow-up: I pointed out where I BELIEVE this discussion should be made heavy. } 
%SW: Yes, see there now
Unlike prior works that use a fixed subspace dimension, the proposed strategies result in a low per-iteration cost in {\sf ANASTAARS} in terms of function evaluations/shots.
Below, for all $i=0,1,\dots$, the $\theta_{k,\ell}^{i}$, $\ell=1,2,\dots$, are realizations of independent random samples of~$\ubar{\theta}_{k,\ell}^i$ of the independent random variables $\ubar{\theta}_k^i$ distributed as $\ubar{\theta}$. Unless otherwise stated, $\Qk=\sqrt{\frac{d}{q}}\Uk\in\rdq$ with $q<d$, where $\Uk$ is a realization of a random matrix $\Urandom_k$ obtained from Haar distribution (see Theorem~\ref{HaarTheor1}), and hence the columns of $\Uk$ are orthonormal.
We define $\mathbf{1}_q:=(1,\dots,1)^\top\in\Rq$, $\mathbf{0}_q=(0,\dots,0)^\top\in\Rq$ and $\hat{q}:=\sqrt{1+\frac{1}{q}}$. $\bm{I}_q\in\rqq$ as the identity matrix. 
$\mathscr{P}^2_q$ is the space of polynomials of degree less than or equal to~$2$ in~$\Rq$.

\subsection{Subspace Linear Models}
The linear model construction (at iteration~$k$) described next is inspired by~\cite[Theorem~3.16]{DzaWildSub2022}. At each point of a poised set\footnote{Poisedness can be understood as a nonzero volume of the convex hull of the interpolation points; see, e.g., \cite[Section 2.2]{LaMeWi2019} for more details. }
$\SSk^q:=\accolade{\szerok,\sonek,\dots,\squk}\subseteq\Rq\cap\mathcal{B}(\bm{0},\dk)$, with $\szerok=\bm{0}$,
let 
$\fsk(\siik;\theta_k^i):=\frac{1}{\pi_k}\sum_{\ell=1}^{\pi_k}f_{\theta_{k,\ell}^{i}}(\xk+\Qk\siik)$. A linear model 
\begin{align*}
	\mhatk(\s) = \aok+\akTransp\s,\quad (\aok,\ak)\in\R\times\Rq,
\end{align*}
is built by fitting values of $\aok,\ak$, such that $\mhatk(\siik)=\fsk(\siik;\theta_k^i)$ for all $\siik\in\SSk^q$. 
That is, $\aok=\fsk(\szerok;\theta_k^0)$ while $\ak=\nabla \mhatk(\s)$ is obtained by solving the $q\times q$ linear system
{\footnotesize{
		\begin{align*}
			\Lk^\top\ak=\bm{\delta}^{f_k(\SSk^q;\theta_k)}:=
			\begin{bmatrix}
				\fsk(\sonek;\theta_k^1) - \fsk(\szerok;\theta_k^0)\\
				\fsk(\stwok;\theta_k^2) - \fsk(\szerok;\theta_k^0) \\
				\vdots \\
				\fsk(\squk;\theta_k^q) - \fsk(\szerok;\theta_k^0)
			\end{bmatrix}
			\in\Rq,
		\end{align*}
}}
where $\Lk:=[\sonek-\szerok\ \cdots\ \squk-\szerok]\in\rqq$.

Recall from Algorithm~\ref{algoStoScalTR} that on any unsuccessful iteration~$k$,, we have $\xkun=\xk$. 
Inspired by the {\it explicit geometric construction} of Haar measure~\cite[Section~1.2]{MeckesElizHaar2019}, we set
\begin{align}\label{QkPlusOne}
	\Q_{k+1}:=\scalebox{0.9}{$\sqrt{\frac{d}{q+1}}$}[\Uk, \bm{\mu}_{k+1}]\in\R^{d\times (q+1)}
\end{align}
after an unsuccessful iteration, 
where $\bm{\mu}_{k+1}$ is a realization of a random vector uniformly distributed on the intersection of the $d$-dimensional sphere and the orthogonal complement $\Uk^{\perp}$. 
%\in\rd\cap \Uk^{\perp}$, 

%i.e., orthonormal to each column of $\Uk$, is a realization of a random vector 
%${\ubar{\bm{\mu}}}_{k+1}\sim
%$\mathcal{U}(\mathbb{S}^{d-1})$, i.e., uniformly distributed on the unit sphere $\mathbb{S}^{d-1}$. 
Analogous to the construction of $\mhatk$, 
using
{\footnotesize{
		\begin{align*}
			\SSkOneqplus:=\accolade{\!\siikOne=
				\begin{bmatrix}
					\hat{q}\siik \\
					0
				\end{bmatrix}\!,
				\s_{k+1}^{q+1}=
				\begin{bmatrix}
					\bm{0}_q \\
					\zeta_{k+1}
				\end{bmatrix}:
				\siik\in\SSk^q, 0<\abs{\zeta_{k+1}}\leq \dk\!
			},
		\end{align*}
}}
we construct 
a $(\!q+1\!)$-dimensional subspace linear model $\mhatkOne$ defined by
\begin{align*}
	\mhatkOne(\s) = \aokOne+\akOneTransp\s,\quad (\aokOne,\akOne)\in\R\times\R^{q+1}.
\end{align*}
Note that
%by seeking values for $\aokOne$ and $\akOne$. 
% such that $\mhatkOne(\siikOne)=\fskOne(\siikOne;\theta_{k+1}^i)$, $i=0,1,\dots,q+1$. However, 
\begin{align}\label{pastnewEq}
	\xkun+\QkOne\siikOne=\xk+\Qk\siik,\quad i=0,1,\dots,q, 
\end{align}
and so the construction of $\mhatkOne$ reuses past points $\xk+\Qk\siik$ and their corresponding function values $\fsk(\siik;\theta_k^i)$, in addition to one new point $\xkun+\QkOne\s_{k+1}^{q+1}=\xk+\sqrt{\frac{d}{q+1}}\bm{\mu}_{k+1}\zeta_{k+1}$ and its corresponding function value  $\fsk(\s_{k+1}^{q+1};\theta_{k+1}^{q+1})$. 
Thus, the resulting linear interpolation system for constructing $\mhatkOne$ %is built by fitting values for 
defined by 
$\aokOne$ and $\akOne^\top:=[(\akOne^q)^\top\ a_{k+1}^1]$ 
is
%such that $\mhatkOne(\siikOne)=\fsk(\siik;\theta_k^i)$, $i=0,1,\dots,q$, and $\mhatkOne(\s_{k+1}^{q+1})=\fsk(\s_{k+1}^{q+1};\theta_{k+1}^{q+1})$, leading to the 
only a 
$(q+1)\times (q+1)$ linear system:
%Let $\delta^{f_{k+1}^{q+1}}:=\fsk(\s_{k+1}^{q+1};\theta_{k+1}^{q+1})-\fsk(\szerok;\theta_k^0)$. Solving the $(q+1)\times (q+1)$ linear system
%{\footnotesize{
		\begin{align}\label{LkPlusSystem}
			\begin{bmatrix}
				\hat{q}\Lk^\top & \bm{0}_q \\
				\bm{0}_q^\top & \zeta_{k+1}
			\end{bmatrix}
			\begin{bmatrix}
				\akOne^q \\
				a_{k+1}^1
			\end{bmatrix}
			=
			\begin{bmatrix}
				\bm{\delta}^{f_k(\SSk^q;\theta_k)} \\
				\delta^{f_{k+1}^{q+1}}
			\end{bmatrix},
		\end{align}
		where $\delta^{f_{k+1}^{q+1}}:=\fsk(\s_{k+1}^{q+1};\theta_{k+1}^{q+1})-\fsk(\szerok;\theta_k^0)$.
		%}}
Solving~\eqref{LkPlusSystem} yields the closed form 
\begin{align*}
	\akOne^q=\frac{\ak}{\hat{q}}, \quad a_{k+1}^1=\frac{\delta^{f_{k+1}^{q+1}}}{\zeta_{k+1}}\ \text{and}\ \aokOne=\aok.
\end{align*}
\subsection{Efficient Quadratic Subspace Models}\label{SecBFrobNorMod}

%\mmnote{We have not even mentioned what an MFN model is at this point. I suspect I even know DFO people who don't know what this means, let alone a more general audience. I believe, for the scope of this paper, it is better to condense this section (III.B) and the next (III.C) into a single paragraph. Example text: ``The interpolation system~\eqref{LkPlusSystem} implies the selection of a linear basis for interpolation. In model-based DFO, other bases of low dimensionality are often considered. As a commonly employed example, one can consider a basis of quadratic polynomials in $d$ dimensions and regularize the resulting (generally) underdetermined system by minimizing the Frobenius norm of the model Hessian (relevant cites go here). Similarly, one can employ a sub-basis of quadratic polynomials that only spans models with a diagonal Hessian (relevant cites go here). A slight adaptation of the reuse strategy proposed in \eqref{QkPlusOne} can be employed in either of these two settings. }
%SW: Matt made a good point and I've tried to provide further context for these models and have combine them into a single section. An alternative approach is to put the details in an appendix, but this didn't end up saving too much space before the numerical results. Please take a close look at my text.

We now consider two approaches to obtain subspace models that are nonlinear and able to address settings where less than a full quadratic ($\frac{1}{2}(q+1)(q+2)$) number of interpolation points are available. In both cases 
the strategy of reusing past interpolation points is similar to the linear case and is not further detailed here. 

The first approach is a minimum Frobenius norm (MFN) model Hessian approach. At iteration~$k$, we define $\SSk^\sigma$ as above, where $\sigma\leq \nu(q):=\frac{1}{2}(q+1)(q+2)$.
Consider the natural basis $\bar{\Phi}^{\nu(q)}:=(\bar{\Phi}^q_L,\bar{\Phi}^q_Q)$ of $\mathscr{P}^2_q$,
%\begin{align*}
%\bar{\Phi}^{\nu(q)}:=(\phi^0(\s),\phi^1(\s),\dots,\phi^{\nu(q)}(\s))=(\bar{\Phi}^q_L,\bar{\Phi}^q_Q)
%\end{align*}
where $\bar{\Phi}^q_L(\s)=(1,s_1,\dots,s_q)$ and $\bar{\Phi}^q_Q(\s)=(\frac{1}{2}s_1^2,s_1 s_2\dots,\frac{1}{2}s_q^2)$, and define, for any $\bar{\Phi}^q(\s):=(\phi^0(\s),\phi^1(\s),\dots,\phi^q(\s))$,
{\footnotesize{
		\begin{align*}
			\bm{M}(\bar{\Phi}^q;\SSk^\sigma)=
			\begin{bmatrix}
				\phi^0(\szerok) & \phi^1(\szerok) & \cdots & \phi^q(\szerok) \\
				\phi^0(\sonek) & \phi^1(\sonek) & \cdots & \phi^q(\sonek)\\
				\vdots & \vdots & \vdots & \vdots\\
				\phi^0(\s^\sigma_k) & \phi^1(\s^\sigma_k) & \cdots & \phi^q(\s^\sigma_k)
			\end{bmatrix}.
		\end{align*}
}}
The MFN $q$-dimensional subspace model $\mhatk$ is
\begin{align*}
	\mhatk(\s)=\alpha_L^\top\bar{\Phi}^q_L(\s)+\alpha_Q^\top\bar{\Phi}^q_Q(\s),\quad\s\in\Rq,
\end{align*}
where $(\alpha_L,\alpha_Q)$ is the solution of the optimization problem
{%\footnotesize{
		\begin{align*}
			& \underset{\alpha_L,\alpha_Q}{\min}\ \frac{1}{2}\normii{\alpha_Q}^2\\
			&\! \text{such that}\quad 
			%\bm{M}(\bar{\Phi}^q_L;\SSk^\sigma)\alpha_L+\bm{M}(\bar{\Phi}^q_Q;\SSk^\sigma)\alpha_Q
			\sum_{i\in\{Q,L\}}\bm{M}(\bar{\Phi}^q_i;\SSk^\sigma)\alpha_i
			=f_k(\SSk^\sigma;\theta_k),
		\end{align*}
	}%}
with $f_k(\SSk^\sigma;\theta_k):=(\fsk(\szerok;\theta^0_k),\dots,\fsk(\s^\sigma_k;\theta_k^\sigma))^\top$.

Assuming the iteration~$k$ unsuccessful, the set $\SSkOne^{\sigma+1}$ is constructed analogously to $\SSkOne^{q+1}$ so that~\eqref{pastnewEq} holds with~$q$ replaced by~$\sigma$. Then the MFN $(q+1)$-dimensional subspace model $\mhatkOne$ is given by
\begin{align*}
	\mhatkOne(\s)=\lambda_L^\top\bar{\Phi}^{q+1}_L(\s)+\lambda_Q^\top\bar{\Phi}^{q+1}_Q(\s),\quad\s\in\R^{q+1},
\end{align*}
where $(\lambda_L,\lambda_Q)$ is the solution of the optimization problem
{%\footnotesize{
		\begin{align*}
			& \underset{\lambda_L,\lambda_Q}{\min}\ \frac{1}{2}\normii{\lambda_Q}^2\\
			&\!\text{such that}\quad 
			%\bm{M}(\bar{\Phi}^{q+1}_L;\SSkOne^{\sigma+1})\lambda_L+\bm{M}(\bar{\Phi}^{q+1}_Q;\SSkOne^{\sigma+1})\lambda_Q
			\sum_{i\in\{Q,L\}}
			\bm{M}(\bar{\Phi}^{q+1}_i;\SSkOne^{\sigma+1})\lambda_i
			=
			\begin{bmatrix}
				f_k(\SSk^{\sigma};\theta_k) \\
				\fsk(\s_{k+1}^{\sigma+1};\theta_{k+1}^{\sigma+1})
			\end{bmatrix},
		\end{align*}
	}%}
so that only $\fsk(\s_{k+1}^{\sigma+1};\theta_{k+1}^{\sigma+1})$ at iteration~$k+1$ is newly computed, while $f_k(\SSk^{\sigma};\theta_k)$ from iteration~$k$ is reused.

%\subsection{Quadratic Subspace Models Using a Diagonal Hessian}
%The strategy of reusing previous interpolation points (e.g., through~\eqref{pastnewEq}) is analogous to the linear case, and therefore not elaborated on further. 
Our second nonlinear model approach uses model Hessians that are diagonal, and hence there are $2q+1$ model parameters.

Let $\accolade{\bm{e}^i_q}_{i=1}^q$ be the standard basis of~$\Rq$. At iteration~$k$, $\SSk^q:=\accolade{\szerok}\cup\SSk^{q,1}\cup\SSk^{q,2}$, with $\szerok=\bm{0}_q$, $\SSk^{q,1}:=\accolade{\siik:=\dk\bm{e}^i_q}_{i=1}^q$ and $\SSk^{q,2}:=\accolade{\s_k^{q+i}:=-\dk\bm{e}^i_q}_{i=1}^q$. %$\SSk^q:=\accolade{\bm{0}_p,\dk\bm{e}^1_q,\dots,\dk\bm{e}^q_q,-\dk\bm{e}^1_q,\dots,-\dk\bm{e}^q_q}$.
%$\SSk^q:=\accolade{\bm{0}_p, \siik:=\dk\bm{e}^i_q, \s_k^{q+i}=-\dk\bm{e}^i_q, i=1,\dots,q}.$
We consider $\Phi^q(\s):=(1,s_1,\dots,s_q,\frac{1}{2}s_1^2,\dots,\frac{1}{2}s_q^2)$ so that
{\footnotesize{
		\begin{align}\label{MPhikq}
			\bm{M}(\Phi^q;\SSk^q)=
			\begin{bmatrix}
				1 & \bm{0}_q^\top & \bm{0}_q^\top\\
				\bm{1}_q & \dk\bm{I}_q & \frac{1}{2}\dk^2\bm{I}_q\\
				\bm{1}_q & -\dk\bm{I}_q & \frac{1}{2}\dk^2\bm{I}_q
			\end{bmatrix}
			\in\R^{(2q+1)\times (2q+1)}. 
		\end{align}
}}
Let 
$f_k(\SSk^q;\theta_k):=$
\scalebox{0.8}{$
	\begin{bmatrix}
		f_k(\SSk^{q,1};\theta_k)\\
		f_k(\SSk^{q,2};\theta_k)
	\end{bmatrix},
	$} 
with $f_k(\SSk^{q,j};\theta_k)$, $j=1,2$, defined as above.
Then solving the linear system
\begin{align*}
	\bm{M}(\Phi^q;\SSk^q)\bm{h}_{2q+1}=f_k(\SSk^q;\theta_k),\ \bm{h}_{2q+1}=(h_0,\dots,h_{2q})^\top,
\end{align*}
leads to a quadratic subspace model $\mhatk$ defined as in~\eqref{quadModEq}, where
{\footnotesize{
		\begin{align*}
			\mathfrak{f}_k=h_0,\ \gkhat=(h_1,\dots,h_q)^\top\ \text{and}\  
			\Hkhat=
			\scalebox{0.8}{$
				\begin{bmatrix}
					h_{q+1} & \cdots & \mathbf{0} \\
					\vdots & \ddots & \vdots \\
					\mathbf{0} & \cdots & h_{2q}
				\end{bmatrix}
				$}.
		\end{align*}
}}
Assuming the iteration~$k$ unsuccessful, we define $\bm{M}(\Phi^{q+1};\SSk^{q+1})$ using~\eqref{MPhikq}, and then
%\begin{align*}
%\SSkOne^{q+1}:=\accolade{\bm{0}_{q+1}}\cup\accolade{\pm\hat{q}\dk\bm{e}^i_{q+1}}_{i=1}^q\cup\accolade{\s_{k+1}^{q+1}}\cup\accolade{\s_{k+1}^{2(q+1)}},
%\end{align*}
%where $\s_{k+1}^{q+1}=-\s_{k+1}^{2(q+1)}=\hat{q}\dk\bm{e}^{q+1}_{q+1}$, with $\accolade{\bm{e}^i_{q+1}}_{i=1}^{q+1}$ being the standard basis of~$\R^{q+1}$.
$\mhatkOne$ by solving
{\footnotesize{
		\begin{align*}
			\bm{M}(\Phi^{q+1};\SSk^{q+1})\bm{\ell}_{2q+3}=
			\begin{bmatrix}
				f_k(\SSk^{q,1};\theta_k)\\
				\fsk(\s_{k+1}^{q+1};\theta_{k+1}^{q+1})\\
				f_k(\SSk^{q,2};\theta_k)\\
				\fsk(\s_{k+1}^{2(q+1)};\theta_{k+1}^{2(q+1)})
			\end{bmatrix}, 
		\end{align*}
}}
where $\s_{k+1}^{q+1}=-\s_{k+1}^{2(q+1)}=\hat{q}\dk\bm{e}^{q+1}_{q+1}\in\R^{q+1}$.
%, with $\accolade{\bm{e}^i_{q+1}}_{i=1}^{q+1}$ being the standard basis of~$\R^{q+1}$.
\section{The {\sf ANASTAARS} Algorithm}
{\sf ANASTAARS} is given by Algorithm~\ref{algoANASTAARS} and inspired by {\sf STARS}, as discussed in Section~\ref{sec2SecB}. The {\sf STARS} instance in Algorithm~\ref{algoStoScalTR} constructs a fixed-dimensional subspace model using newly generated interpolation sets and associated newly computed estimates of function values in every iteration.
On the other hand, Algorithm~\ref{algoANASTAARS} explicitly employs an adaptive subspace selection strategy, designed to occasionally (in particular, following every unsuccessful iteration) avoid such entirely new computations. 
%In contrast to previous work involving random subspaces with fixed dimensions, ANASTAARS introduces an adaptive subspace selection strategy. Instead of generating a completely new poised set of interpolation points at each iteration, the proposed method updates the model by generating only a few or even a single new interpolation point, reusing past points (and their corresponding function values) from lower-dimensional subspaces in such a way that the resulting set remains poised.
%Unlike Algorithm~\ref{algoStoScalTR} where function estimates are always recomputed at each iteration using new samples of the noisy objective
%for an appropriate sample size~$\pi_k$, $\fok=\frac{1}{\pi_k}\sum_{\ell=1}^{\pi_k}f_{\theta_\ell^{\bm{0}}}(\xk)$, where the~$\theta_\ell^{\bm{0}}$ are realizations of independent random samples of~$\ubar{\theta}$.
\begin{algorithm}[htb]
\caption{New {\sf ANASTAARS} algorithm.}
\label{algoANASTAARS} 
{%\footnotesize{
\textbf{[0] Initialization}\\
\hspace*{4mm}Fix \scalebox{0.8}{$\gamma>1$}, \scalebox{0.8}{$\ef,\eta_1\in(0,1)$}, \scalebox{0.8}{$\eta_2,\dmax,r>0$}, \scalebox{0.8}{$\!\delta_0\!\in\! (0,\dmax)$}, \scalebox{0.8}{$q_{\max}\!\in\![\![2,d]\!]$},
%\\ \hspace*{4mm}
\scalebox{0.8}{$q_0\!\in\!\intbracket{1,q_{\max}}$}. Select \scalebox{0.8}{$\bm{{\x}}_0\in\rd$}, set \scalebox{0.8}{$F_{\rm flag}=0$}, \scalebox{0.8}{$k \gets 0$}\\ 
\hspace*{4mm}and \scalebox{0.8}{$q\gets q_0$}.\\
\textbf{[1] Construction of subspace model}\\
\hspace*{4mm}If \scalebox{0.8}{$F_{\rm flag}=0$} or \scalebox{0.8}{$(q+1)>q_{\max}$}: reset $q\gets q_0$, generate \scalebox{0.8}{$\Qk\!=\!\sqrt{\frac{d}{q}}\Uk\in\rdq$}
%\\ \hspace*{6.7mm}
via Haar measure. Generate a poised set\\ \hspace*{4mm}\scalebox{0.8}{$\SSk\subset\R^q$}.
%\\ \hspace*{6.7mm}
Build a $q$-dimensional subspace model \scalebox{0.8}{$\mhatk:\R^q\to\R$} using \scalebox{0.8}{$\Qk$} and \scalebox{0.8}{$\SSk$}. \\
\hspace*{4mm}Otherwise: generate \scalebox{0.8}{$\mu_k\in\bm{U}_{k-1}^\perp$} from \scalebox{0.8}{$\mathcal{U}(\mathbb{S}^{d-1})$}, define \scalebox{0.8}{$\Uk=[\bm{U}_{k-1},\mu_k]$} and 
%\\ \hspace*{6.7mm}
\scalebox{0.8}{$\Qk=\sqrt{\frac{d}{q+1}}\Uk\in\R^{d\times (q+1)}$} (see~\eqref{QkPlusOne}).\\ 
\hspace*{4mm}Generate a new point \scalebox{0.8}{$\s_k^{q+1}\in \R^{q+1}$}
%\\ \hspace*{6.7mm}
and obtain a poised set \scalebox{0.8}{$\SSk\subset\R^{q+1}$} using \scalebox{0.8}{$\mathbb{S}_{k-1}$} and \scalebox{0.8}{$\s_k^{q+1}$} (see Section~\ref{Sec3ModelBuild}). \\
%\\ \hspace*{6.7mm}
\hspace*{4mm}Build a \scalebox{0.8}{$(q+1)$}-dimensional subspace model \scalebox{0.8}{$\mhatk\!:\R^{q+1}\!\!\to\!\R$} using \scalebox{0.8}{$\Qk$}, \scalebox{0.8}{$\SSk$}, 
%\\ \hspace*{6.7mm}
function estimates associated to \scalebox{0.8}{$\mathbb{S}_{k-1}$},\\ 
\hspace*{4mm}and at \scalebox{0.8}{$\s_k^{q+1}$} (see Section~\ref{Sec3ModelBuild}).
%\\ \hspace*{6.7mm}
Set \scalebox{0.8}{$q\gets q+1$} and \scalebox{0.8}{$F_{\rm flag}=0$}. \\
\textbf{[2] Step calculation}\\
\hspace*{4mm}Compute \scalebox{0.8}{$\sk\approx {\argmin}\left\{ \mhatk(\s) : \; \s\in\R^q, \normii{\s}\leq \dk\right \}$} satisfying~\cite[(2.1)]{DzaWildSub2022}.\\
\textbf{[3] Estimate computation}\\
\hspace*{4mm}Compute \scalebox{0.8}{$\ef$}-accurate estimates \scalebox{0.8}{$\fok\approx f(\xk)$} and \scalebox{0.8}{$\fsk\approx f(\xk+\Qk\sk)$} and 
%\\ \hspace*{4mm}
obtain an estimate of noise level at \scalebox{0.8}{$\xk$}\\ 
\hspace*{4mm}through its standard deviation \scalebox{0.8}{$\epsok$};
%\\ \hspace*{4mm}
if available, use samples from previous iterations. \\
\textbf{[4] Updates}\\
\hspace*{4mm}Compute \scalebox{0.8}{$\tilde{\rho}_k=\frac{\fok-\fsk+r\epsok}{\mhatk(\bm{0})-\mhatk(\sk)}$}\\
\hspace*{4mm}If \scalebox{0.8}{$\tilde{\rho}_k\geq \eta_1$} and \scalebox{0.8}{$\normii{\gkhat}\geq \eta_2\dk$} (\textbf{success}):
%\\ \hspace*{6.7mm}
set \scalebox{0.8}{$\bm{\x}_{k+1}=\xk+\Qk\sk$} and \scalebox{0.8}{$\dkun=\min\accolade{\gamma\dk,\dmax}$}.  \\
\hspace*{4mm}Otherwise (\textbf{failure}): \scalebox{0.8}{$\bm{\x}_{k+1}=\xk$},  \scalebox{0.8}{$\dkun=\gamma^{-1}\dk$} and \scalebox{0.8}{$F_{\rm flag}=1$}.\\
\hspace*{4mm}Update the iteration counter \scalebox{0.8}{$k \gets k+1$} and go to~\textbf{[1]}.\\
}%}
\end{algorithm}
Indeed, if iteration~$k-1$ is successful or its $q$-dimensional subspace model is such that $q+1>q_{\max}$ for some $q_{\max}\in \intbracket{2,d}$, then a $q_0$-dimensional subspace model, with $q_0\in \intbracket{1,q_{\max}}$, is constructed at iteration~$k$ using a new interpolation set $\SSk$ and its associated function estimates. Otherwise, a $(q+1)$-dimensional subspace model is constructed using a set $\SSk\subset\R^{q+1}$ obtained from $\mathbb{S}_{k-1}\subset\R^{q}$ combined with a new interpolation point $\s_k^{q+1}\in \R^{q+1}$, and their associated function estimates as detailed in Section~\ref{Sec3ModelBuild}. We note that the condition $q+1>q_{\max}$ prevents the attempted construction of $d'$-dimensional subspace models, with $d'>d$, following an  unsuccessful iteration. 

As an additional feature, and inspired by noise-aware numerical optimization (e.g.,~\cite{LarMenShiQAOA2024,menickelly2023latency,MMSWMX23}), {\sf ANASTAARS} incorporates an estimate~$\epsok$ of the noise level at~$\xk$ into a modified ratio test $\tilde{\rho}_k$ of the reduction in the estimates and the reduction in the model. 
This alternative ratio, seen in Step 4 of Algorithm~\ref{algoStoScalTR}, is employed to determine whether an iteration is successful or not. 
We estimate $\epsok$ via the sample standard deviation; that is, for the computed number~$\pi_k>1$ of samples, $\epsok^2:=\frac{1}{\pi_k-1}\sum_{\ell=1}^{\pi_k}(f_{\theta_\ell^{\bm{0}}}(\xk)-\fok)^2$, where  $\fok=\frac{1}{\pi_k}\sum_{\ell=1}^{\pi_k}f_{\theta_\ell^{\bm{0}}}(\xk)$.
%\mmnote{Based on how little the motivation for this $\rho$ test is actually presented here, it may make sense to proportionally ease off the gas in the intro in discussing noise-awareness. Will circle back to see how this reads.}
%, and the~$\theta_\ell^{\bm{0}}$ are realizations of independent random samples of~$\ubar{\theta}$.
%Unlike {\sf STARS} where function estimates are always recomputed at each iteration using new samples of the noisy objective, available samples from previous iterations are reused by {\sf ANASTAARS}, inspired by~\cite{audet2019stomads,}
%for an appropriate sample size~$\pi_k$, $\fok=\frac{1}{\pi_k}\sum_{\ell=1}^{\pi_k}f_{\theta_\ell^{\bm{0}}}(\xk)$, where the~$\theta_\ell^{\bm{0}}$ are realizations of independent random samples of~$\ubar{\theta}$.

\section{Numerical experiments}
The practical performance of a variant of Algorithm~\ref{algoANASTAARS} using MFN subspace quadratic models (introduced in Section~\ref{SecBFrobNorMod}) is illustrated through experiments conducted on standard QAOA benchmarks. This variant, referred to as {\sf ANASTAARS-QD2}, uses $q$-dimensional subspace models, with $q\in \intbracket{q_0,q_{\max}}$, where $q_0=2$ and $q_{\max}=d$. 
We set the other algorithmic parameters as $r=1$, $\gamma=2$, $\eta_1=0.01$, $\eta_2=0.9$, $\dmax=5$, and $\delta_0=1$. 

We compare {\sf ANASTAARS-QD2} with {\bl{\sf STARS-QD2}, a variant of {\sf STARS} using MFN $2$-dimensional subspace quadratic models, and} other methods previously tested in the literature for optimizing VQAs. More precisely, {\sf ANAS-}{\sf TAARS-QD2} is {\bl also} compared with {\sf PyBOBYQA}~\cite{PyBOYQA2019}, the implicit filtering method referred to as {\sf ImFil}~\cite{Kelley2011} (each as wrapped in {\sf scikit-quant}~\cite{ScikitQuant}), {\sf NOMAD}~\cite{Le09b}, 
%the method of Simultaneous Perturbation  Stochastic Approximation ({\sf SPSA})~\cite{SPSA1992}, 
{\sf ANATRA}~\cite{LarMenShiQAOA2024}, and {\sf NEWUOA}~\cite{Po2006}. We note that while a variant of {\sf NOMAD} is also wrapped in {\sf scikit-quant}, we have chosen to use its version~4 because it is more recent and offers significant improvements. 
Moreover, for the same reasons provided in~\cite[Section~5.2]{LarMenShiQAOA2024}, we intentionally omitted the optimizer {\sf SnobFit}~\cite{Huyer2008} wrapped in {\sf scikit-quant}. We {\bl note that unlike~\cite{LarMenShiQAOA2024}, no comparison was made with the method of Simultaneous Perturbation  Stochastic Approximation ({\sf SPSA})~\cite{SPSA1992}, not only due to its unavailability in Qiskit~v0.46.1 (used in our experiments), but also because of its poor performance compared to the aforementioned optimizers for VQAs.}
%chose to employ Qiskit's implementation of~{\sf SPSA}.

{\sf PyBOBYQA} is an extension of the widely used DFO optimizer {\sf BOBYQA}~\cite{bobyqa}. It is a model-based DFO optimizer that incorporates several heuristics aimed at enhancing the robustness of {\sf BOBYQA} in the presence of noise.
{\sf ImFil} is a DFO method with a relatively intricate implementation that fundamentally operates as an inexact quasi-Newton approach using gradient estimates computed via central finite differences with initially large difference parameters. 
%When an iteration yields an objective value at the finite difference stencil center that is not better than any of the corresponding forward or backward evaluation points, the algorithm registers a {\it stencil failure} and refines the finite difference parameter. 
Intentionally designed to handle noisy cost functions, {\sf ImFil} is a competitive method in general noise settings~\cite{LarMenShiQAOA2024}.
{\sf NOMAD} is an implementation of the {\sf MADS}~\cite{AuDe2006} algorithm, which, unlike model-based methods, does not build explicit gradient or higher-order models. Instead, {\sf NOMAD} explores a mesh of trial points and updates the incumbent if improvement is observed. 
Although not originally designed for noisy problems, {\sf NOMAD} could potentially exhibit robustness to noise due to its model-free nature. 
%Recent stochastic extensions of {\sf MADS} exist, but are not yet included in {\sf NOMAD}.
%{\sf SPSA} is a simple method that estimates a stochastic gradient using a two-point finite difference approximation. This approach is particularly appealing, especially in high-dimensional settings, because it may enable meaningful optimization progress with only two function evaluations, unlike model-based methods, which typically require $\mathcal{O}(d)$ evaluations just to initiate the optimization process.
{\sf ANATRA} is a noise-aware, model-based trust-region algorithm designed for noisy DFO problems, such as those arising in VQAs. The key feature of {\sf ANATRA} is its attempt to explicitly incorporate dynamic estimates of noise into its $\rho$ test, as we mimic in {\sf ANASTAARS-QD2}. 
{\sf NEWUOA} is an iterative model-based DFO algorithm using interpolation models built with $2d+1$ points. It employs a trust-region approach where only one interpolation point per iteration is altered. 
%It ensures numerical stability and efficiency by carefully choosing initial models, maintaining independence among interpolation points, and performing stable updates. 
%A computational work of $\mathcal{O}\left((3d+1)^2\right)$ per iteration makes the method suitable for high-dimensional problems.

In our first experiments we simulate QAOA MaxCut circuits with $p=5$ layers in Qiskit~\cite{QiskitContributor2023}, resulting in a set of $2p=10$ free parameters $(\gamma_1,\dots,\gamma_5,\beta_1,\dots,\beta_5)^\top=:\x$ in the $(d=10)$-dimensional Problem~\eqref{stochProb}. 
We use the QASM simulator in Qiskit to simulate an ideal (state-vector) execution of the QAOA circuit in our experiments.
By using the MaxCut cost function values suggested by QASM, we compute the shot-averaged estimates $f^{\bm{j}}_k=\frac{1}{B}\sum_{\ell=1}^{B}f_{\theta_\ell^{\bm{j}}}(\xk)$, $\bm{j}\in\accolade{\bm{0},\s}$, used by Algorithm~\ref{algoANASTAARS} {\bl and {\sf STARS-QD2}}. Here $B:=\pi_k\in\accolade{50,100,500,1000}$ denotes the per-evaluation shot count. 
We experiment with the MaxCut problem both on a toy graph and on standard benchmark Chv\'atal graphs, the former having a MaxCut value of~$6$ and the latter having a MaxCut value of $20$.
%\textcolor{red}{This section is written based on the numerical section of published ANATRA: not sure I found a definition of Toy or a matrix there. And the QASM sentence below is extracted from Section 5.4(ANATRA) which is related to the ``Tests on VQA problems.}

\begin{figure}[htb!]
\centering
\includegraphics[scale=0.2]{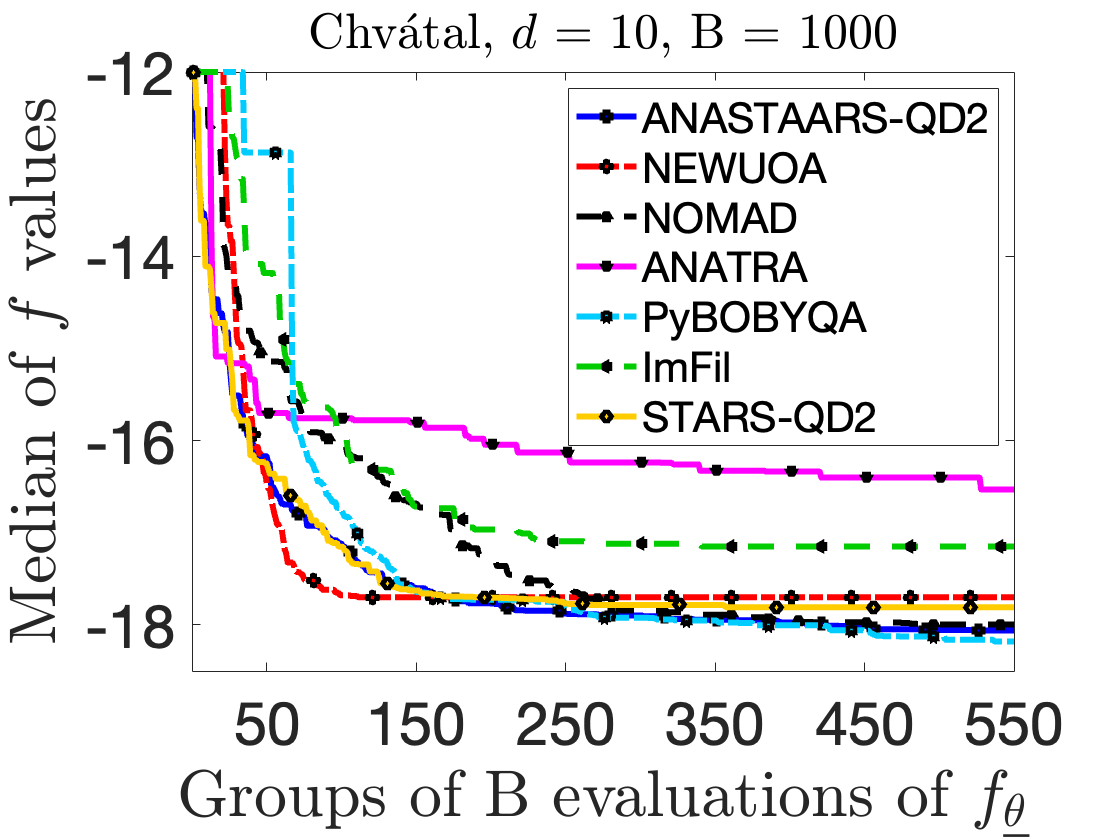}
\includegraphics[scale=0.2]{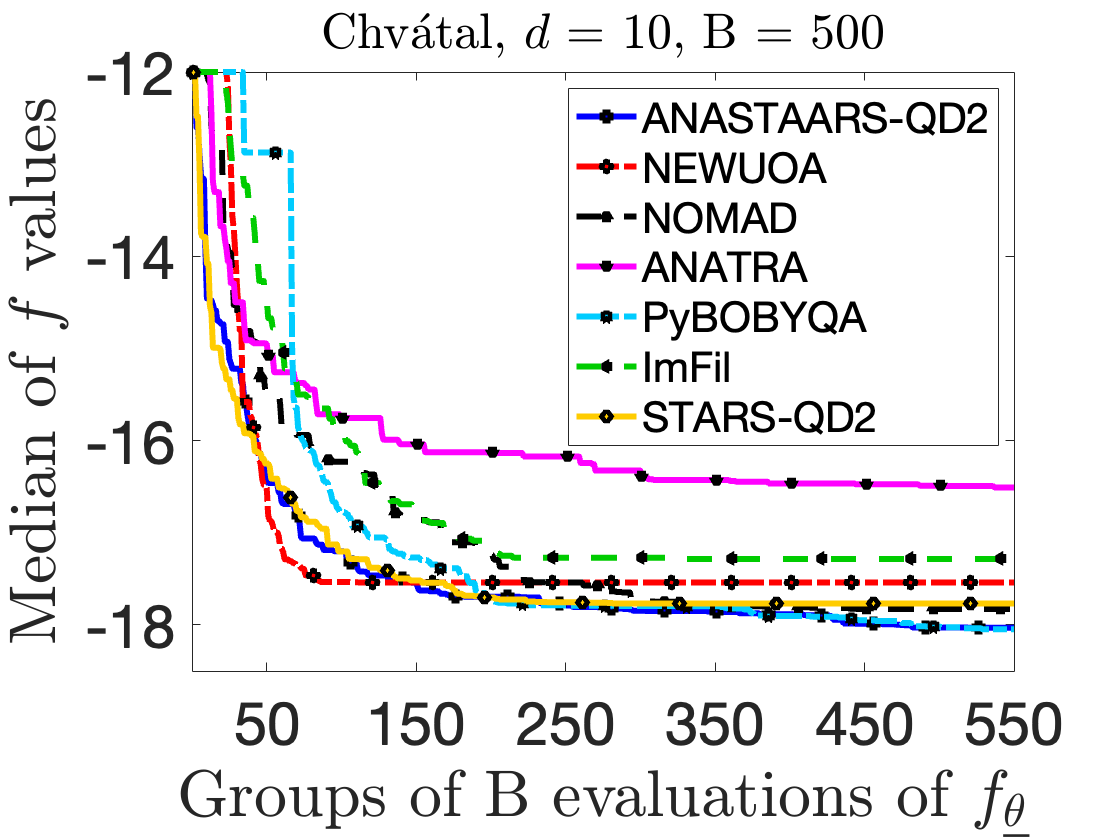}
\includegraphics[scale=0.2]{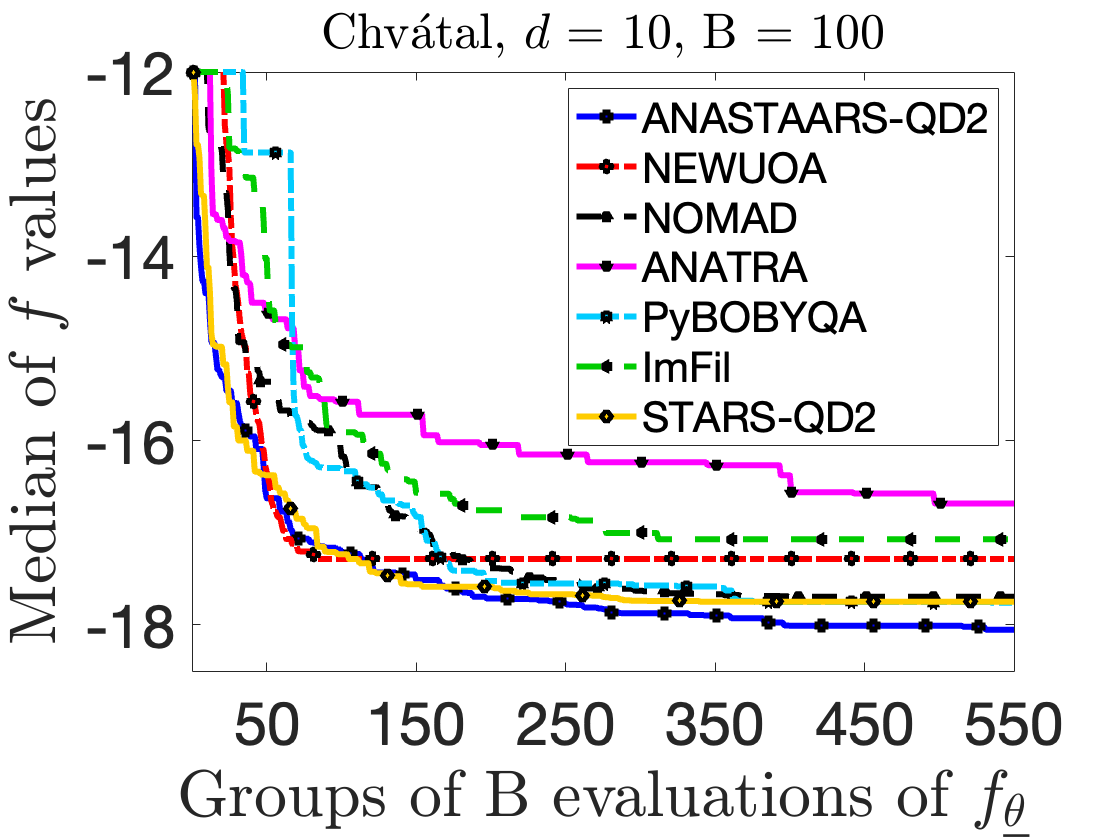}
\includegraphics[scale=0.2]{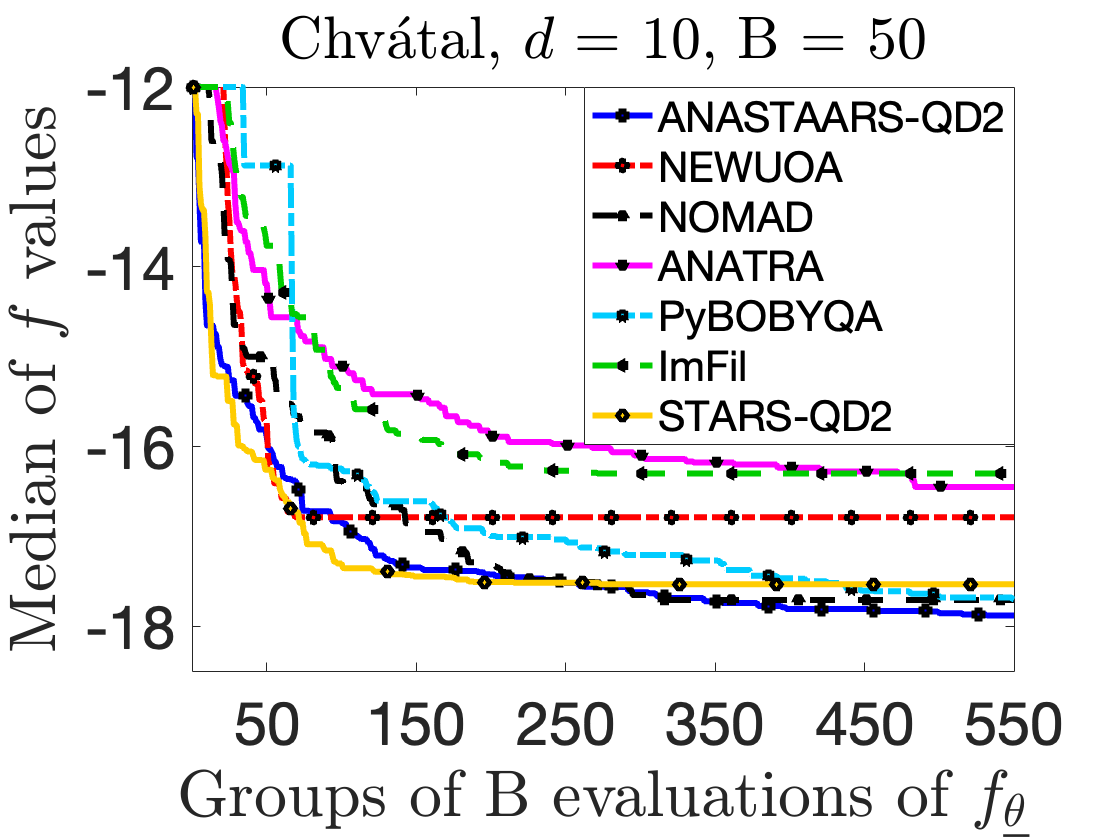}
\caption{\small{Median optimizer trajectories for the Chv\'atal graph with {$d=~\!\!\!10$}.} \label{chvatal_dim10}}
\end{figure}

\begin{figure}[ht!]
\centering
\includegraphics[scale=0.2]{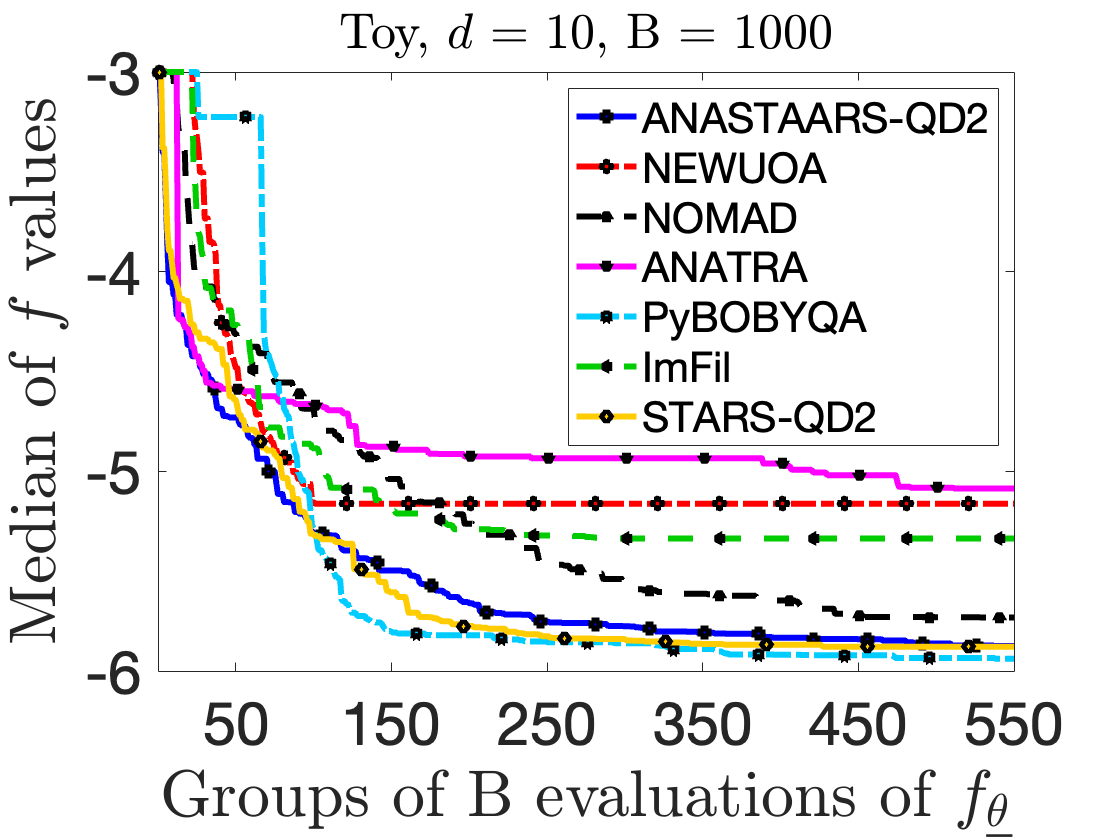}
\includegraphics[scale=0.2]{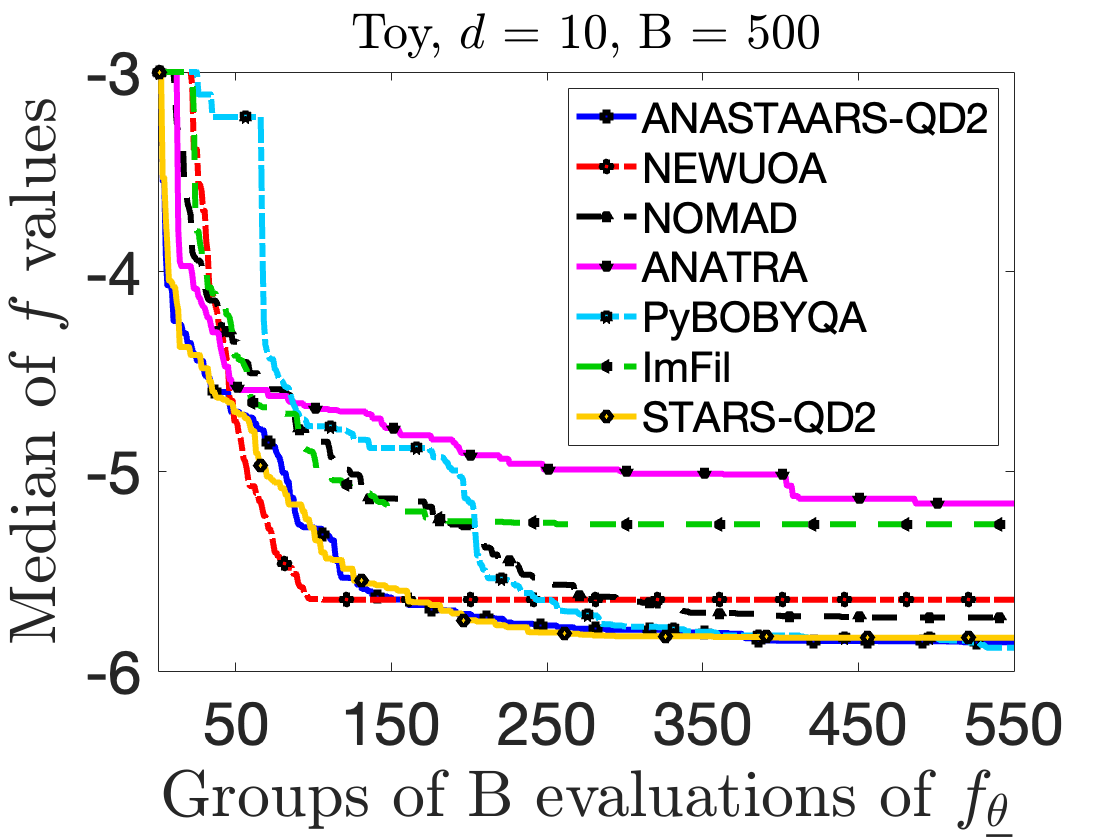}
\includegraphics[scale=0.2]{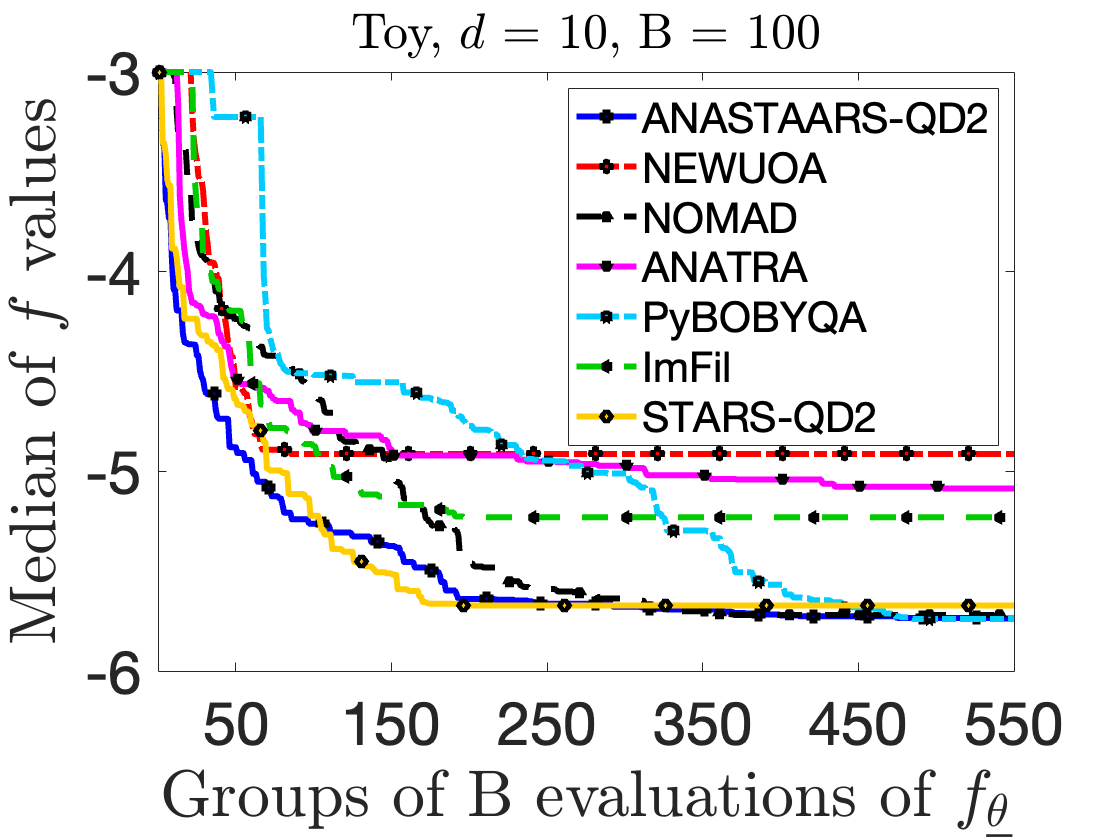}
\includegraphics[scale=0.2]{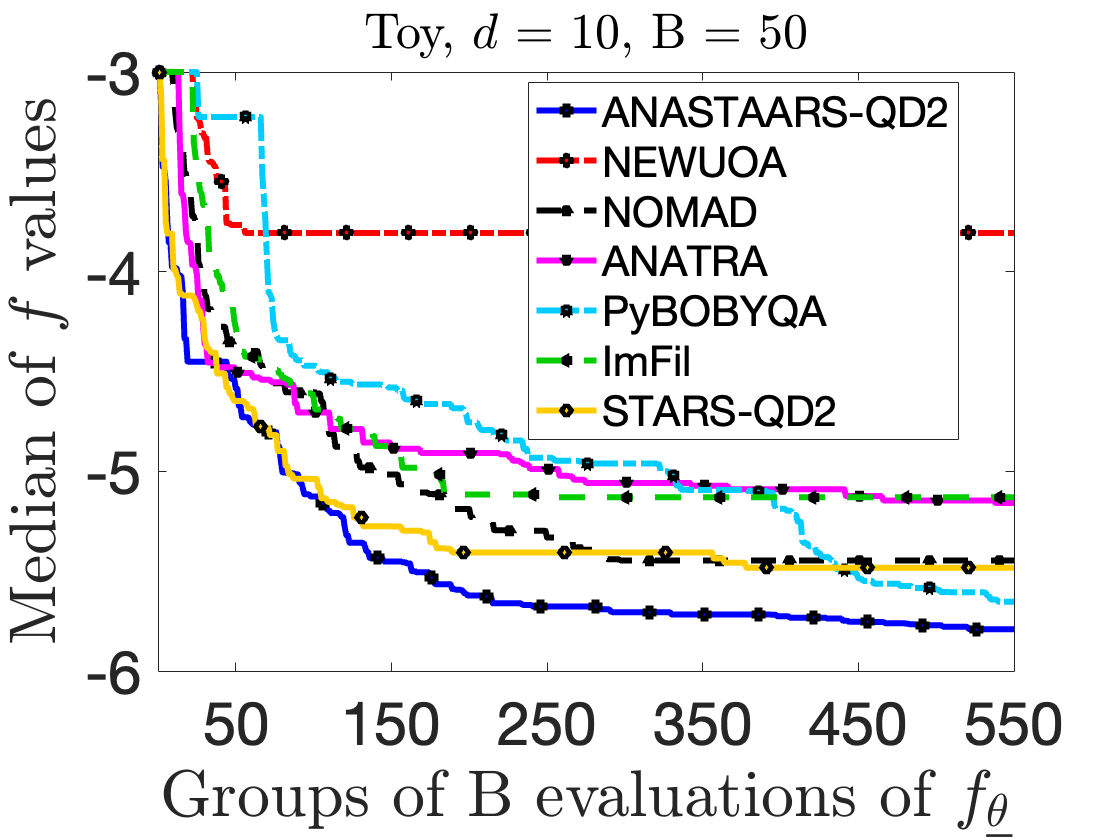}
\caption{\small{Median optimizer trajectories for the toy graph with $d=10$.}}
\label{Toy_dim10}
\end{figure}

\begin{figure}[ht!]
\centering
\includegraphics[scale=0.25]{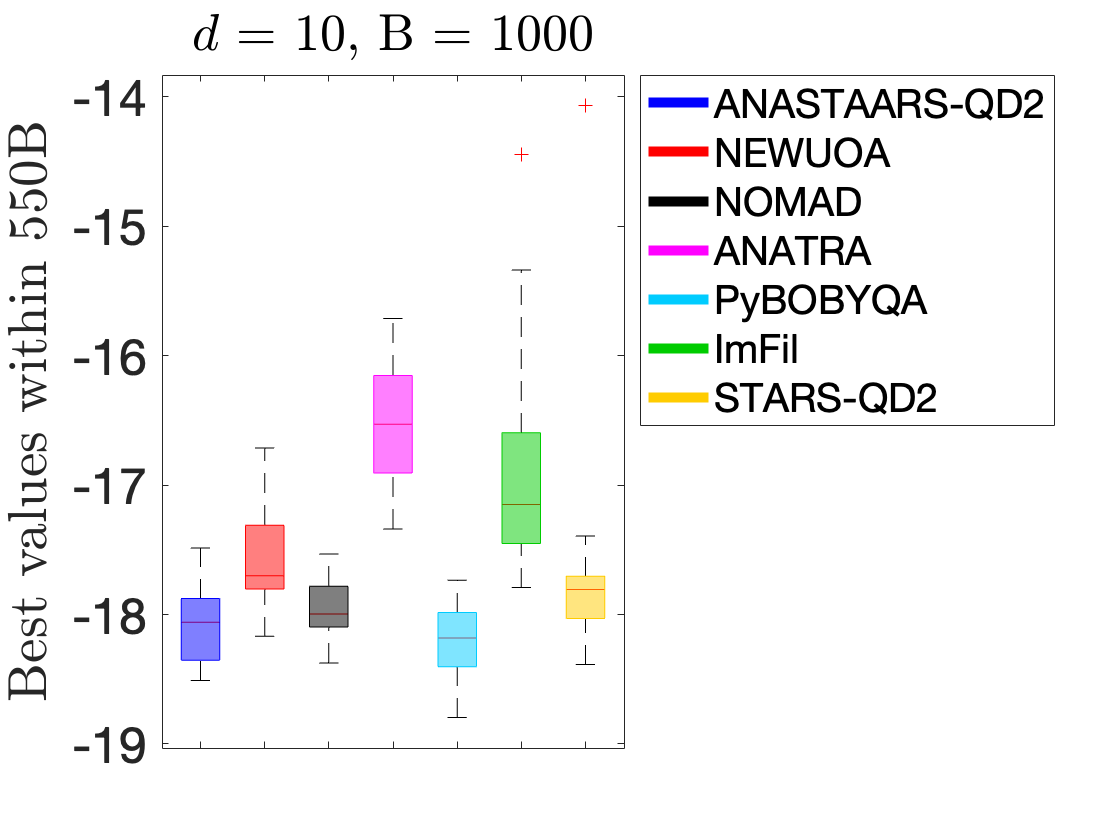}
\caption{\small{Distribution across the 30 trials for the Chv\'atal graph with $d=10$ and $B=1000$.} \label{chvatal_550S_1000}}
\end{figure}

Figures~\ref{chvatal_dim10} and~\ref{Toy_dim10} depict the median performance over~$30$ trials for each optimizer on the Chv\'atal and toy graphs, respectively.
%\textcolor{red}{I cite ANATRA paper here: ``We experiment with the MaxCut problem both on a toy graph with MaxCut value of 6 and on the Chv\'atal 
%graph, a standard benchmark that has a MaxCut value of 20. In our first set of QAOA experiments, illustrated in 
%Figure 4, we employed the QASM simulator in Qiskit to simulate \textbf{an ideal} execution of the QAOA circuit.''}
These figures show that {\sf ANASTAARS-QD2} is able to 
%match excellent initial reduction achieved by {\sf SPSA} \textit{and} 
match the best long-term performance achieved by 
%{\sf ANATRA} and 
{\sf PyBOBYQA} for these $10$-dimensional problems. {\bl Unlike {\sf STARS-QD2}, {\sf ANASTAARS-QD2} does not easily get stuck. This is likely due to the latter's adaptive subspace strategy, which---following unsuccessful iterations---permits the use of higher-dimensional, potentially improved or more accurate models, thereby facilitating continued progress toward better solutions.} 
For the largest shot count and total number of shots, Figure~\ref{chvatal_550S_1000} depicts the variability across the 30 trials and illustrates that even at this large budget of shots, {\sf ANASTAARS-QD2} is among the best optimizers in terms of trial quantiles.

\begin{figure}[p!]
\centering
\includegraphics[scale=0.14]{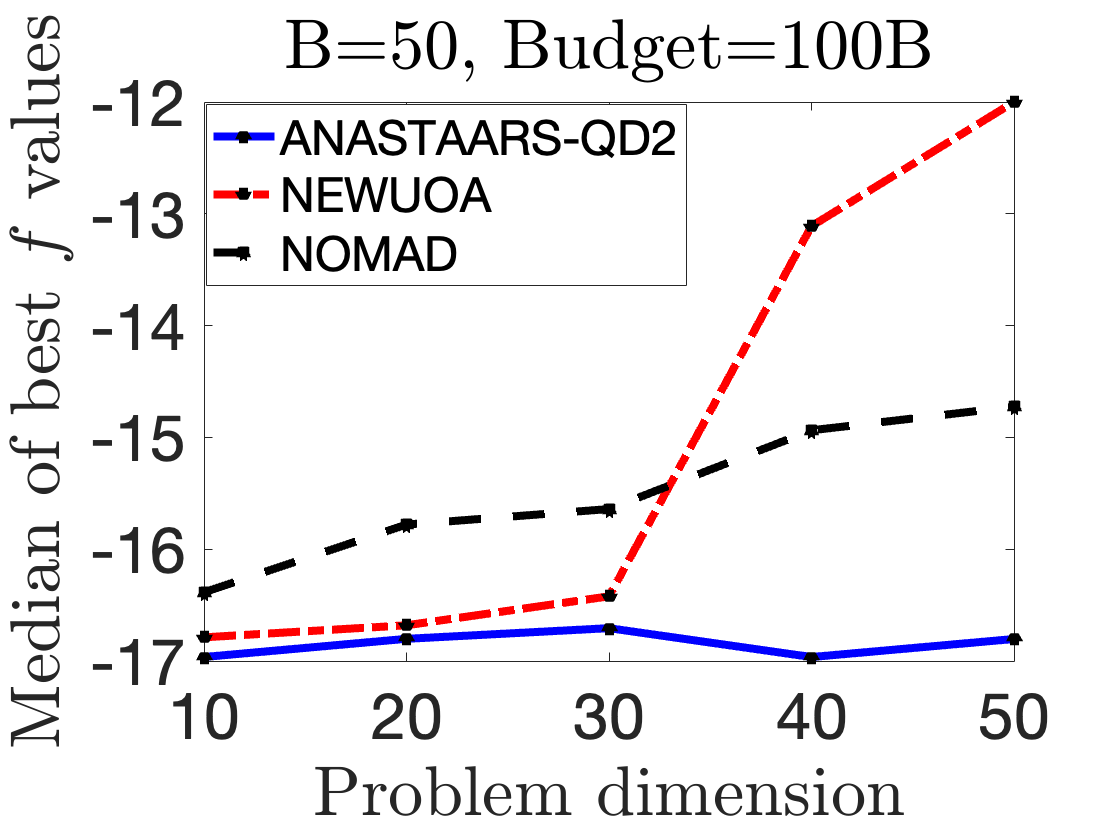}
\includegraphics[scale=0.14]{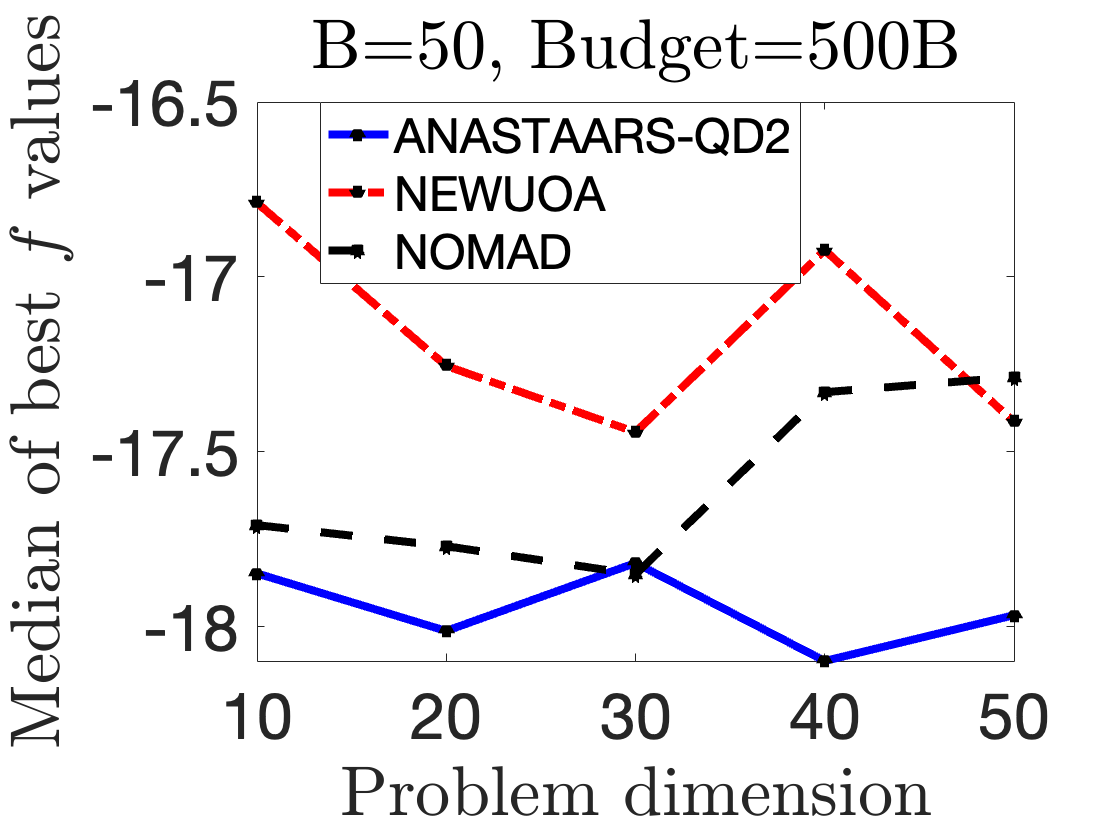}
\includegraphics[scale=0.14]{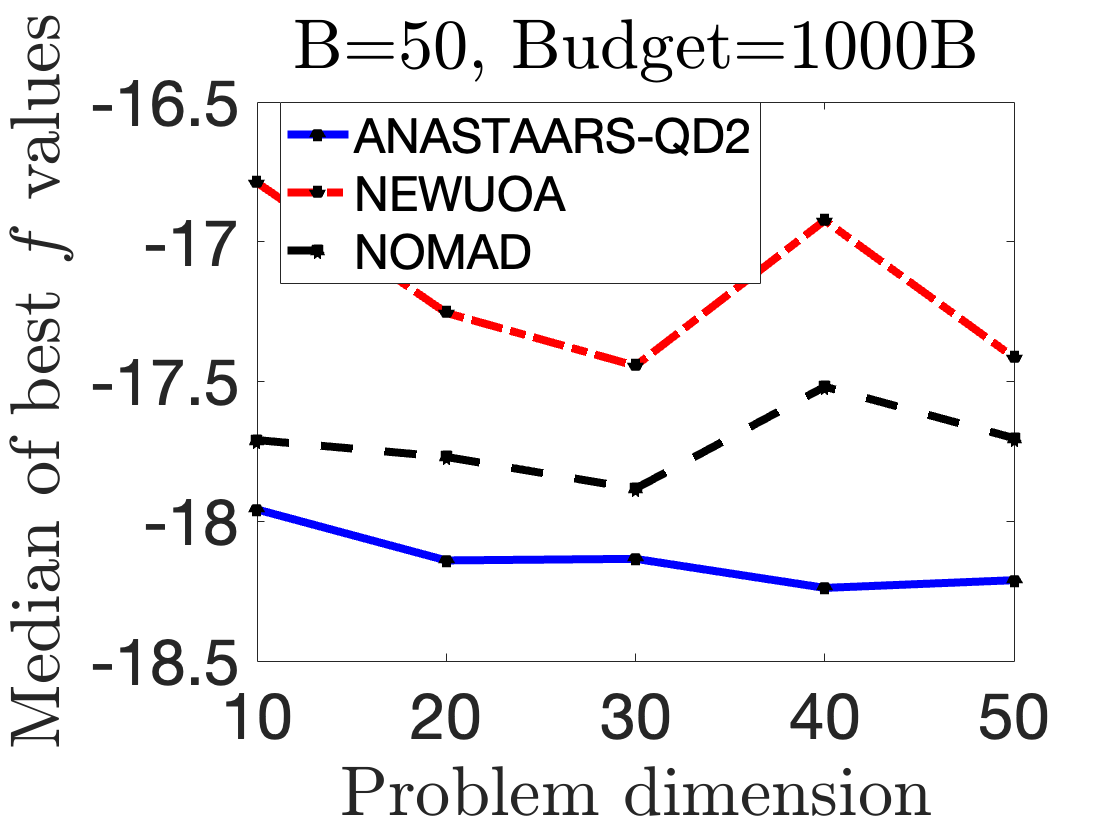}
\includegraphics[scale=0.14]{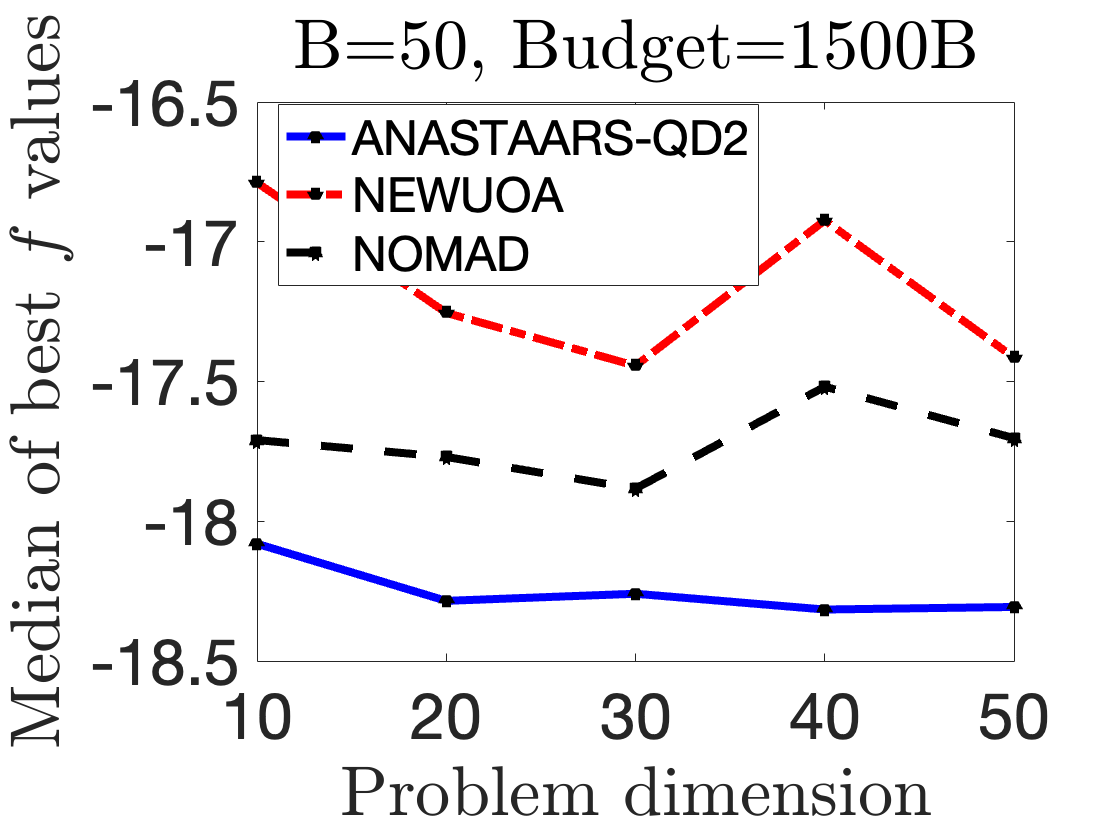}
\includegraphics[scale=0.14]{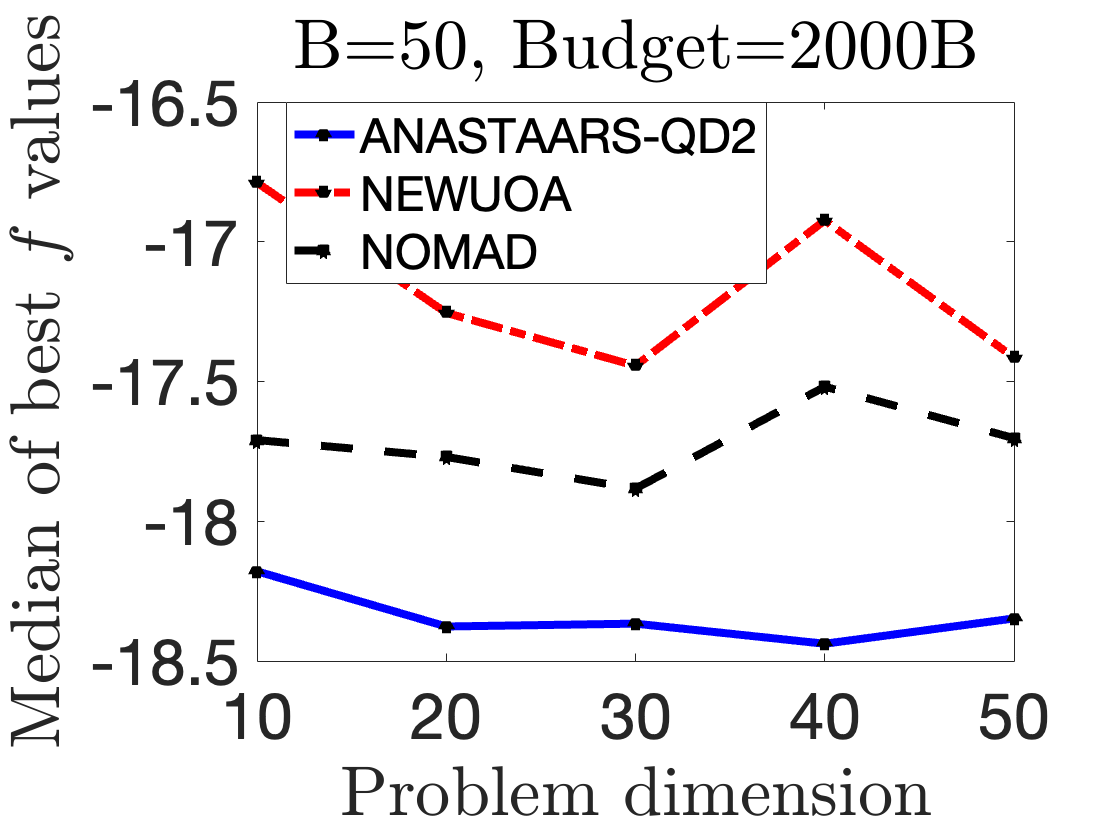}
\includegraphics[scale=0.14]{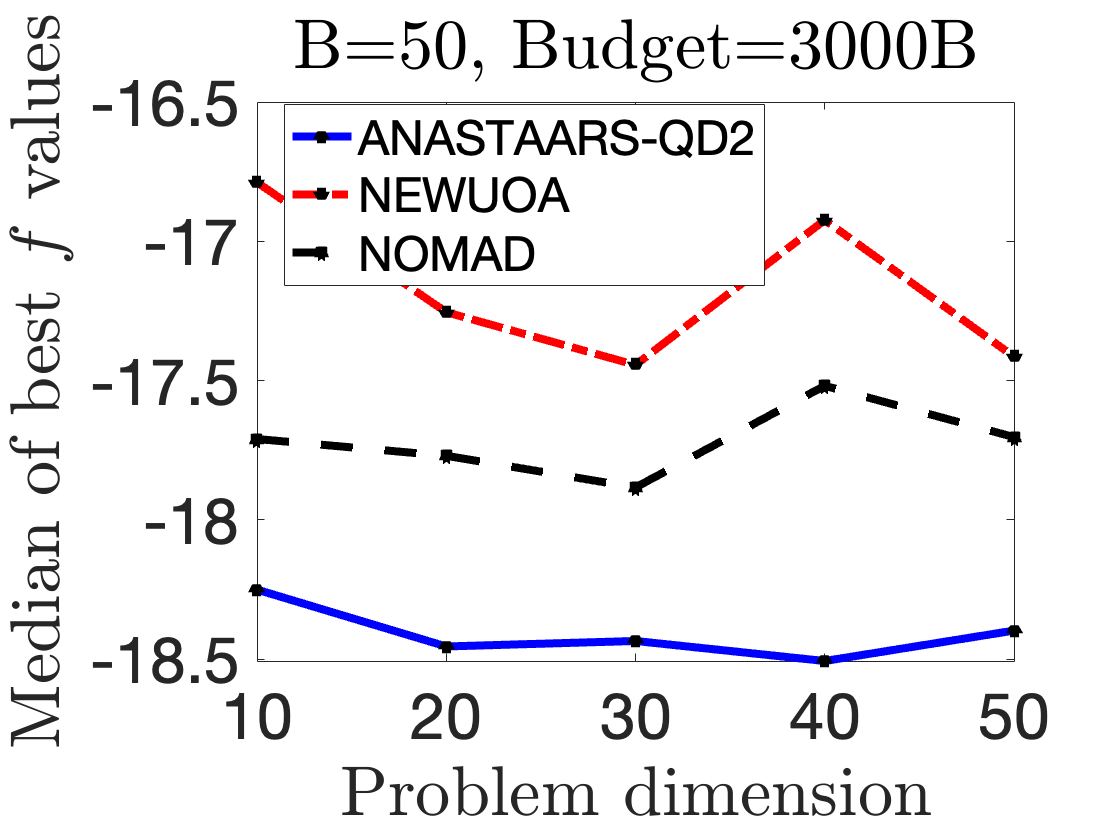}
\caption{\small{Chv\'atal scalability with 50 shots.}} 
%I wouldn't include NEWUOA in these figures. It didn't perform well on the 10-dimensional problems above. But Py-BOBYQA was one of the best (if not THE BEST) above. It has to be included here. \textcolor{red}{ In this case, this means ANATRA should be included here as well except that we can't run it for higher dimensions if I'm correct? Here I simply excluded PyBOBYQA due to its worse performance for limited budget (from the above graphs). The goal of these last graphs is to fix the budget and play with the dimensions, per our discussions back then. We can therefore remove large budgets here if you're fine with that. This said, I would have liked to try PyBOBYQA here as well, but sorry, can't before Monday ... a lot going on :) For example, we can select the graphs corresponding to a budget of 100S (see the conv. graphs above)}
%\textcolor{blue}{MM: FWIW, I'm with Jeff here. This is a strange downselect that minimally needs some words. PyBOBYQA would have been the optimizer to use over both NEWUOA and NOMAD here. }\textcolor{red}{Well, if someone would like to run PyBOBYQA now and share the data for all these dimensions and budgets, I'm happy to generate the graphs. Otherwise, I can only do this later :)}}
\label{Chavatal50_budgets}
\end{figure}

\begin{figure}[p!]
\centering
\includegraphics[scale=0.14]{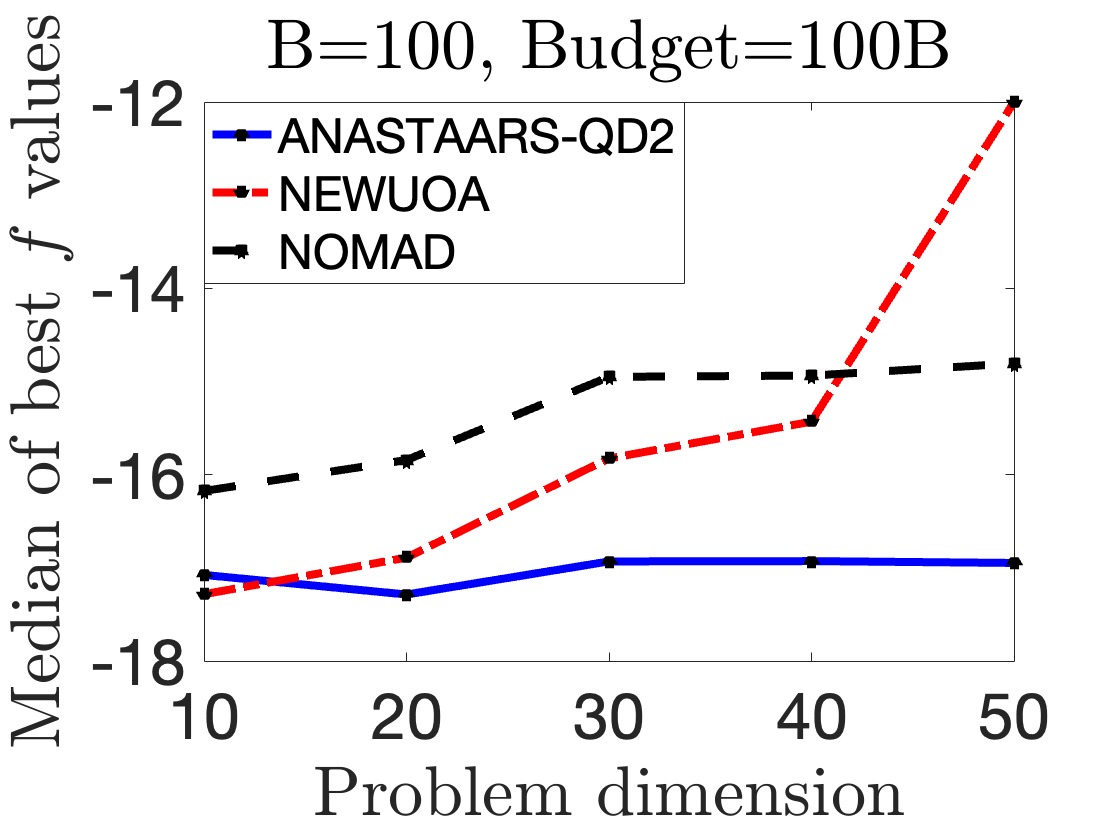}
\includegraphics[scale=0.14]{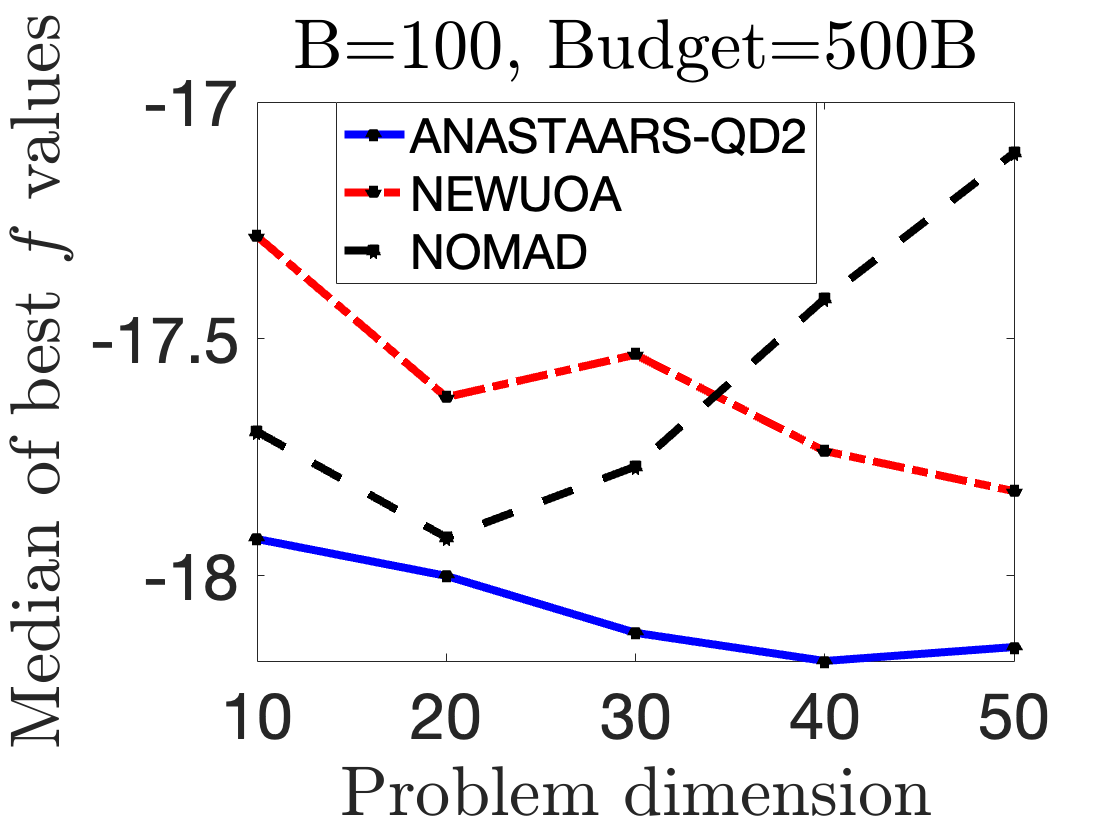}
\includegraphics[scale=0.14]{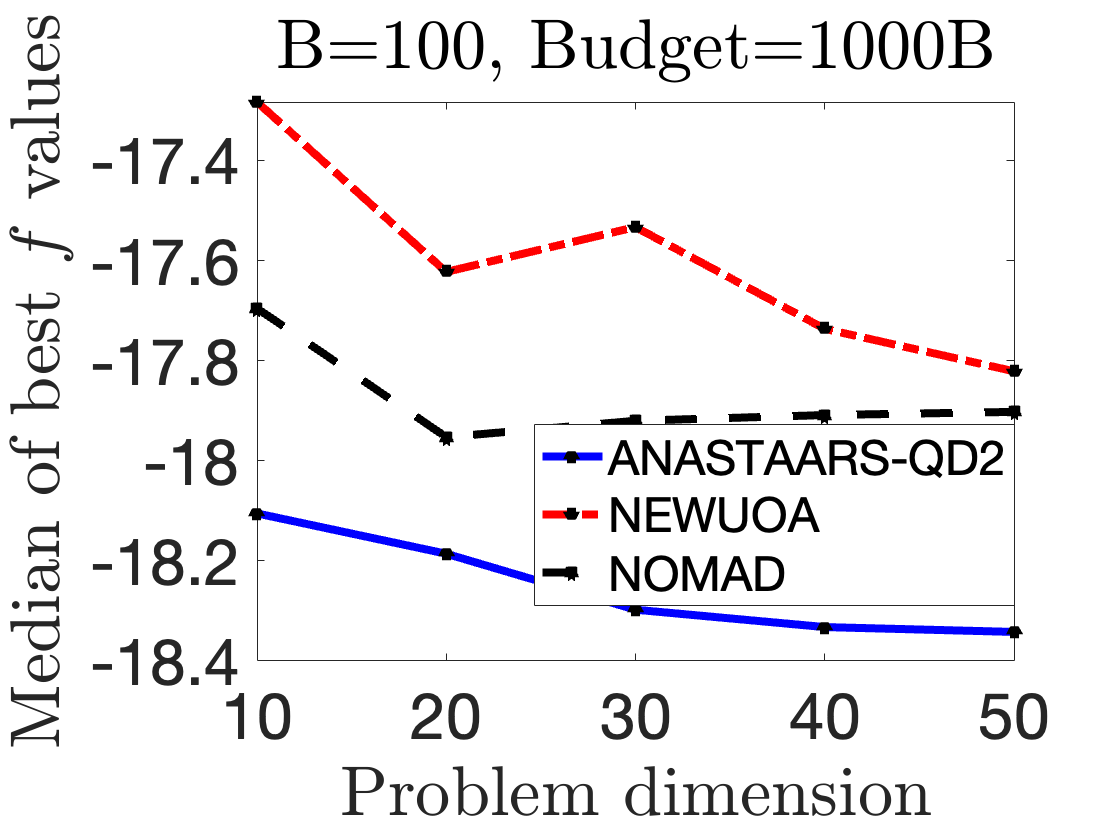}
\includegraphics[scale=0.14]{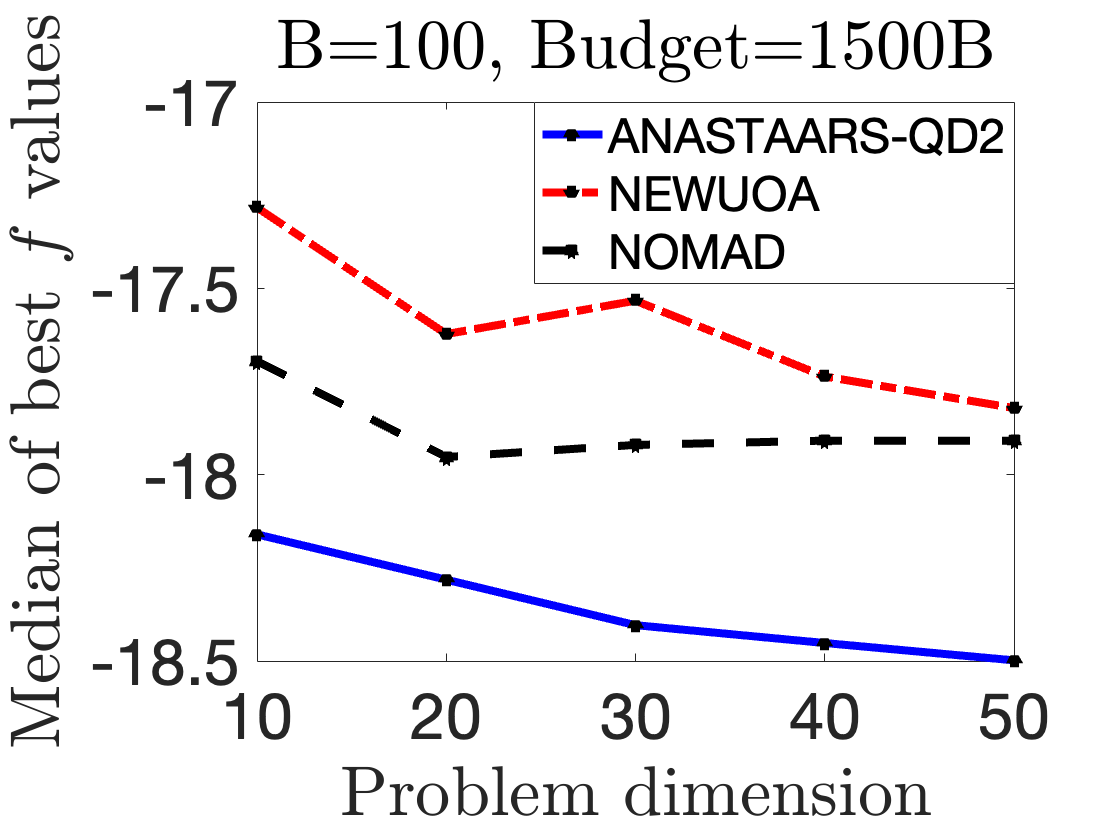}
\includegraphics[scale=0.14]{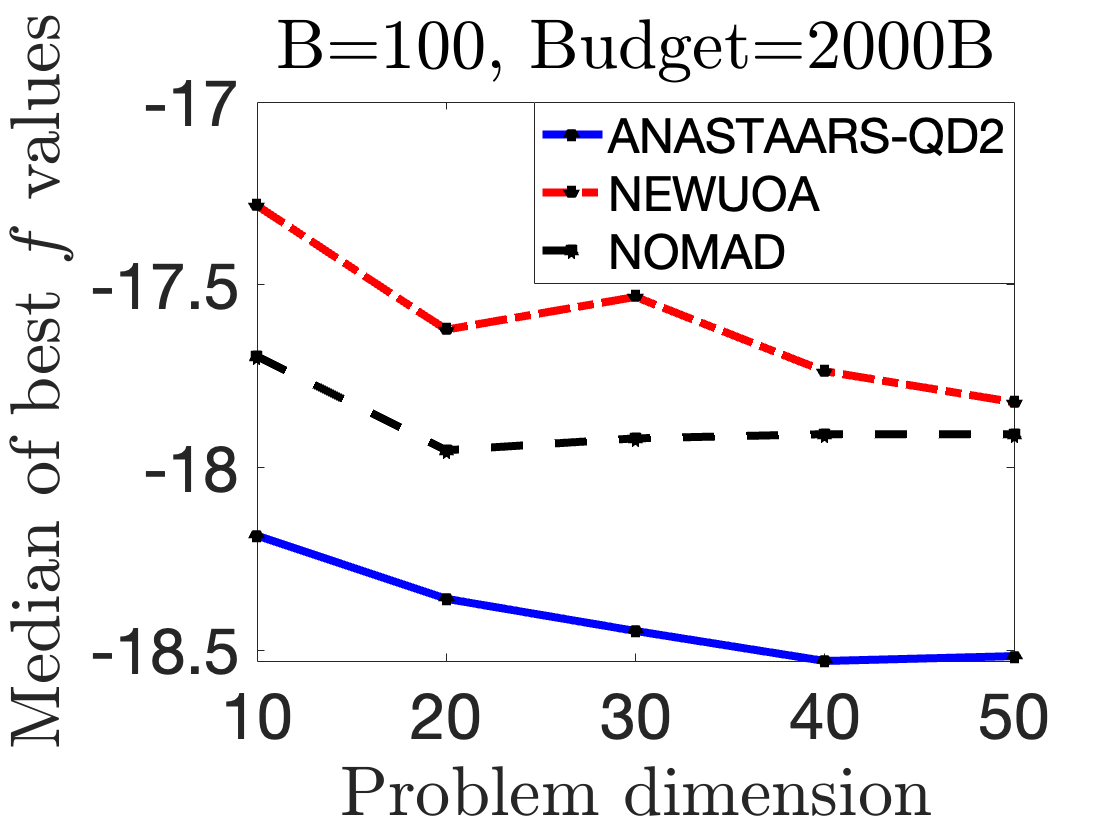}
\includegraphics[scale=0.14]{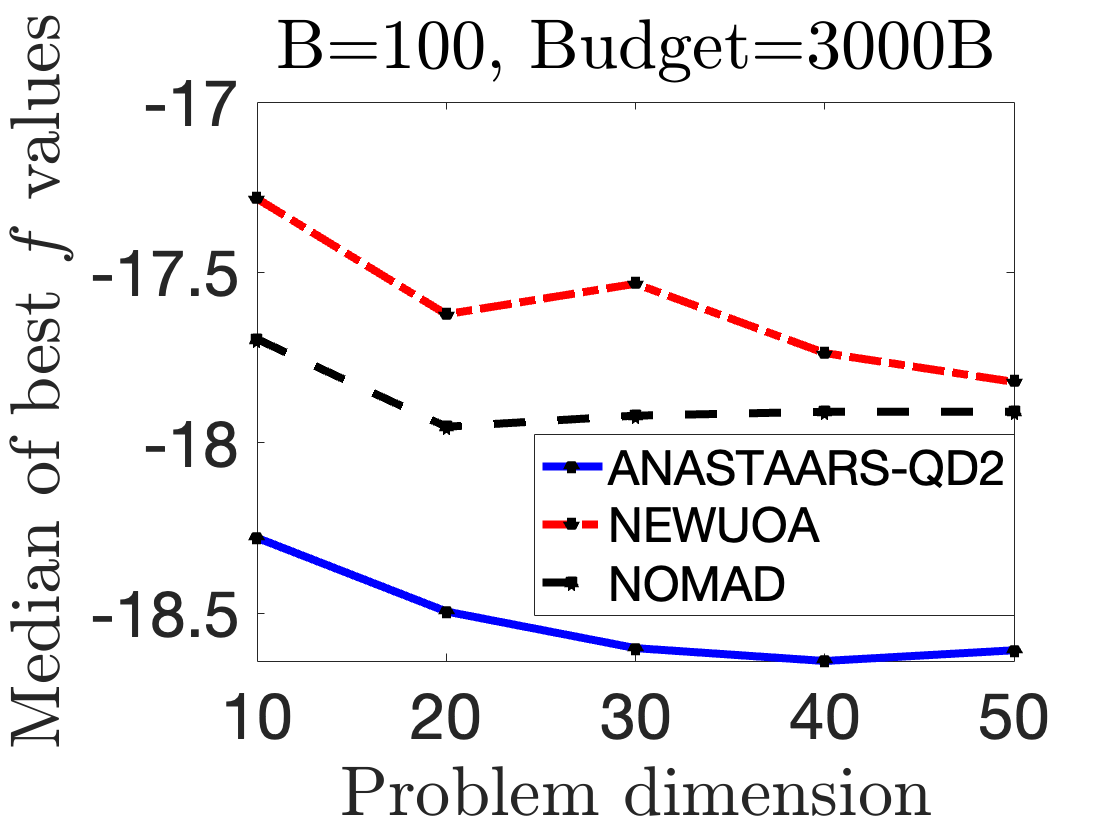}
\caption{\small{Chv\'atal scalability with 100 shots.}}
\label{Chavatal100_budgets}
\end{figure}

\begin{figure}[p!]
\centering
\includegraphics[scale=0.14]{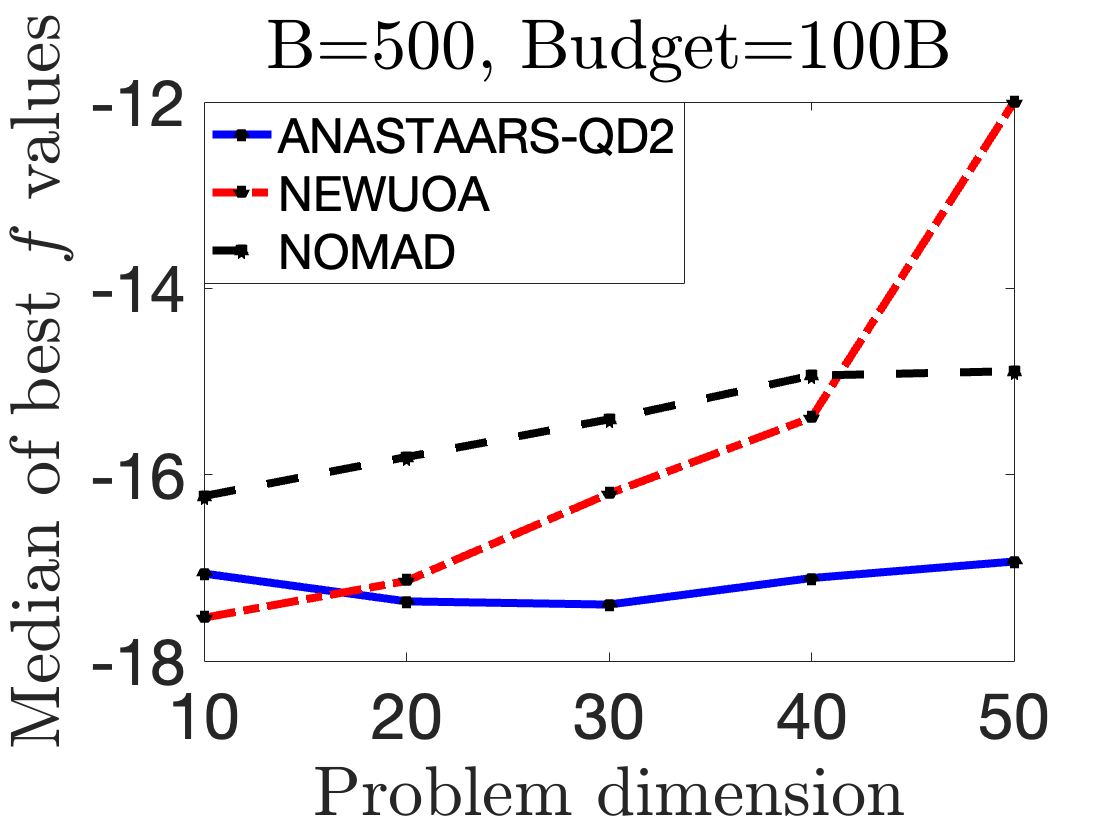}
\includegraphics[scale=0.14]{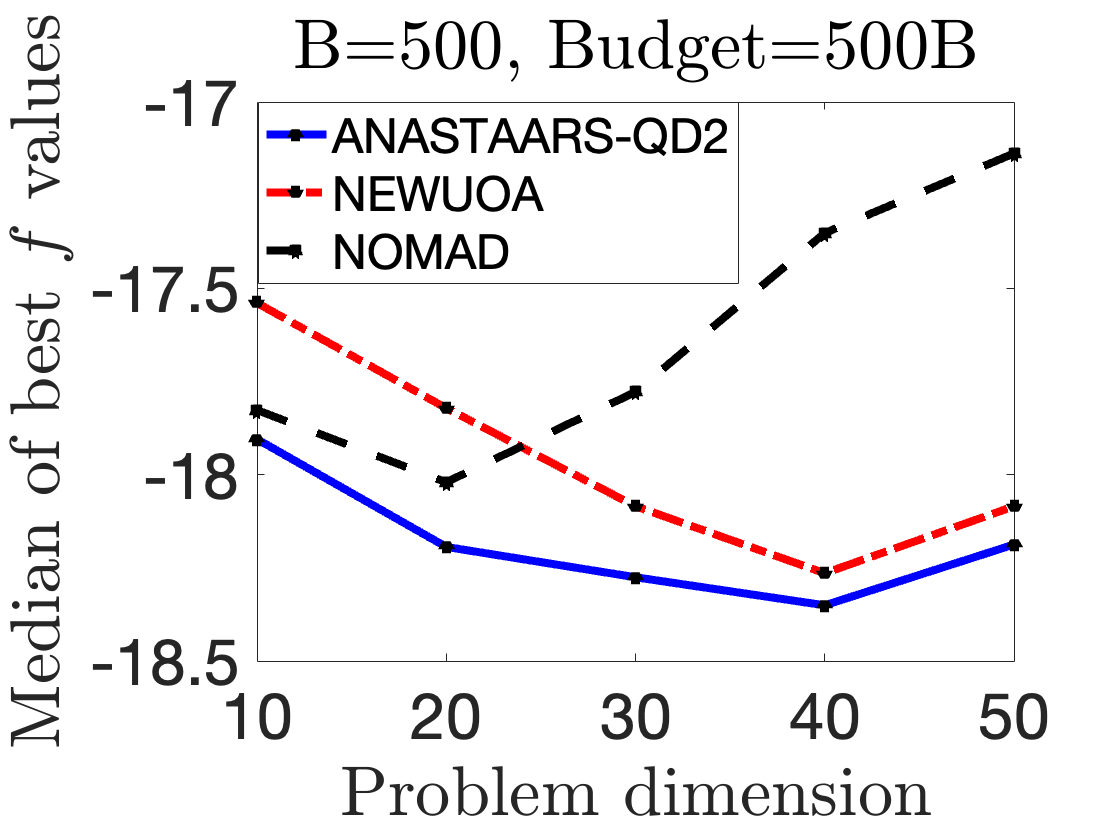}
\includegraphics[scale=0.14]{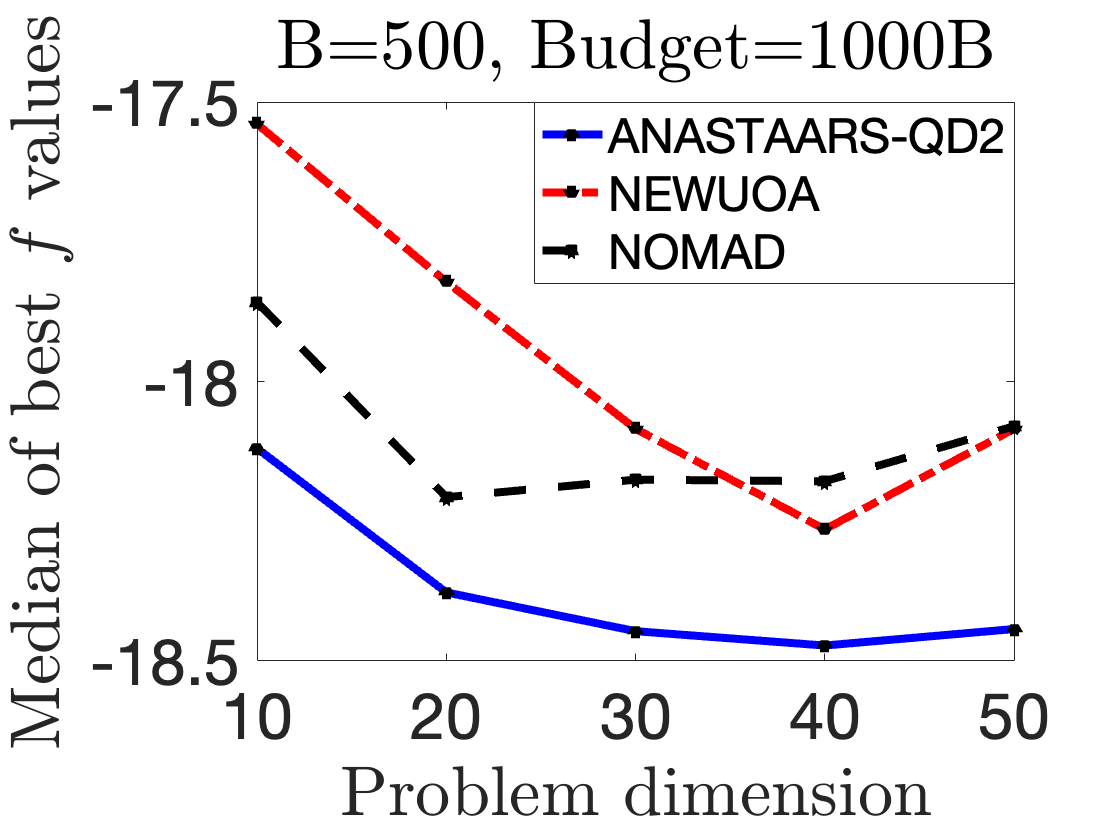}
\includegraphics[scale=0.14]{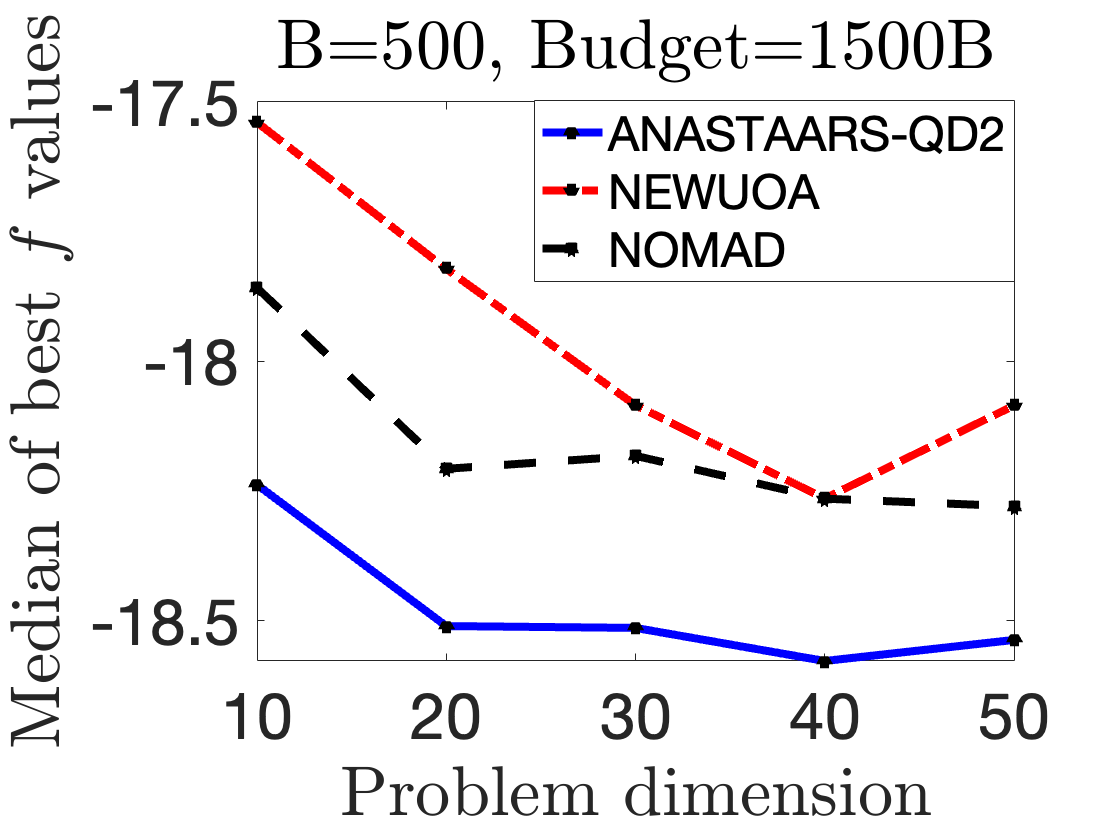}
\includegraphics[scale=0.14]{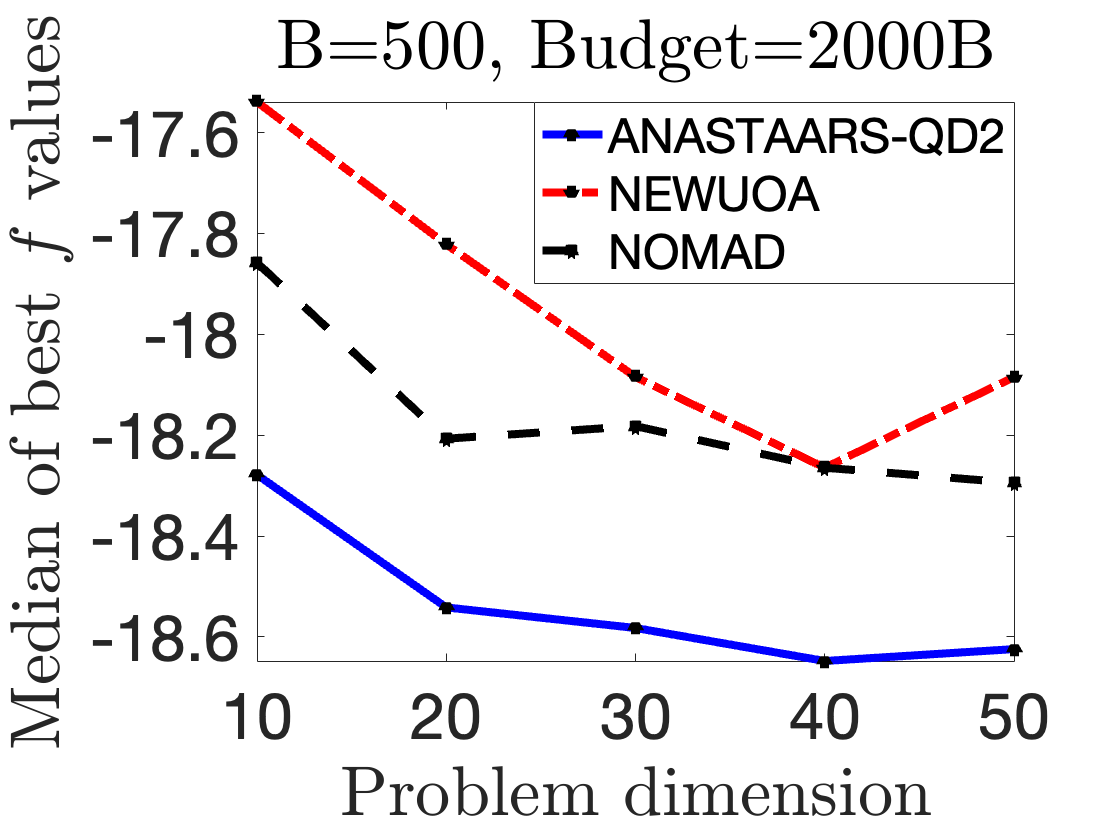}
\includegraphics[scale=0.14]{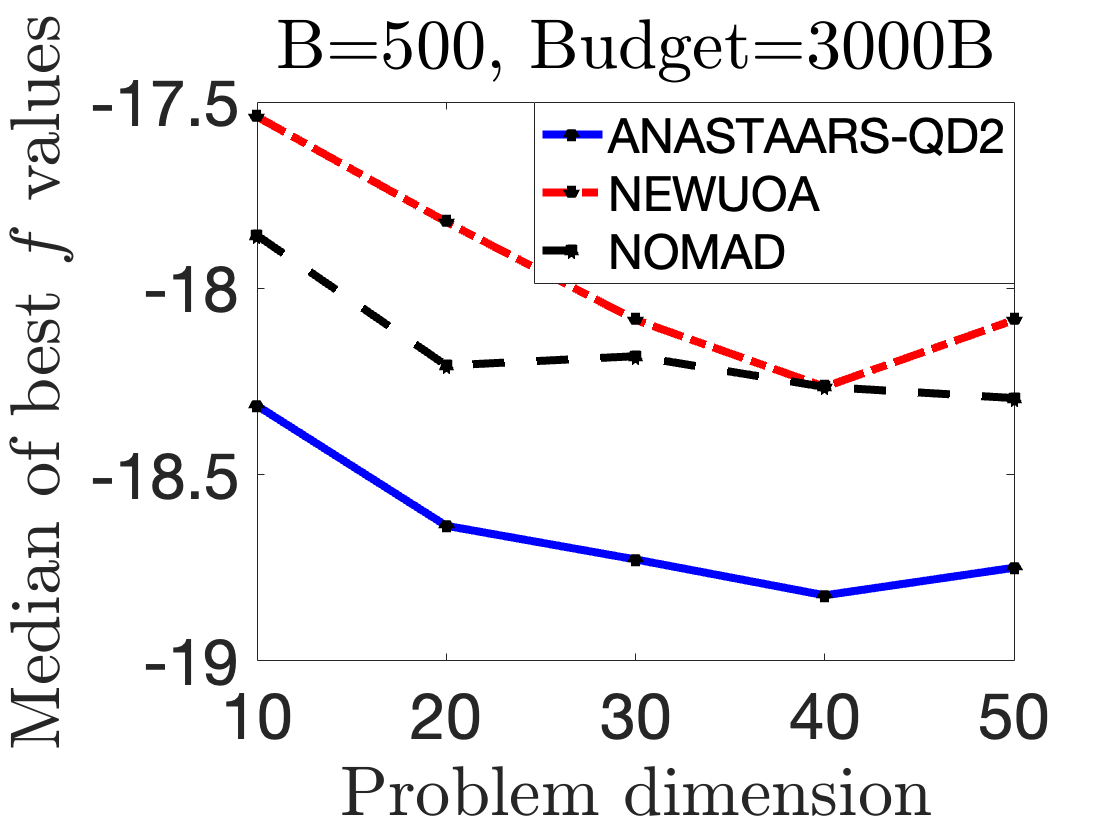}
\caption{\small{Chv\'atal scalability with 500 shots.}}
\label{Chavatal500_budgets}
\end{figure}

\begin{figure}[p!]
\centering
\includegraphics[scale=0.14]{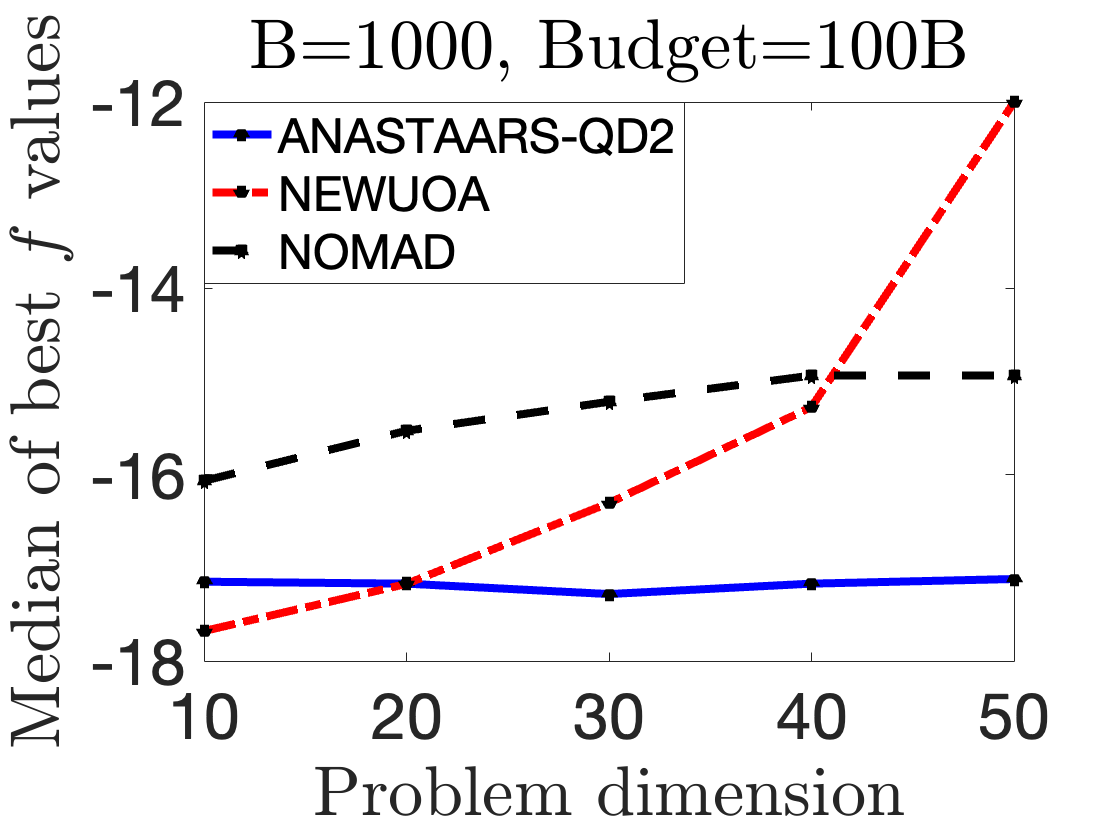}
\includegraphics[scale=0.14]{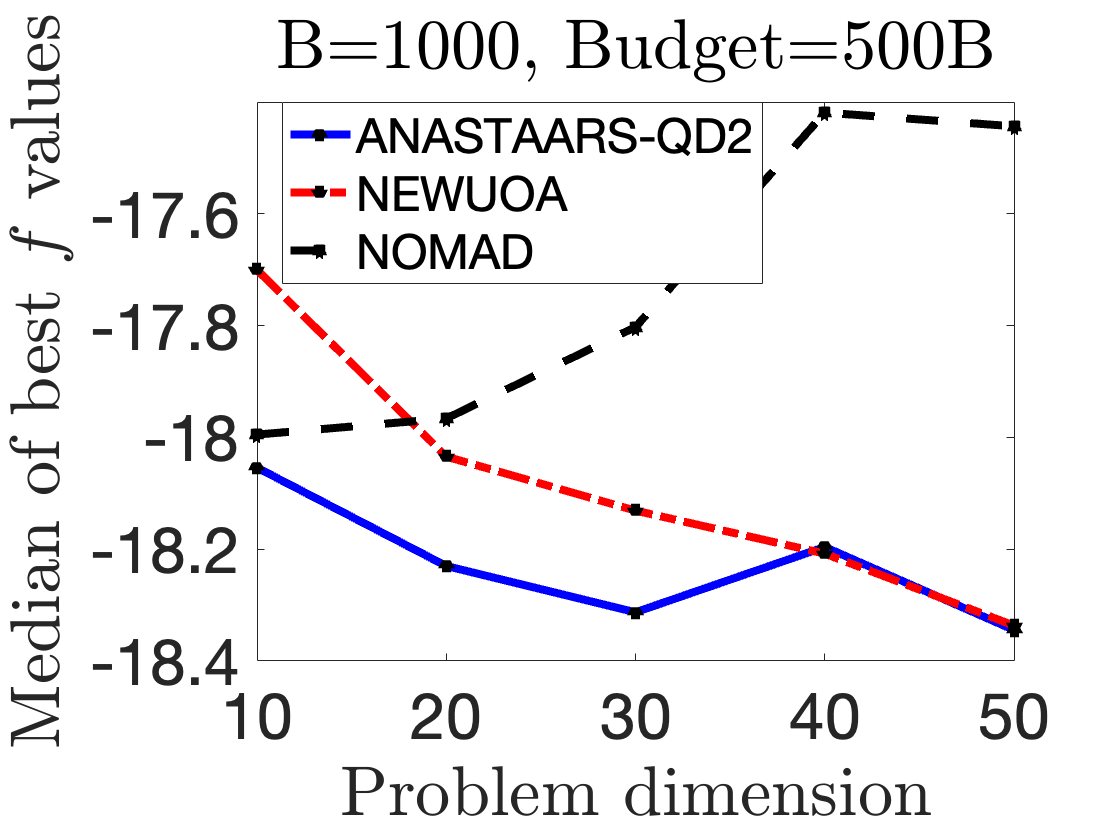}
\includegraphics[scale=0.14]{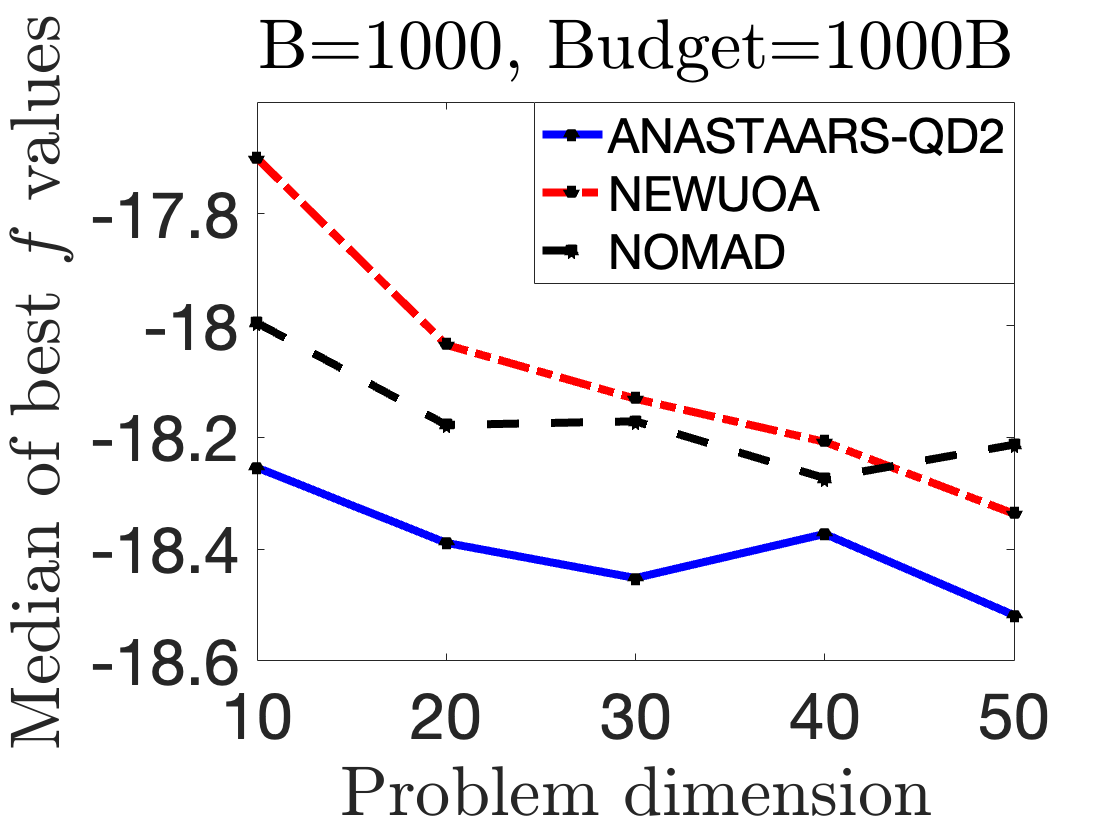}
\includegraphics[scale=0.14]{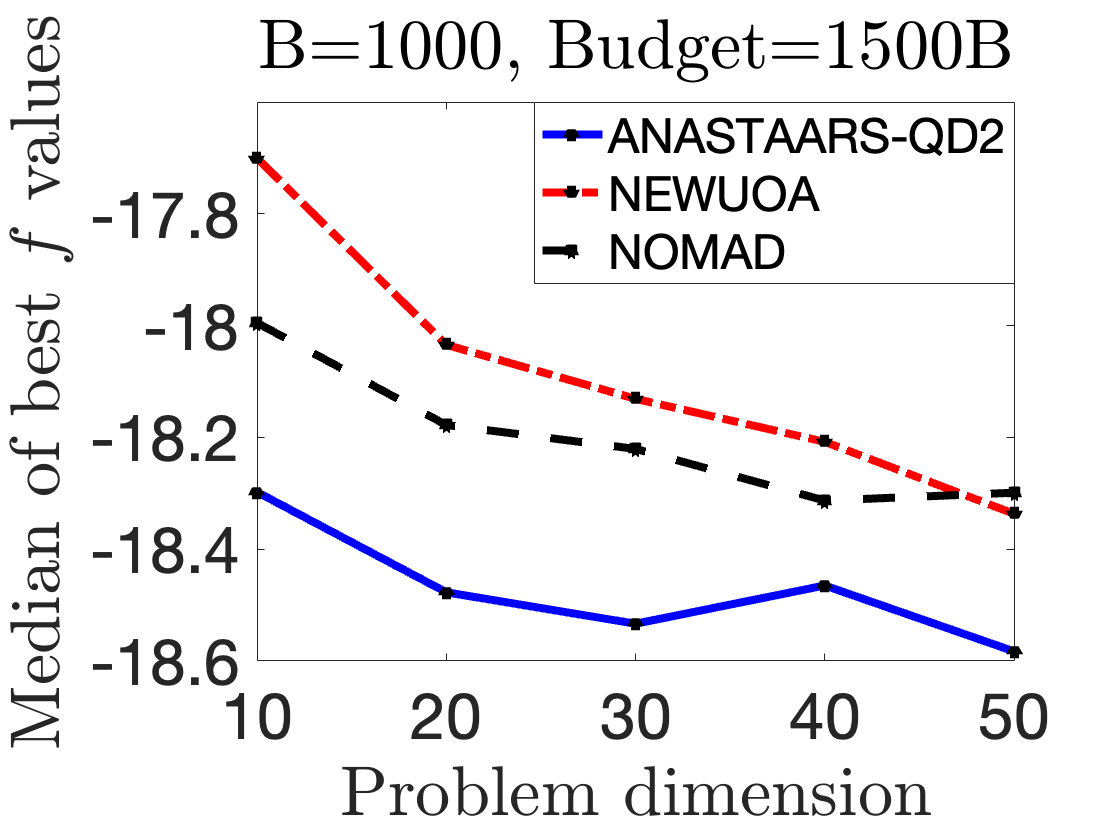}
\includegraphics[scale=0.14]{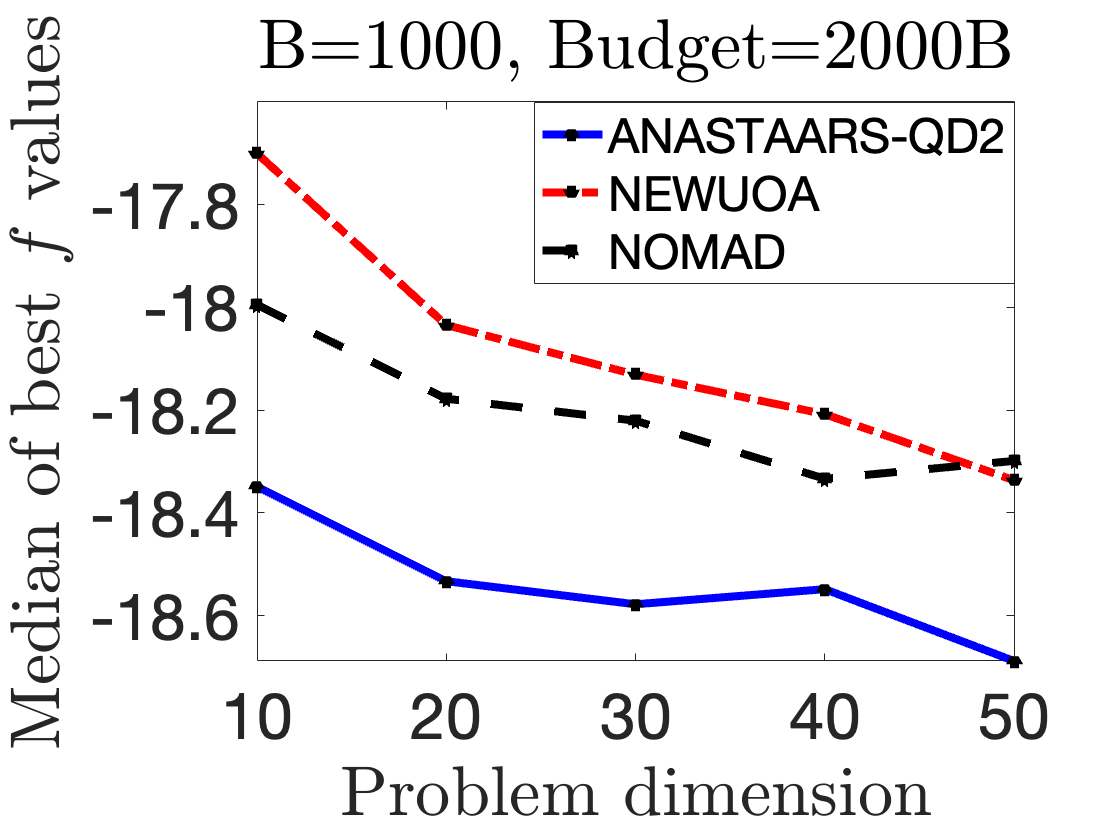}
\includegraphics[scale=0.14]{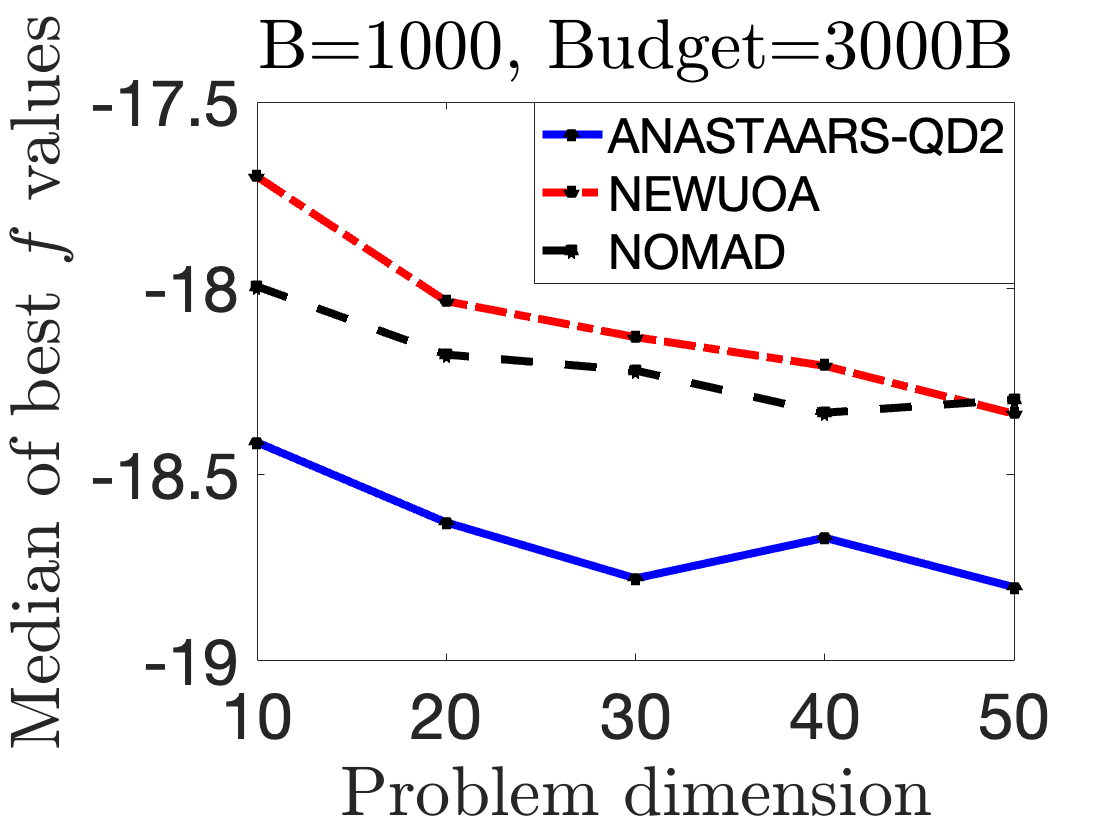}
\caption{\small{Chv\'atal scalability with 1000 shots.}}
\label{Chavatal1000_budgets}
\end{figure}

\begin{figure}[p!]
\centering
\includegraphics[scale=0.14]{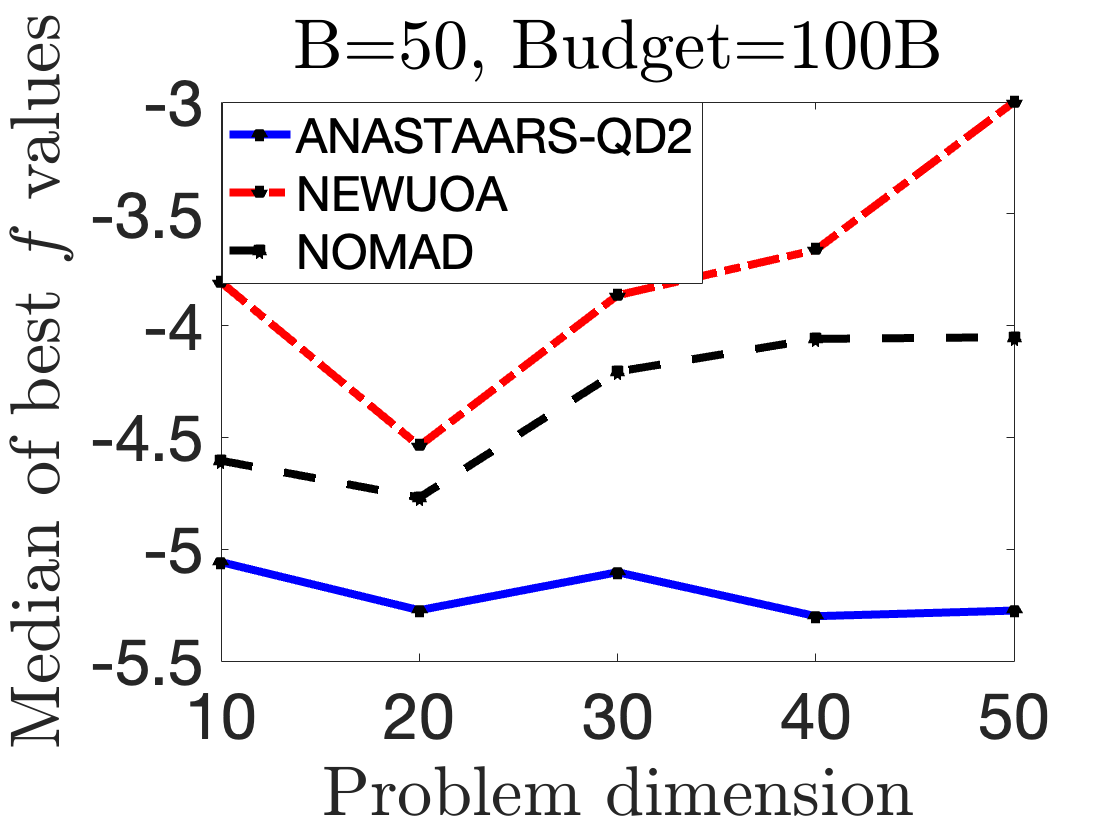}
\includegraphics[scale=0.14]{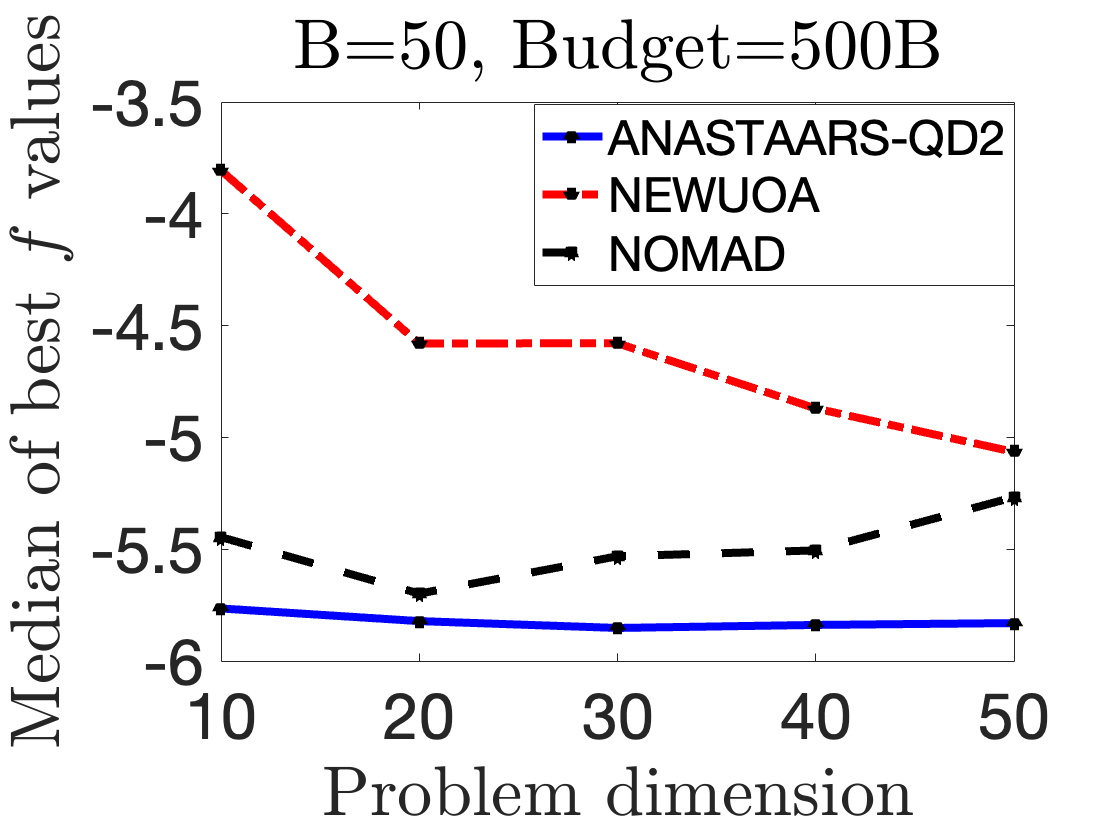}
\includegraphics[scale=0.14]{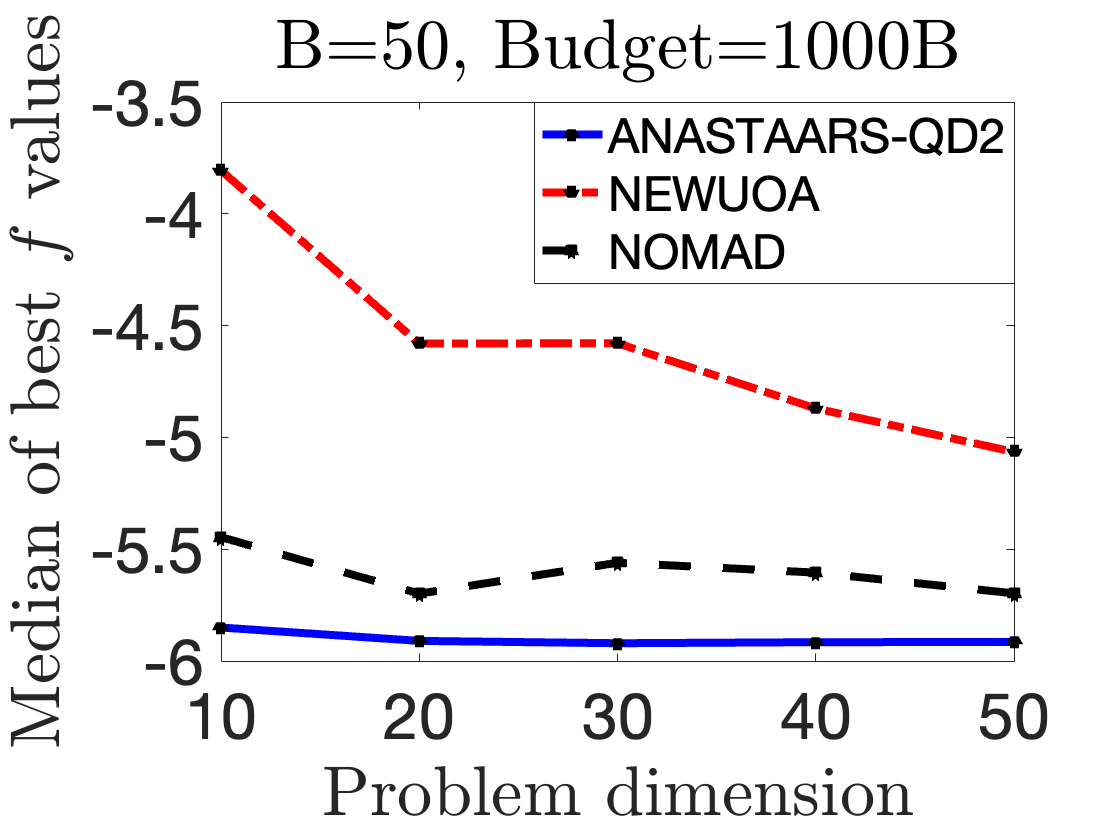}
\includegraphics[scale=0.14]{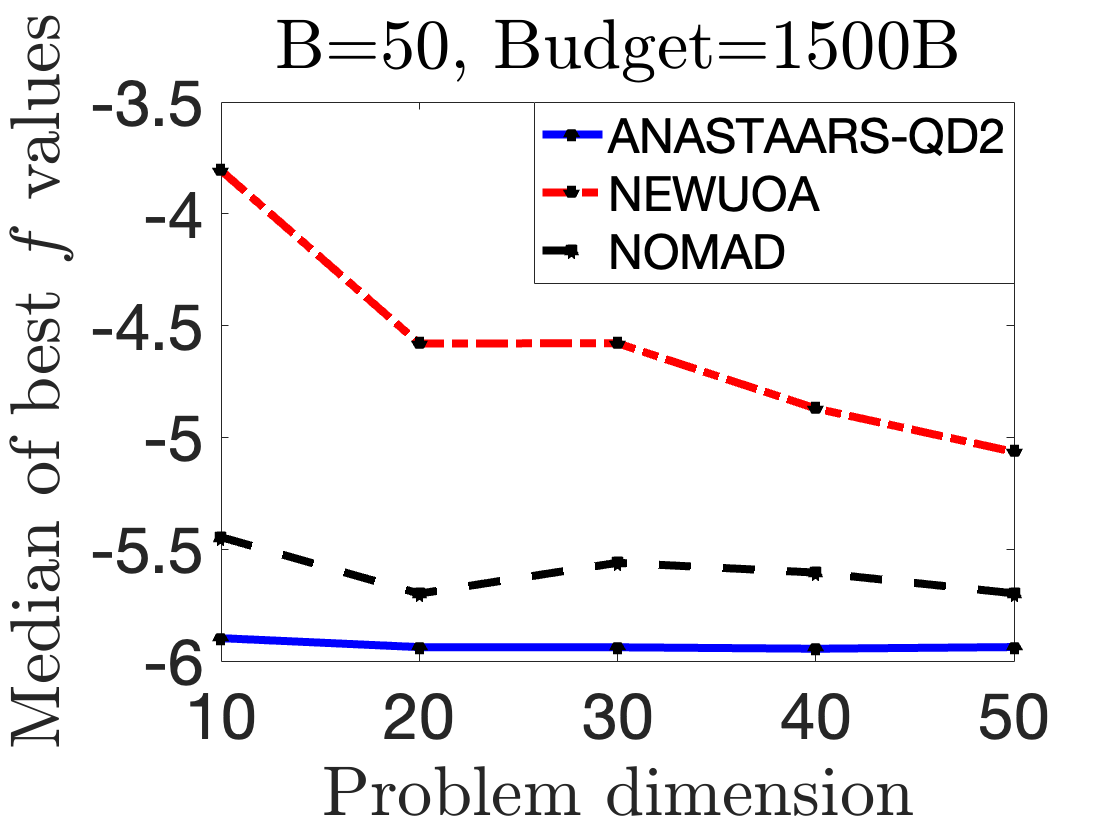}
\includegraphics[scale=0.14]{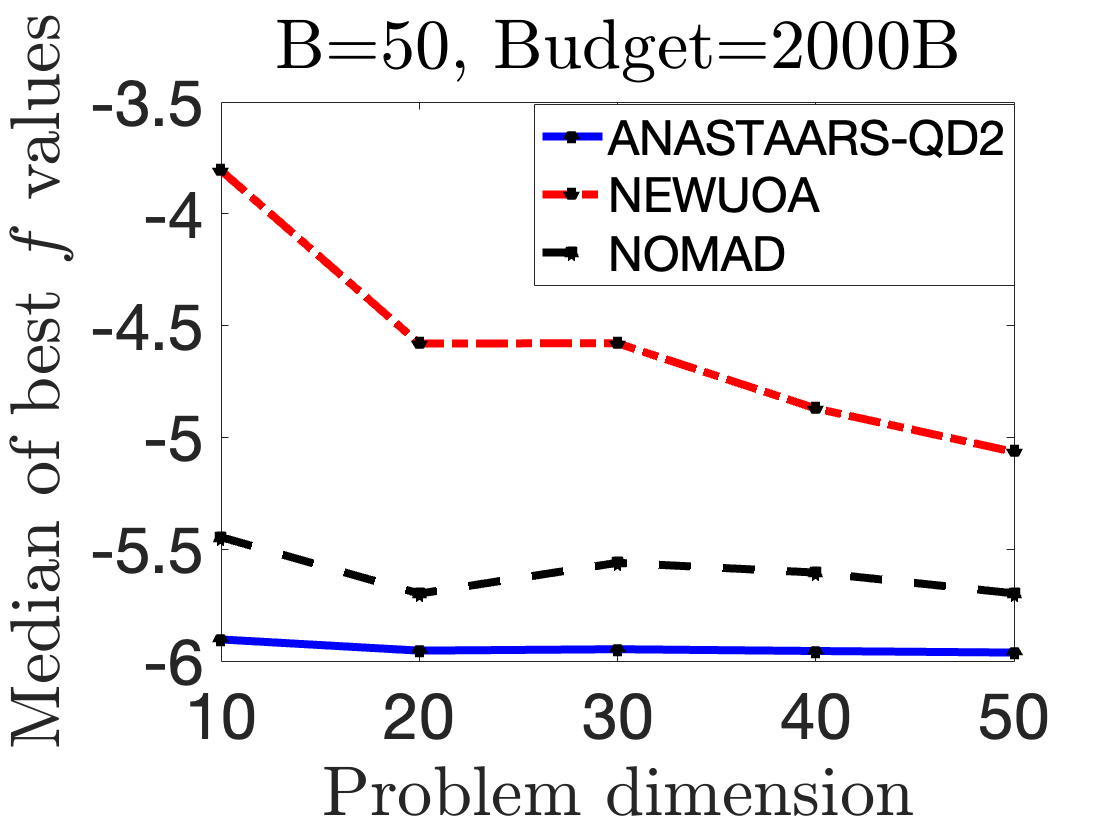}
\includegraphics[scale=0.14]{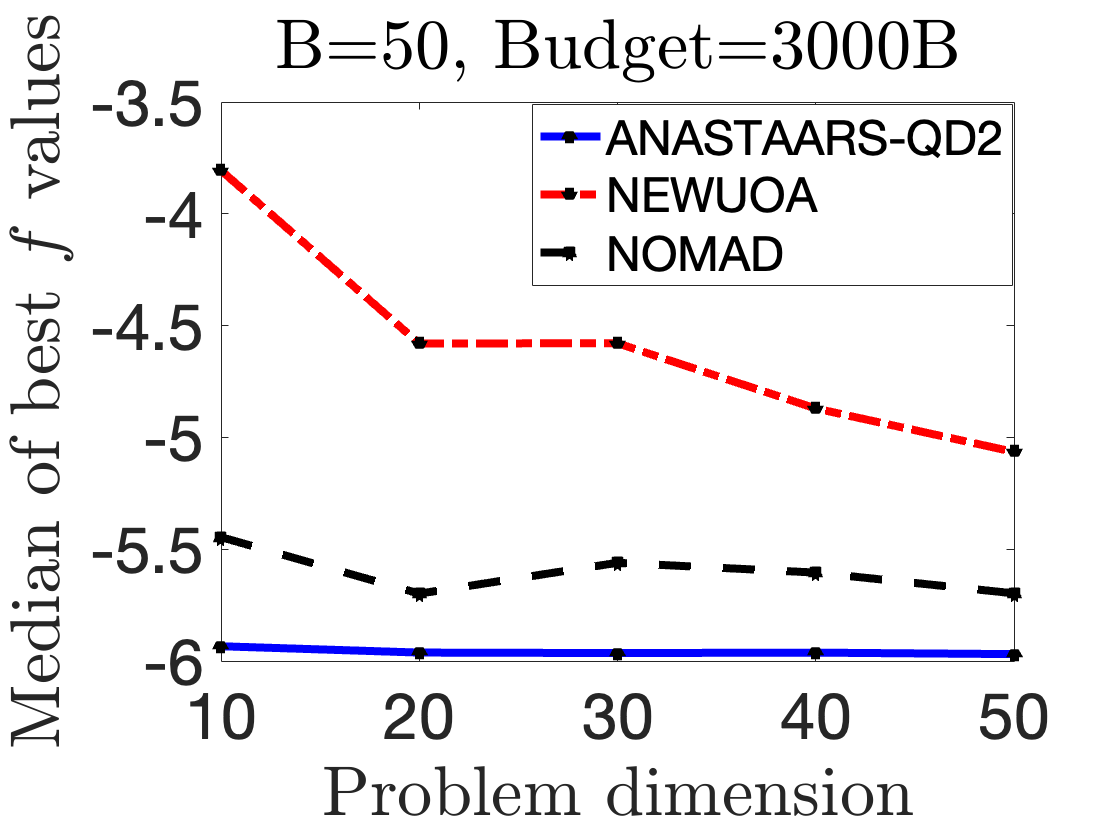}
\caption{\small{Toy scalability with 50 shots.}}
\label{Toyl50_budgets}
\end{figure}

\begin{figure}[p!]
\centering
\includegraphics[scale=0.14]{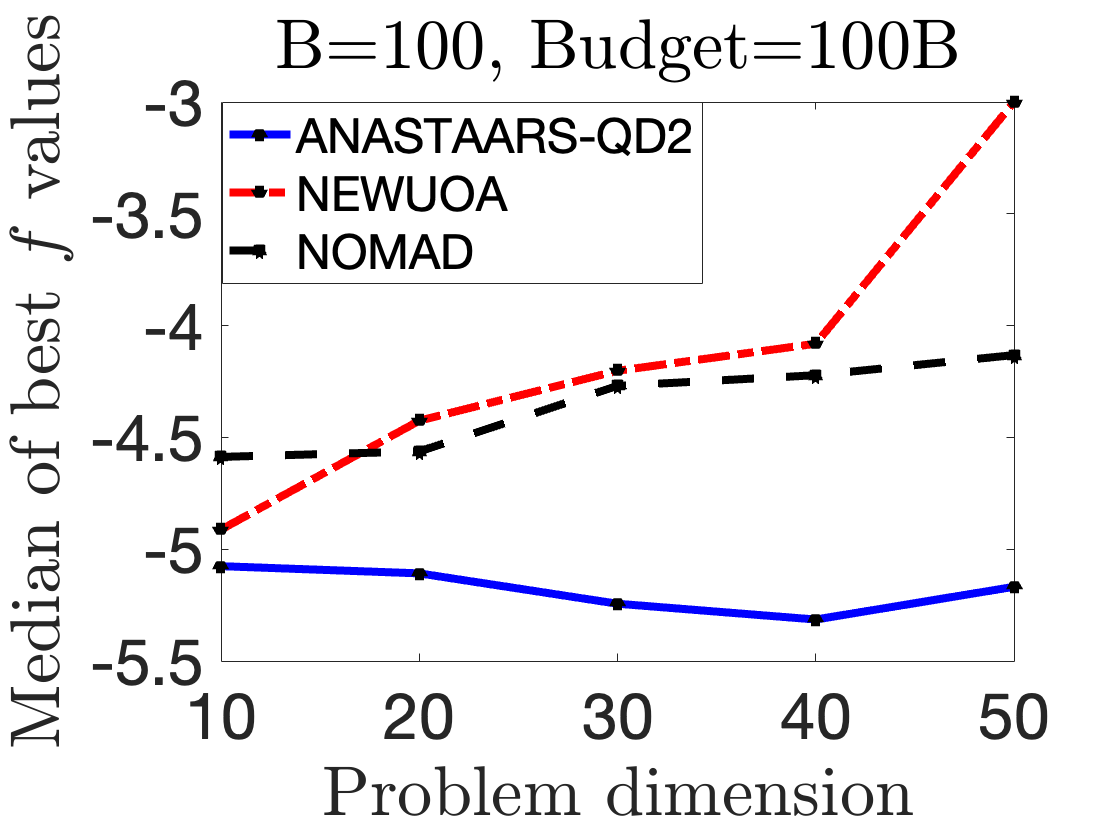}
\includegraphics[scale=0.14]{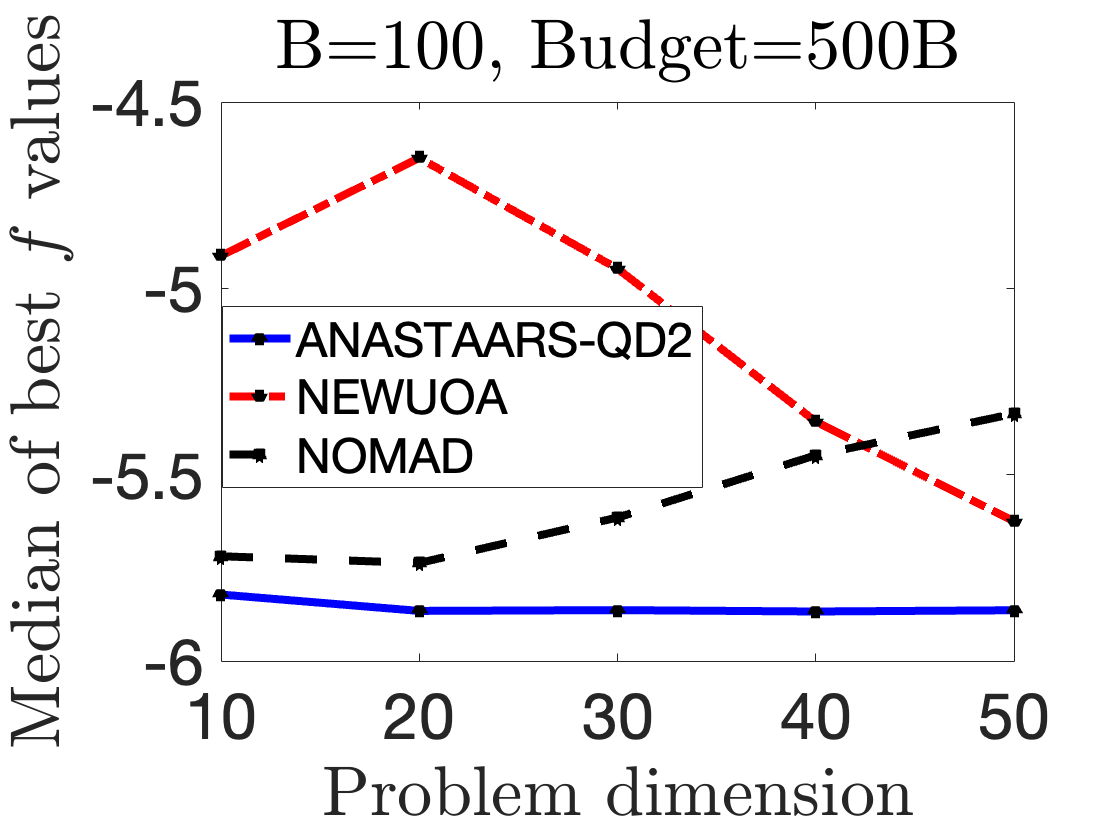}
\includegraphics[scale=0.14]{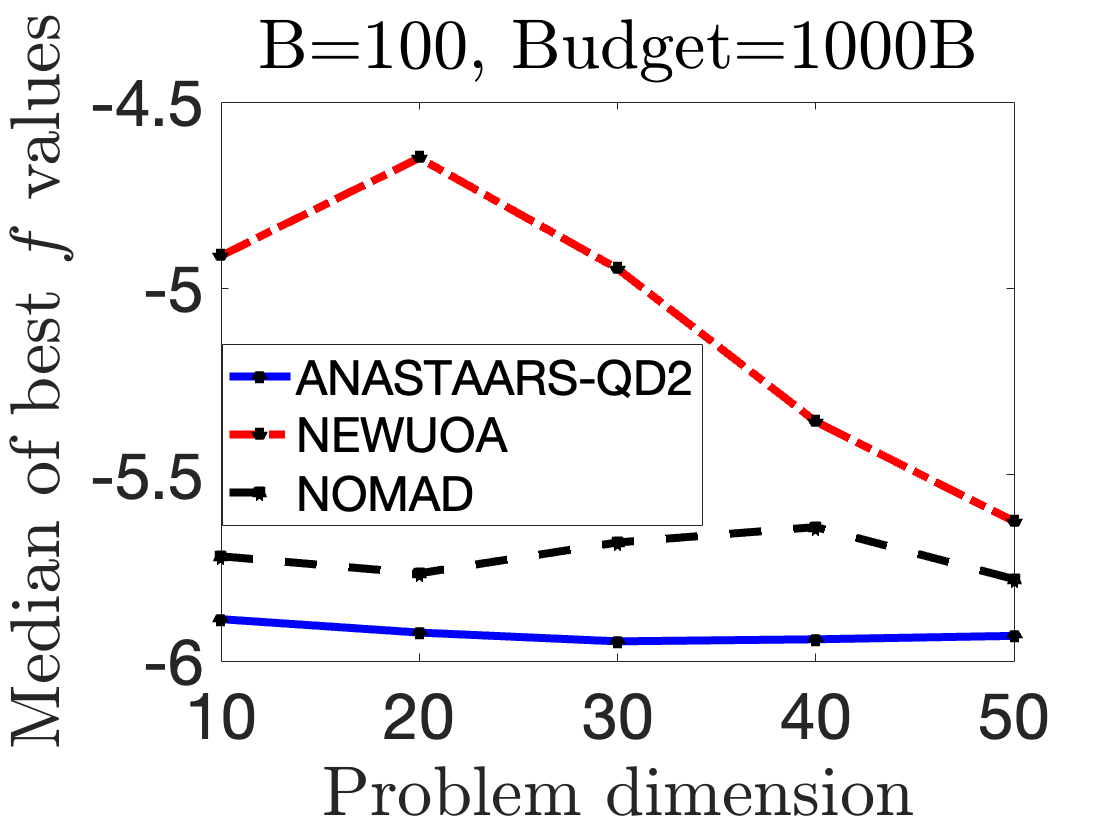}
\includegraphics[scale=0.14]{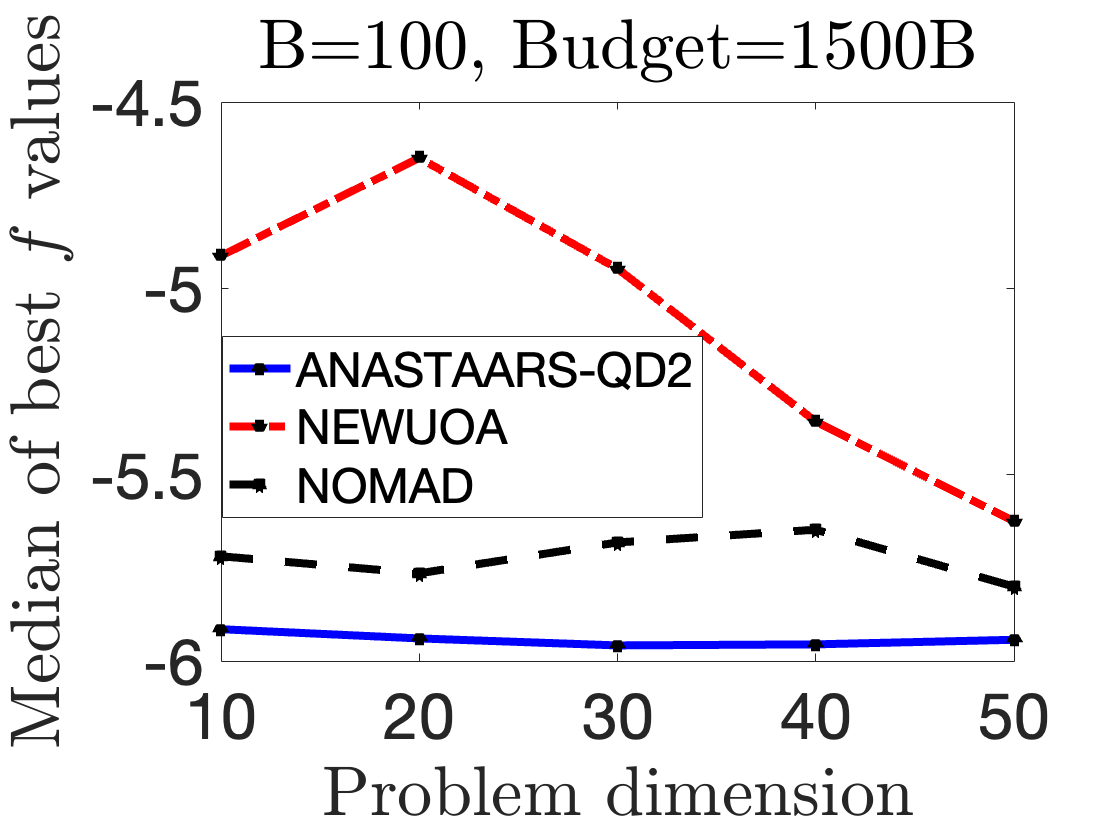}
\includegraphics[scale=0.14]{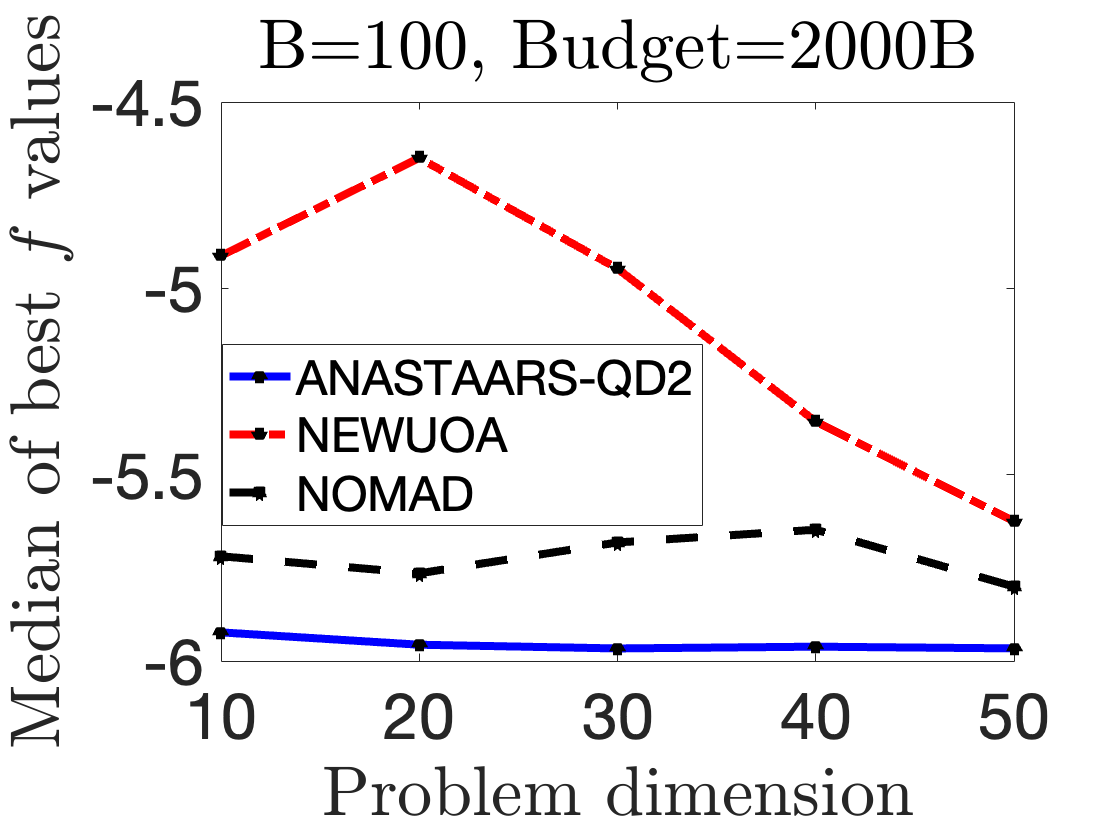}
\includegraphics[scale=0.14]{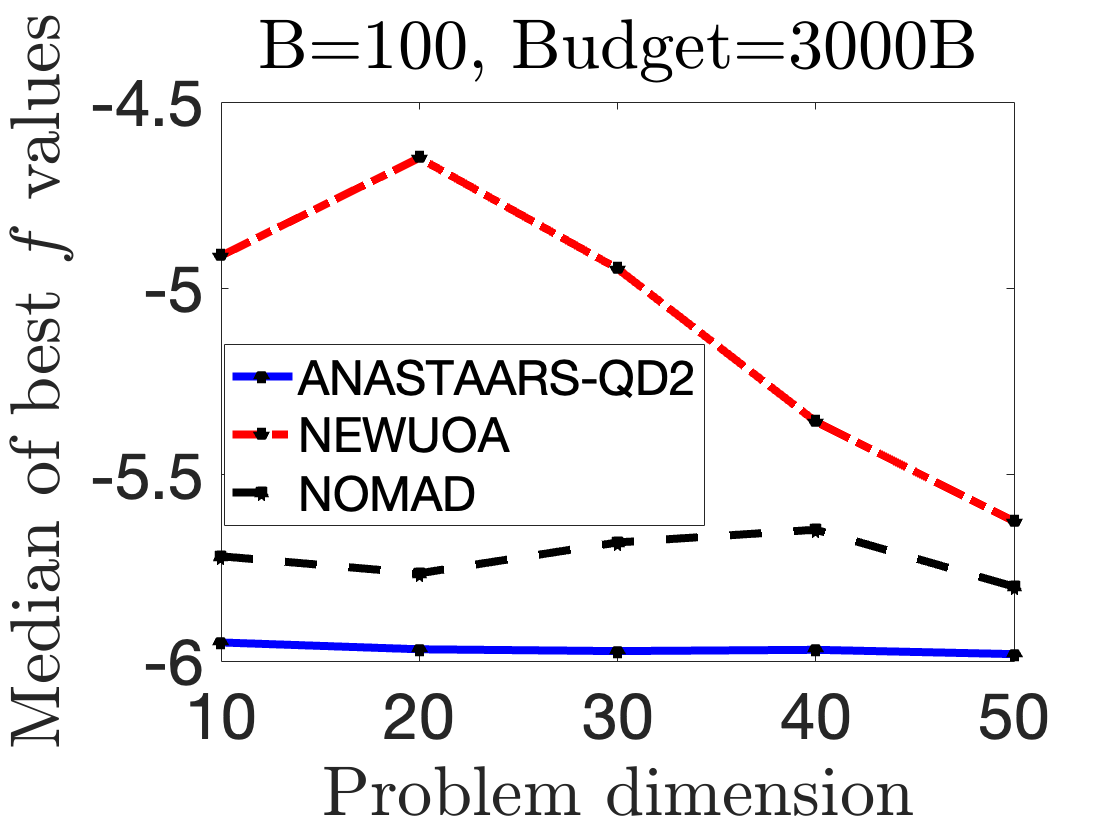}
\caption{\small{Toy scalability with 100 shots.}}
\label{Toy100_budgets}
\end{figure}

\begin{figure}[p!]
\centering
\includegraphics[scale=0.14]{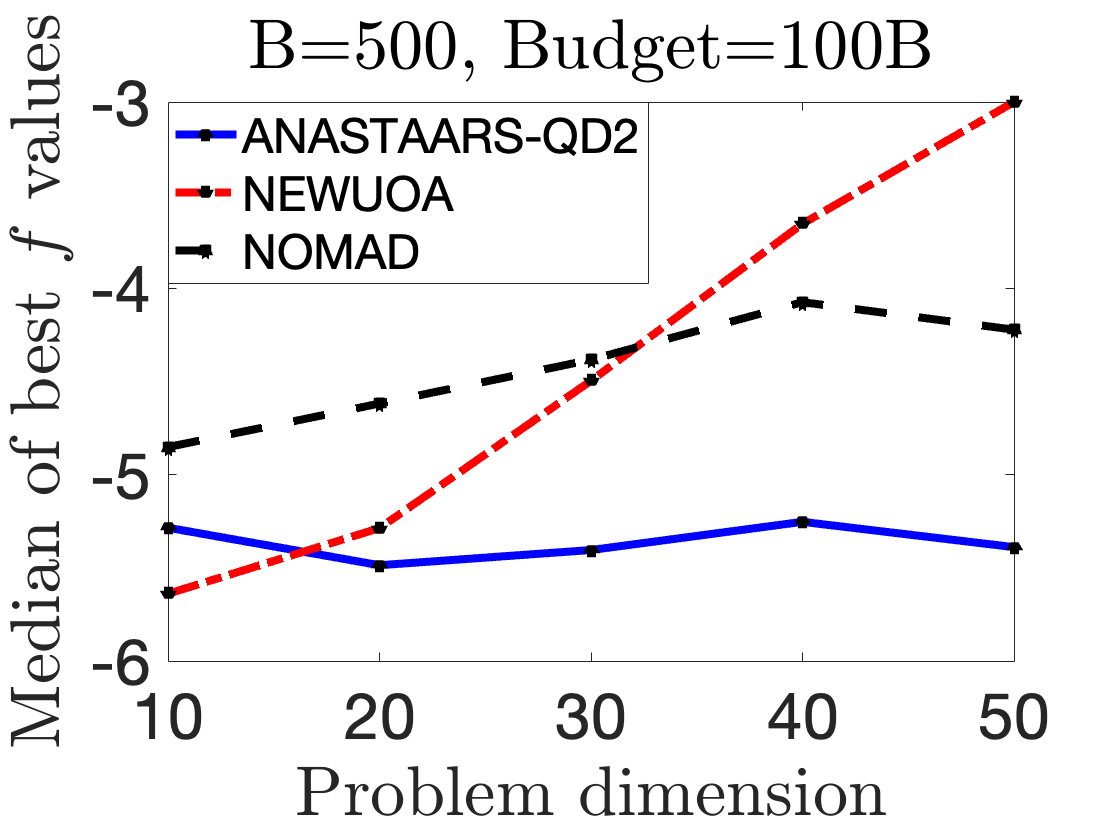}
\includegraphics[scale=0.14]{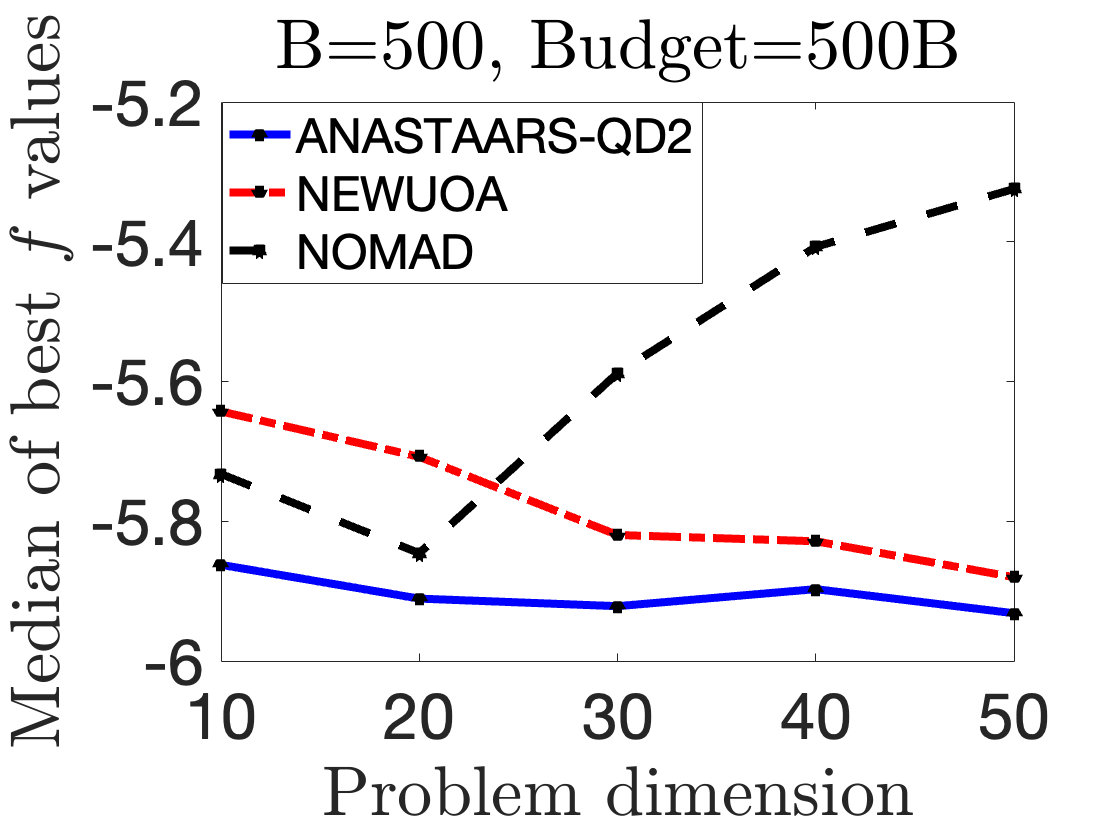}
\includegraphics[scale=0.14]{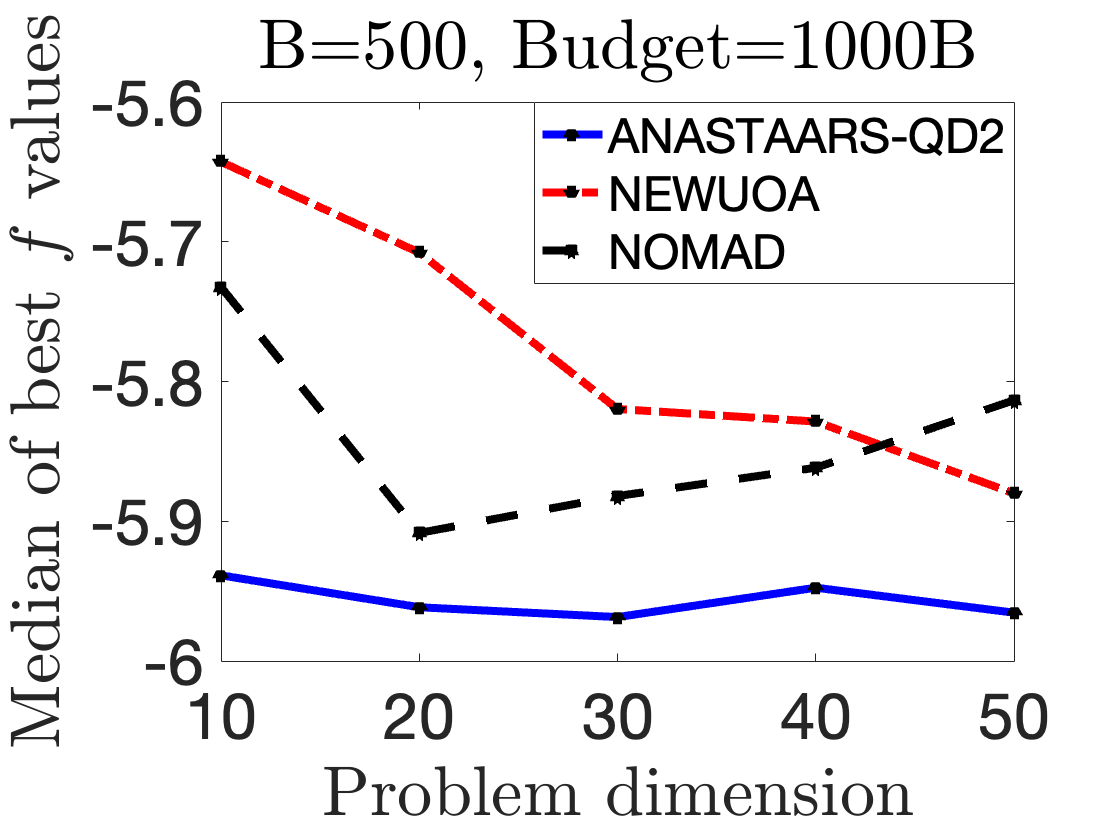}
\includegraphics[scale=0.14]{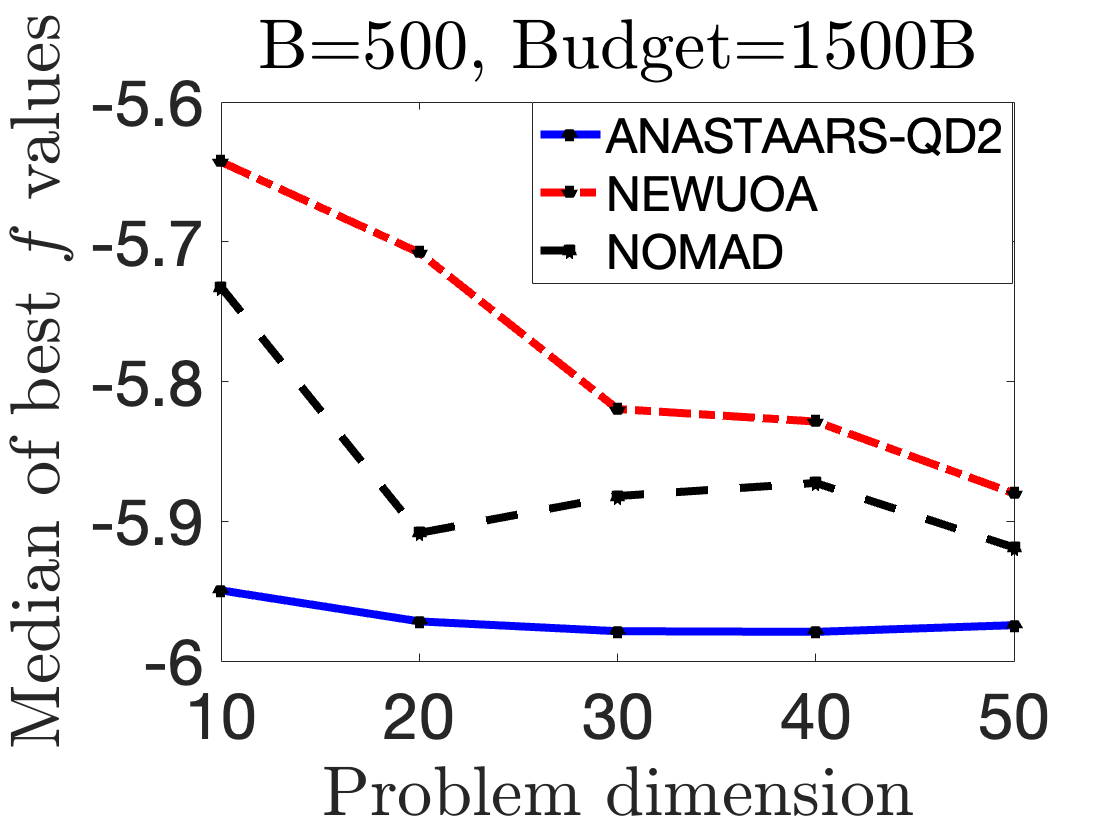}
\includegraphics[scale=0.14]{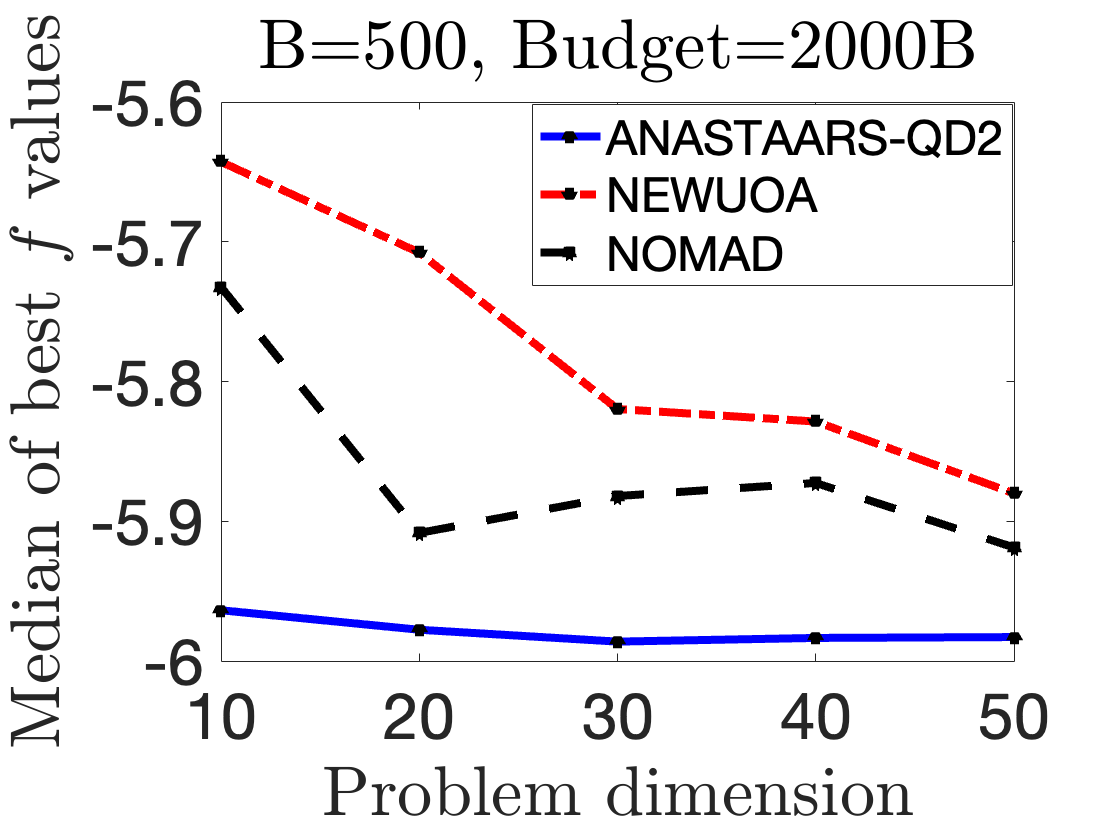}
\includegraphics[scale=0.14]{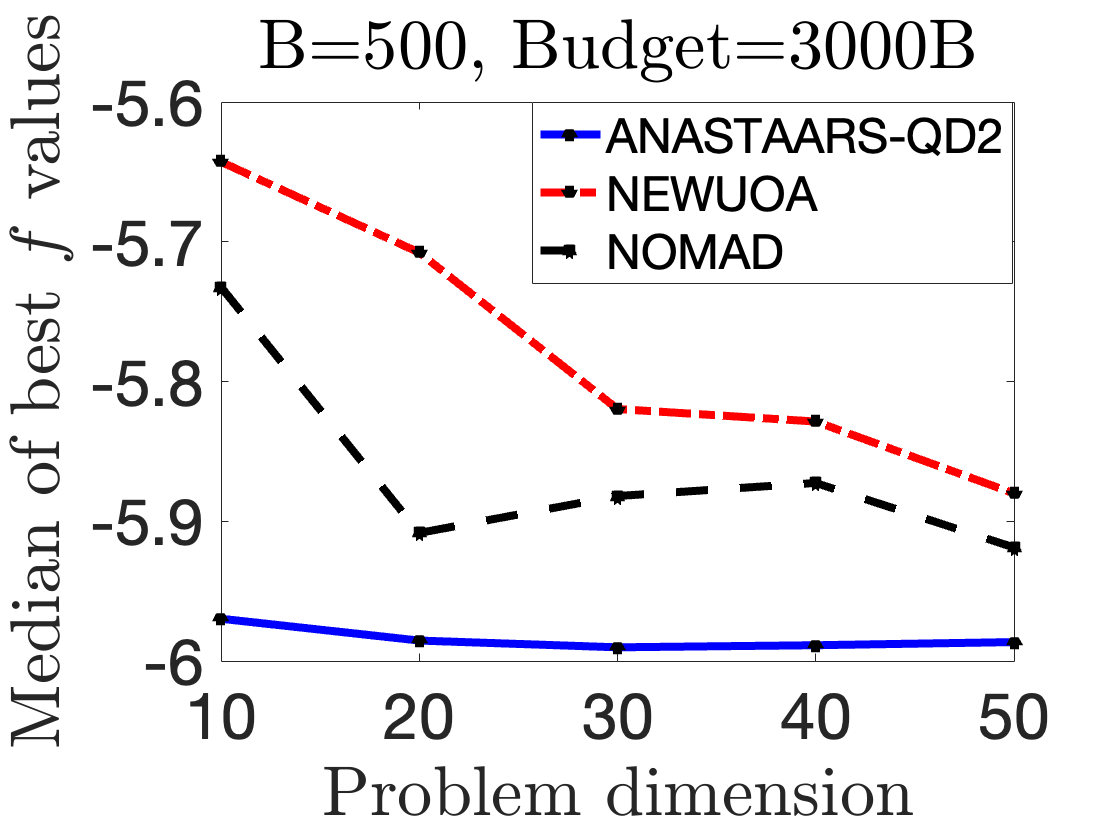}
\caption{\small{Toy scalability with 500 shots.}}
\label{Toy500_budgets}
\end{figure}

\begin{figure}[p!]
\centering
\includegraphics[scale=0.14]{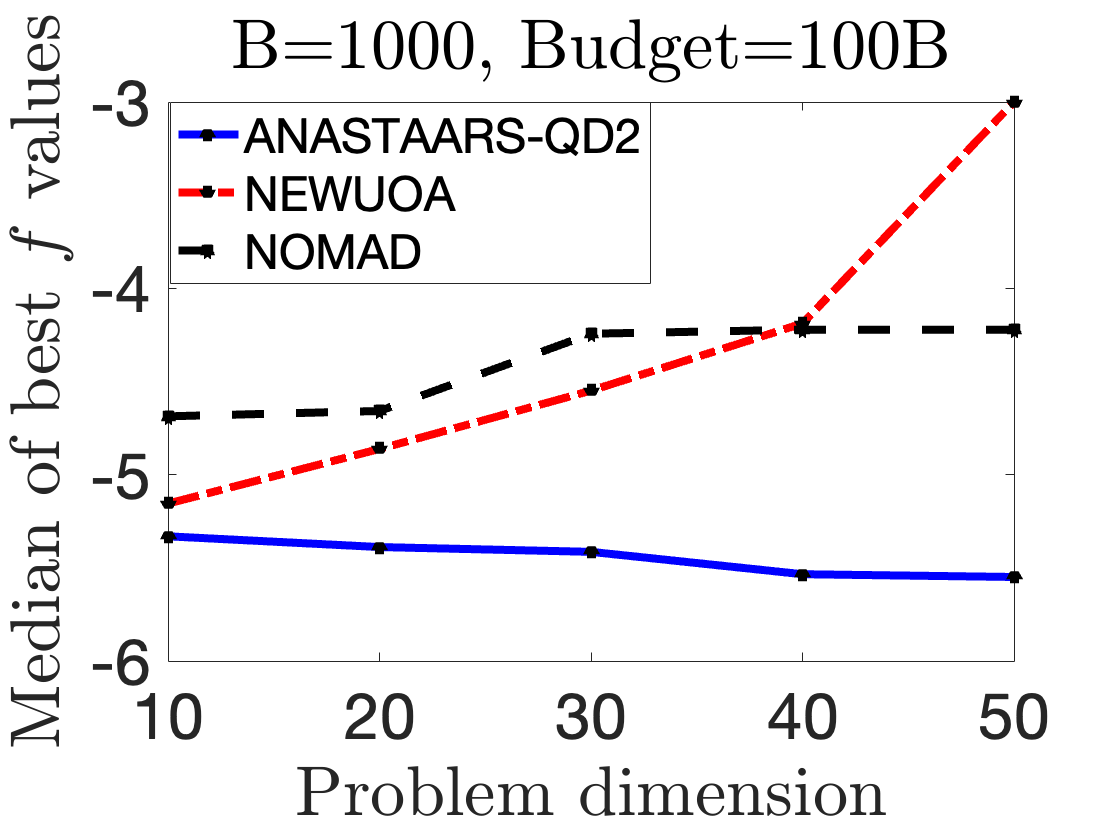}
\includegraphics[scale=0.14]{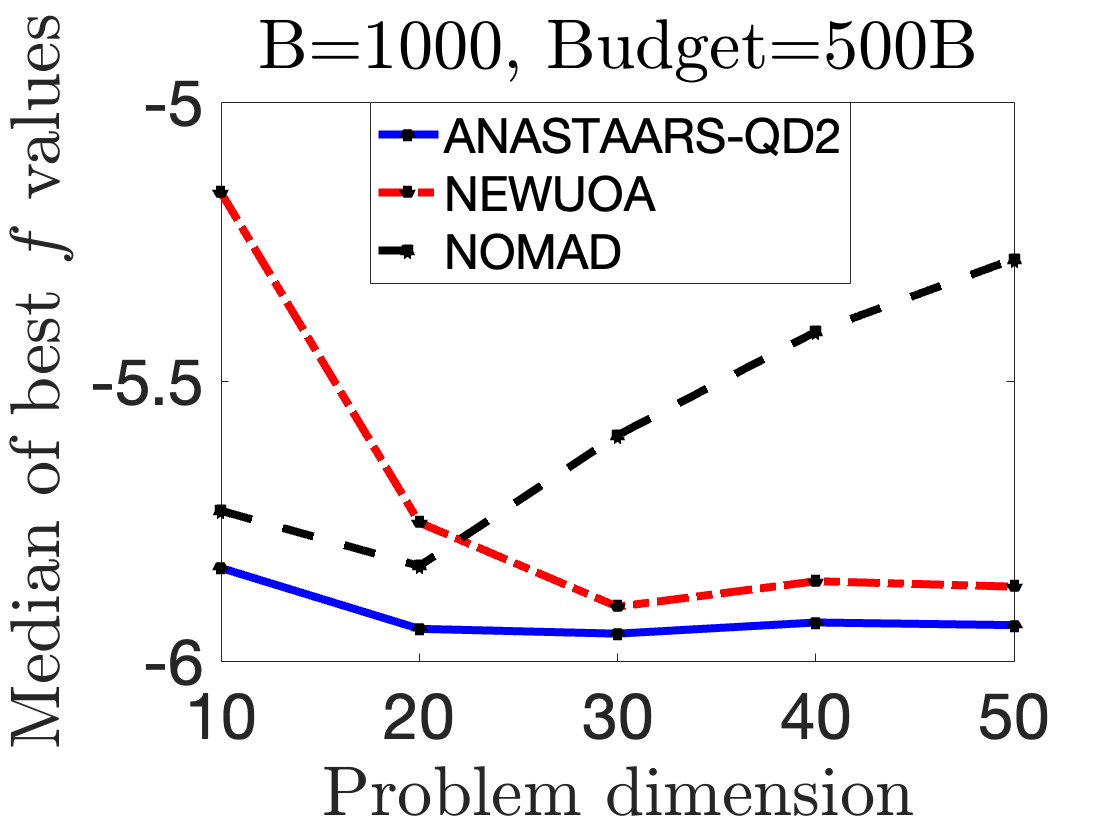}
\includegraphics[scale=0.14]{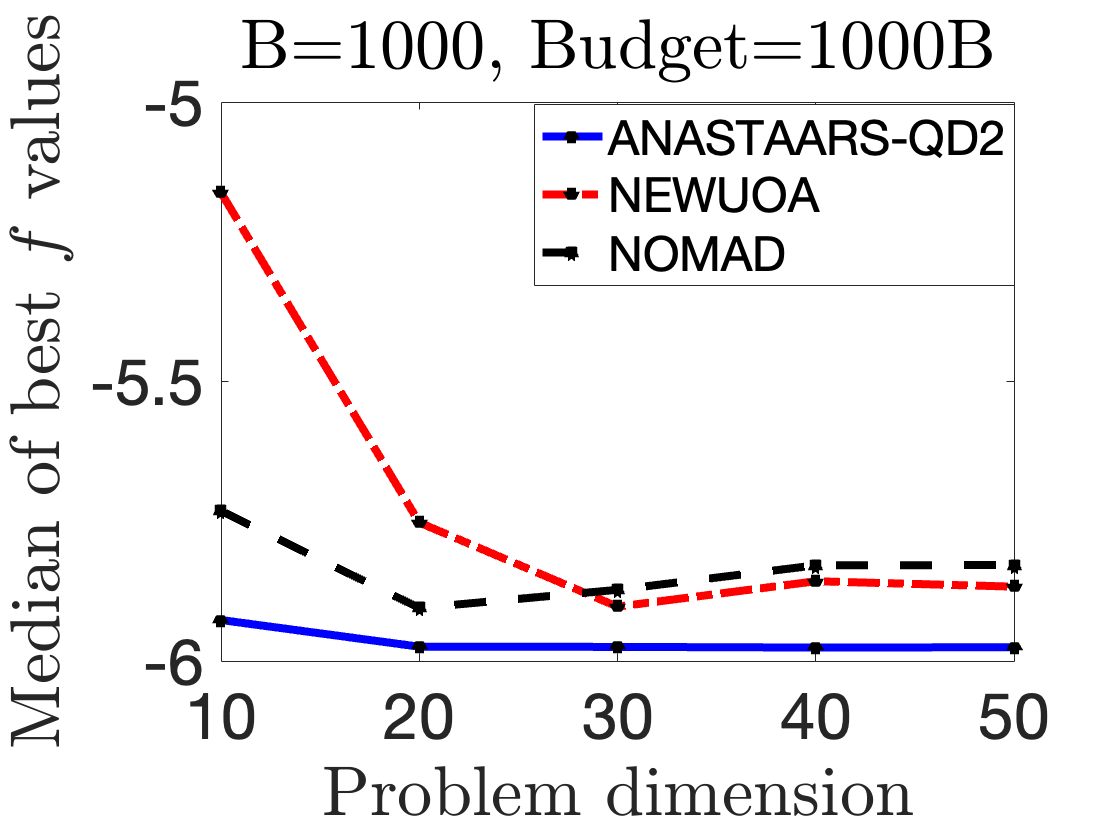}
\includegraphics[scale=0.14]{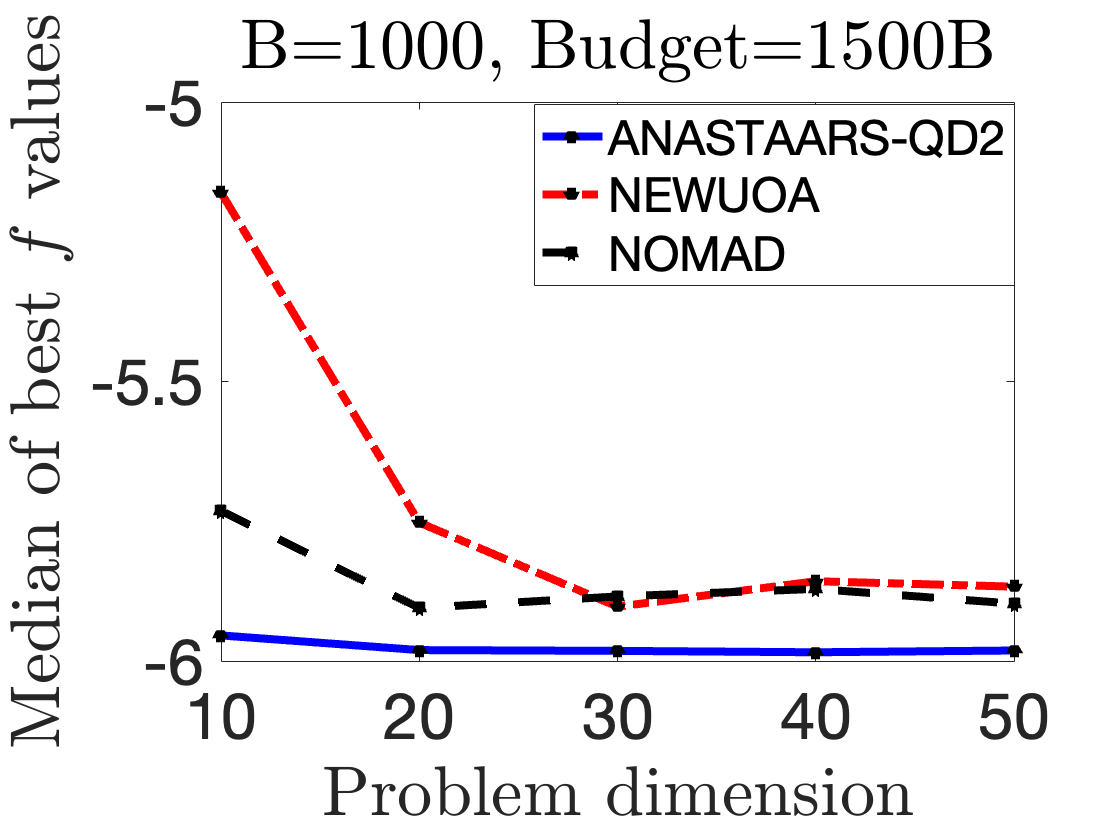}
\includegraphics[scale=0.14]{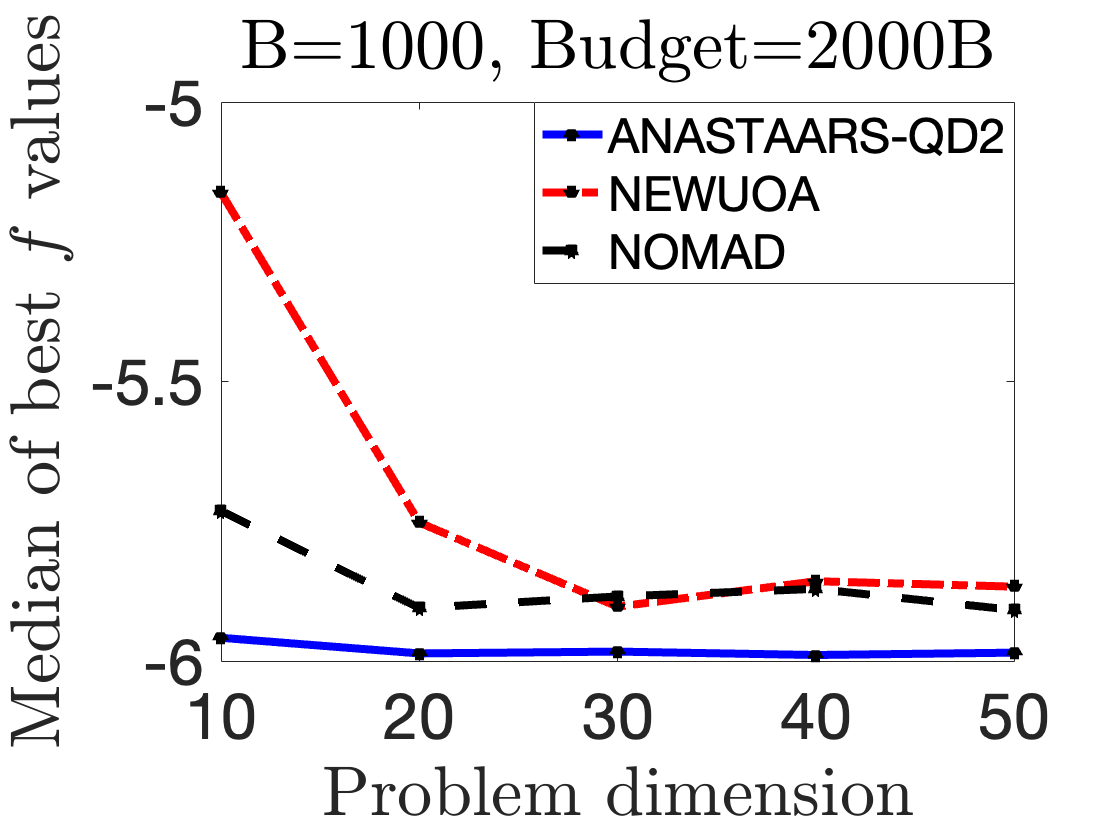}
\includegraphics[scale=0.14]{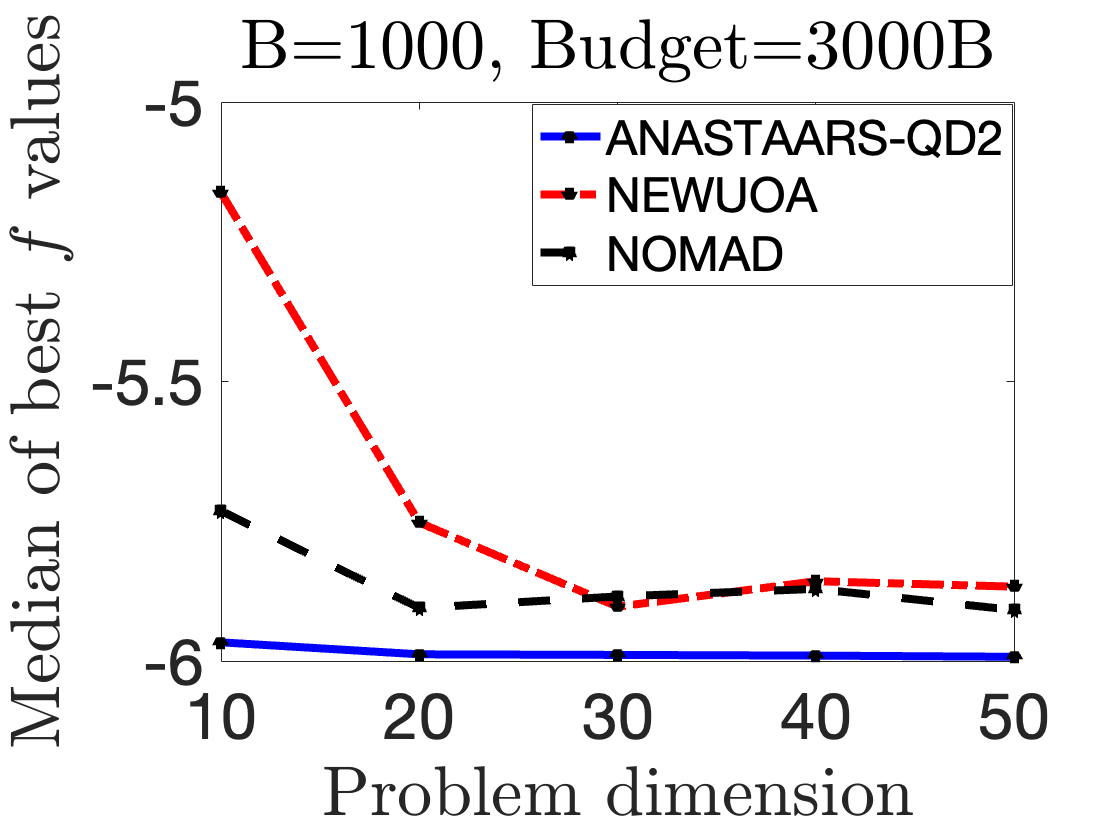}
\caption{\small{Toy scalability with 1,000 shots.}}
\label{Toy1000_budgets}
\end{figure}

Our next set of experiments is aimed at testing optimizer scalability as the number~$p$ of QAOA layers increases. We show the behavior for $p=5,\ldots, 25$, corresponding to dimensions $d=10, \ldots, 50$. Following an initial downselection based on the scalability of the various methods, we report the performance of {\sf NEWUOA} and {\sf NOMAD} in addition to that of  {\sf ANASTAARS-QD2}.
For different numbers of shots, Figures~\ref{Chavatal50_budgets}--\ref{Chavatal1000_budgets} show the median performance for the Chv\'atal graph, and Figures~\ref{Toyl50_budgets}--\ref{Toy1000_budgets} show the median performance for the toy graph.

In the smallest budget cases (upper left in each figure), the lead of {\sf ANASTAARS-QD2} increases as the dimension increases. In the largest budget cases (lower left in each figure), {\sf ANASTAARS-QD2} maintains its lead over the other two optimizers as the dimension grows. 

The results suggest that {\sf ANASTAARS-QD2} offers a way to consider larger-scale QAOA problems than are currently studied. By displaying consistent behavior across different numbers of shots used in the cost function estimates, the results also suggest that {\sf ANASTAARS-QD2} can be reliably deployed in different shot regimes.


%--------------------------Toy----------------------


\clearpage
\bibliography{dzahini-bibliography}
\bibliographystyle{abbrvnat}

%-------------------------------------------------------------

% This can be deleted in the version that appears, it is only needed for the Argonne PANDA submission and ArXiv submission:
\vfill
\framebox{\parbox{.90\linewidth}{\scriptsize The submitted manuscript has been
created by UChicago Argonne, LLC, Operator of Argonne National Laboratory
(``Argonne''). Argonne, a U.S.\ Department of Energy Office of Science
laboratory, is operated under Contract No.\ DE-AC02-06CH11357.  The U.S.\
Government retains for itself, and others acting on its behalf, a paid-up
nonexclusive, irrevocable worldwide license in said article to reproduce,
prepare derivative works, distribute copies to the public, and perform publicly
and display publicly, by or on behalf of the Government.  The Department of
Energy will provide public access to these results of federally sponsored
research in accordance with the DOE Public Access Plan
\url{http://energy.gov/downloads/doe-public-access-plan}.}}
\end{document}